%% file: paper.tex
\documentclass[breakable]{bytedance_seed}

\usepackage[toc,page,header]{appendix}

\usepackage{microtype}
\usepackage{graphicx}
\usepackage{subcaption}
\usepackage{booktabs} % for professional tables

\usepackage{hyperref}

\usepackage{amsmath}
\usepackage{amssymb}
\usepackage{mathtools}
\usepackage{amsthm}

\theoremstyle{plain}

\theoremstyle{definition}

\theoremstyle{remark}

\def\ours{\textsc{Ditron}} % put the abb name here

\usepackage{url}

\usepackage{soul}
\usepackage{ulem}
\usepackage{wrapfig}

\usepackage{xcolor}
\usepackage{tcolorbox}
\usepackage{listings}
\usepackage[utf8]{inputenc}
\usepackage{xcolor}
\newcommand{\hlcode}[1]{\colorbox{yellow!30}{#1}}
\usepackage[textsize=tiny]{todonotes}

\usepackage{minitoc}

\title{DITRON: Distributed Multi-level Tiling Compiler for Parallel Tensor Programs}

\author[1,3]{Size Zheng}
\author[1]{Xuegui Zheng}
\author[1]{Hanshi Sun}
\author[1]{Qi Hou}
\author[1,*]{Wenlei Bao}
\author[1,4]{Shiyu Li}
\author[1,5]{Haojie Duanmu}
\author[1]{Jin Fang}
\author[1,4]{Chenli Xue}
\author[1]{Chenhui Huang}
\author[1]{Yuanqiang Liu}
\author[1,2]{Renze Chen}
\author[1,*]{Ningxin Zheng}
\author[1]{Dongyang Wang}
\author[1]{Li-Wen Chang}
\author[4]{Liqiang Lu}
\author[2]{Yun Liang}
\author[3]{Jidong Zhai}
\author[1]{Xin Liu}

\affiliation[1]{ByteDance Seed}
\affiliation[2]{Peking University}
\affiliation[3]{Tsinghua University}
\affiliation[4]{Zhejiang University}
\affiliation[5]{Shanghai Jiao Tong University}

\contribution[*]{Work done at ByteDance Seed}
\abstract{
The scaling of large language models (LLMs) is currently bottlenecked by the rigidity of distributed programming. While high-performance libraries like CuBLAS and NCCL provide optimized primitives, they lack the flexibility required for rapidly evolving model architectures. Conversely, existing tensor compilers fail to address the complex memory hierarchy of distributed clusters effectively. To bridge this gap, we propose DITRON, a scalable tile-level compiler that democratizes high-performance distributed kernel development. DITRON introduces a novel hierarchical programming abstraction spanning Core, Device, and Task levels to map tensor programs efficiently onto heterogeneous distributed hardware. This abstraction allows DITRON to support diverse parallelism strategies while abstracting away the complexity of inter-node and intra-node communication. 

Evaluated across large-scale clusters, DITRON achieves performance parity with or exceeding expert-tuned CUDA libraries, delivering speedups of $6\%–30\%$ on isolated kernels and $5\%–30\%$ on end-to-end inference in vLLM. Furthermore, DITRON demonstrates strong portability, achieving significant speedups on both NVIDIA and AMD platforms.
\ours{} has been deployed at the enterprise level for both training and inference. It achieves an MFU improvement of over 10\% in training tasks, saving approximately 500,000 GPU hours of training cost per month. For inference tasks, it delivers an end-to-end gain of over 20\% and has been applied to cloud service inference and edge inference scenarios.
}

\correspondence{Size Zheng at \email{zheng.size@bytedance.com}}

\checkdata[Project Page]{\url{https://github.com/ByteDance-Seed/Triton-distributed}}

\begin{document}
\maketitle

%不需要目录就注释掉 注意目录不要和第一页放在一块 要有\newpage
%\newpage
%\tableofcontents
%\newpage

\input{contents/1-introduction}
\input{contents/2-background}

\input{contents/3-ditron}

\input{contents/4-experiments}

\input{contents/5-conclusion}

% \clearpage

\bibliographystyle{plainnat}
\bibliography{reference}

% \clearpage

% \beginappendix

% \input{sections/appendix}

\newpage
\appendix
\onecolumn
\section{Swizzle}
\label{appendix:swizzle}
\input{contents/appendix-swizzle}

\section{Evaluation Setup and Workload Configurations}
\label{appendix:a}

\input{contents/appendix-shape}


\section{Detailed Experiment Results}
\label{appendix:experiments}

\input{contents/appendix-experiments}

\section{Code Examples}
\label{appendix:code-example}
\input{contents/appendix-code}

\input{contents/appendix-primitives}

\end{document}

%% file: contents/1-introduction.tex
\section{Introduction}
\label{sec:intro}

% The speed of large language model (LLM) training and inference depends on the running speed of distributed programs; only with the support of high-speed libraries can models with diverse architectures be put into practical application~\cite{llama3, gemini, gemma2, deepseek-v3, qwen-max, gpt4, claude}. In the past, compiler work (such as Triton~\cite{triton} and TileLang~\cite{tilelang}) and libraries (such as CUTLASS/CuTeDSL~\cite{cutlass} and ThunkderKitten~\cite{thunderkittens}) focus on finding ways to improve kernel performance on a single device, while communication in distributed clusters relied on separate libraries~\cite{nccl} or compilers~\cite{mscclpp}. As the scale of the cluster expands, communication overhead gradually surpasses computation overhead (according to~\citet{flux}, communication can take $20\%-80\%$ runtime in training and inference), diluting the benefits gained from improved computation kernel performance. Therefore, higher requirements are now placed on LLM researchers. They must not only be able to propose new algorithms, but also rapidly implement their algorithms in distributed code to prove that the algorithms are scalable and effective.
The rapid evolution of Large Language Models (LLMs) places immense pressure on the underlying distributed systems. Only with high-speed, scalable distributed execution can models with diverse and emerging architectures be practically deployed~\cite{llama3, gemini, gemma2, deepseek-v3, qwen-max, gpt4, claude}. Historically, optimization efforts have been bifurcated: compilers like Triton~\cite{triton} and TileLang~\cite{tilelang} focus on single-device kernel optimization , while distributed communication relies on rigid libraries like NCCL~\cite{nccl} or domain-specific communication compilers~\cite{mscclpp}. However, as cluster scales expand, communication overhead has emerged as the dominant bottleneck—accounting for $20\%–80\%$ of runtime in training and inference~\cite{flux}. This shifts the burden onto researchers, who must now possess the dual capability of designing novel algorithms and implementing them as highly optimized distributed kernels to prove scalability.

% Two streams of work have been proposed to address this issue: distributed domain-specific langauge (DSL) compilers and distributed CUDA libraries. On one hand, DSL compilers~\cite{coconet, pallas, dist-enisum, tilelink} provide various high-level abstractions for users to describe parallel programs and then translate the DSL to low-level code. On the other hand, distributed CUDA libraries~\cite{nccl, flux, comet, deepep} provide highly-optimized operators tuned by experts.
To address this challenge, the community has largely relied on two approaches: distributed Domain-Specific Language (DSL) compilers and distributed CUDA libraries. DSL compilers~\cite{coconet, pallas, tilelink} offer high-level abstractions for specifying parallel programs , while distributed libraries~\cite{nccl, flux, comet, deepep} provide highly optimized, expert-tuned operators.
% Two lines of work have been proposed to tackle this challenge: distributed domain-specific language (DSL) compilers and distributed CUDA libraries. On one hand, DSL compilers~\cite{coconet, pallas, dist-enisum, tilelink} offer a rich set of high-level abstractions that allow users to specify parallel programs, with the compiler subsequently translating these DSL specifications into low-level executable code. On the other hand, distributed CUDA libraries~\cite{nccl, flux, comet, deepep} supply highly optimized operators that are meticulously tuned by domain experts.

However, both approaches face significant limitations. First, flexibility is compromised. Distributed libraries are inherently non-programmable, blocking the exploration of new architectures. Meanwhile, DSL compilers often restrict users to operator-level granularity~\cite{coconet, dist-enisum}, lacking the expressiveness for fine-grained optimizations. Second, scalability is often limited by rigid tiling assumptions. Frameworks like Pallas~\cite{pallas} and TileLink~\cite{tilelink} enable tile-level programming but often lack a unified multi-level tiling abstraction necessary to pipeline execution across the complex hierarchy of a large-scale cluster. Third, the lack of portable primitives makes porting these frameworks to emerging hardware backends prohibitively expensive.

An ideal compiler should prioritize usability in its design without sacrificing performance. But this standard has not yet been achieved by any other existing work.
In detail, we argue that distributed tensor programming should adhere to three core design principles.
\uline{First, it should provide a flexible programming interface that aligns with the conventions of mainstream tensor compilers.} As noted in a recent report~\cite{kernelevolve}, tile-level compilers such as Triton~\cite{triton} have surpassed CUDA to become the dominant kernel programming framework in the industry. Extending existing tile compilers to support distributed scenarios inherently holds advantages over developing a new programming language from scratch. Such extensions should ensure that legacy kernels can be converted to distributed versions with minimal code modifications. To this end, we adopt Triton’s tile programming model and implement distributed extensions atop it.

\uline{Second, it should embody a scalable programming paradigm that accommodates clusters of arbitrary scales and problem shapes.}
Hardware configurations vary significantly across different tasks. A distributed system may incorporate high-bandwidth interconnects such as NVLink, hardware-accelerated in-switch computing capabilities like NVSwitch, as well as low-bandwidth links including PCIe and Ethernet. Any a priori assumptions may render the compiler impractical for real-world system deployment. To enable seamless adaptation to such heterogeneous systems, we advocate for supporting multi-level tiling in the programming model, where fine-grained tiles are mapped to high-speed connections and coarse-grained tiles to low-speed ones.

\uline{Third, it should provide a unified set of primitives that is portable across diverse hardware backends.}
Different hardware backends feature distinct hardware topologies and underlying technologies. Consequently, low-level programming models entail drastically different optimization strategies and programming paradigms. To enable seamless support for multiple backends, we introduce a suite of hardware-agnostic primitives. Integrating a new hardware backend can then be achieved by instantiating these primitives and implementing corresponding translation rules for them.

Guided by these design insights, we propose \ours{}, a hierarchical distributed tensor compiler. \ours{} is structured into three layers: front-end, mid-end, and back-end.
At the front-end, \ours{} implements three hierarchical tiling levels: core-level tiling, device-level tiling, and task-level tiling. Core-level tiling enables users to leverage small static-shaped vectors or matrices to invoke hardware acceleration units such as Tensor Memory Accelerators (TMA) and Tensor Cores while maintaining full functional and performance compatibility with the existing Triton programming language.
Device-level tiling unlocks the potential of hardware DMA/RDMA engines, which are optimized for high-throughput large-chunk data transfers. This tiling level adopts dynamic shapes and supports runtime shape computation, a critical capability for dynamic model architectures like Mixture-of-Experts (MoE).
Task-level tiling goes a step further by enabling model-level tiling: it fuses the entire model into a single kernel, which is then deployed and executed across distributed clusters to maximize hardware resource utilization.
% At the mid-end, \ours{} maps operator-level collective communications into tile-level semantics and enables compute-communication overlapping. \ours{} supports all major parallelism in LLM training and inference, including tensor parallel, sequence parallel, expert parallel, and pipeline parallel. Communication collectives such as AllGather, ReduceScatter, AllToAll, and AllReduce are broken into small chunks dynamically and each chunk correlates to set of tiles of computation kernels, while computation kernels are inserted with distributed events to work cooperatively with communication kernels.

At the mid-end, \ours{} translates operator-level collective communications into tile-level semantics and enables the overlapping of computation and communication. \ours{} natively supports all major parallelism paradigms for LLM training and inference, including tensor parallelism, sequence parallelism, expert parallelism, and pipeline parallelism. Collective communication primitives such as AllGather, ReduceScatter, AllToAll, and AllReduce are dynamically partitioned into chunks, with each chunk mapped to a set of tiles from the corresponding computation kernels. Meanwhile, distributed synchronization events are embedded into computation kernels to enable seamless coordination with communication kernels.
% At the backend, \ours{} provides a set of primitives, mainly including address mapping, data accessing, and synchronization. These primitives are hardware-neutral and can be translated to assembly for any hardware backend by adding corresponding translation rules.
% We have implemented the translation rules for NVIDIA backend and AMD backend, enabling support for a wide range of GPUs.

At the back-end, \ours{} offers a suite of hardware-agnostic primitives, encompassing address mapping, data access, and synchronization mechanisms. Being hardware-agnostic, these primitives can be translated into hardware-specific assembly code for any target backend by implementing corresponding translation rules.
We have instantiated these translation rules for both NVIDIA and AMD backends, thereby enabling \ours{} to support a broad spectrum of GPUs across these two platforms.

% The major difference between distributed programming and single-device programming lies in the existence of non-uniform memory access (NUMA), which can be viewed as two additional levels of the memory hierarchy: scale-up remote memory and scale-out remote memory (shown in Figure~\ref{fig:intro}).
% % The major difference between distributed programming and single-device programming is the existence of non-uniform memory access (NUMA), which can be viewed as another two levels of memory hierarchy: scale-up remote memory and scale-out remote memory (shown in Figure~\ref{fig:intro}). 
% % To understand how NUMA effects change programming model, 
% We summarize distributed NUMA programming model as the following:

Through excessive validation and evaluation, \ours{} is able to achieve $1.27\times - 19.18\times$ average speedup to vendor-provided non-overlapping kernels across a wide range of workloads, and is even $6\%-30\%$ faster than expert-tuned overlapping libraries implemented in CUDA.
It also renders $5\% - 30\%$ end-to-end performance improvement when integrated with vLLM~\cite{vllm} for large batch size. 
On AMD GPUs the speedup to RocmBLAS+RCCL ranges from $1\%-38\%$; on PCIe GPUs the average speedup to CuBLAS+NCCL is $8.33\times$.

%% file: contents/2-background.tex
\section{Background and Related Work}

\begin{figure*}[!t]
    \centering
    \includegraphics[width=\textwidth]{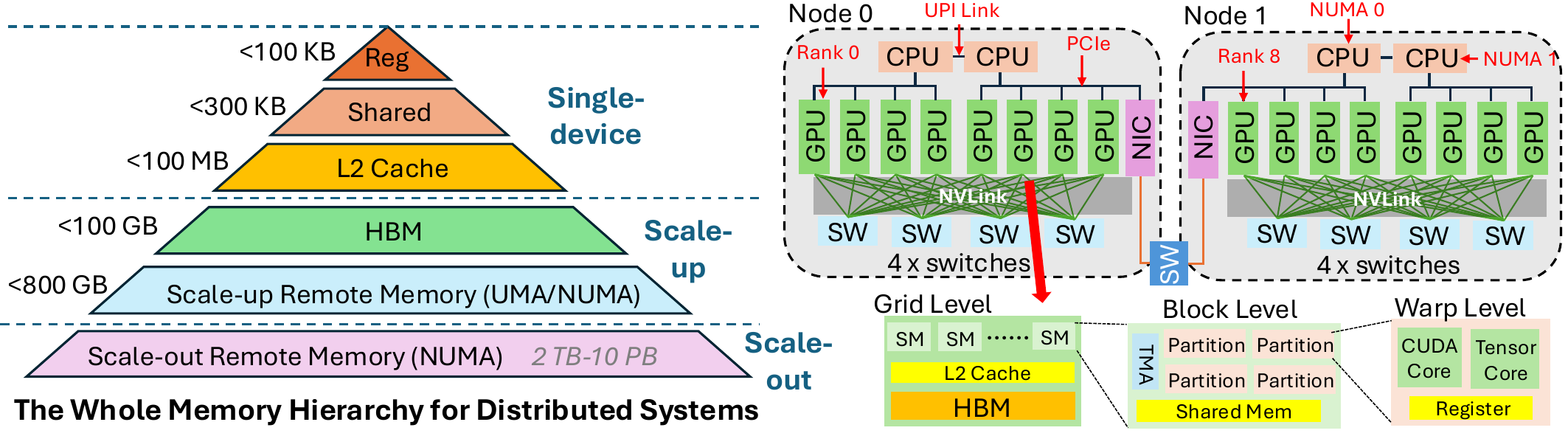}
    \caption{GPU Distributed Memory Hierarchy and Cluster Overview.}
    \label{fig:background}
\end{figure*}

To understand the design rationale behind \ours{}, we first characterize the hardware hierarchy of modern distributed clusters and then analyze how existing programming models fail to align with this hierarchy.

\subsection{The Hierarchy of Distributed Hardware}
The memory and compute hierarchy of a distributed GPU cluster is inherently non-uniform. As illustrated in Figure~\ref{fig:background}, effective distributed programming requires managing data movement across three distinct domains, each with vastly different bandwidths and latencies.

For in-device domain (core-level), it involves data movement between HBM, L2 Cache, and registers. Optimization here relies on maximizing data reuse in SRAM and leveraging specialized compute units like Tensor Cores. Programs for this domain are often composed of instructions for grid-level, block-level, and warp-level computation and communication.
For scale-up domain (intra-node), modern GPU~\cite{hopper} nodes connect GPUs via high-bandwidth fabrics like NVLink. This domain supports Unified Memory Access (UMA) via load/store instructions, but also introduces NUMA characteristics where remote access latency is non-negligible. Crucially, this domain offers specialized hardware features such as NVLink Sharp~\cite{nvswitch} for in-network reduction, which standard software often overlooks.
For scale-out domain (inter-node), communication between nodes relies on Ethernet or InfiniBand. This domain is strictly NUMA, where data transfer requires explicit coordination via direct memory access (DMA) engines and network interface cards (NICs). The bandwidth gap here is significant: while HBM offers several TB/s bandwidth, inter-node links often operate at tens of GB/s bandwidth.

The fundamental challenge in distributed LLM workloads is that these levels are not isolated; a single operation (e.g., a distributed matrix multiplication) often spans all three domains simultaneously.

\subsection{Limitations of Existing Programming Models}
Despite the hierarchical nature of hardware, existing software stacks largely fail to provide a unified abstraction that captures these nuances.

\textbf{Hand-tuned Libraries:}
Previous work~\cite{flux, comet, deepep, tokenweave} attempts to bridge the gap by providing hand-tuned kernels for computation-communication overlapping. For instance, libraries like FLUX~\cite{flux} and COMET~\cite{comet} integrate communication primitives directly into CUTLASS kernels. However, their programming interfaces are often obscure and cumbersome to use, requiring deep knowledge of assembly (PTX) or complex C++ template metaprogramming . This rigidity makes it difficult for researchers to adapt these kernels to new model architectures.

\textbf{Compilers and DSLs:}
Compilers such as CoCoNet~\cite{coconet} and DistEinsum~\cite{dist-enisum} provide operator-level DSLs for distributed programming.
Recently, frameworks like Pallas~\cite{pallas}, TileLink~\cite{tilelink} and IRIS~\cite{iris} have attempted to extend tiling concepts to distributed settings. While promising, they often struggle with the complexity of multi-level hierarchies. For example, TileLink primarily focuses on the scale-up abstraction but fails to expose the controls necessary for efficient scale-out communication.

%% file: contents/3-ditron.tex
\section{\ours{} System Design}

\begin{figure*}[!t]
\begin{center}
%\framebox[4.0in]{$\;$}
% \fbox{\rule[-.5cm]{0cm}{4cm} \rule[-.5cm]{4cm}{0cm}}
\includegraphics[width=\textwidth]{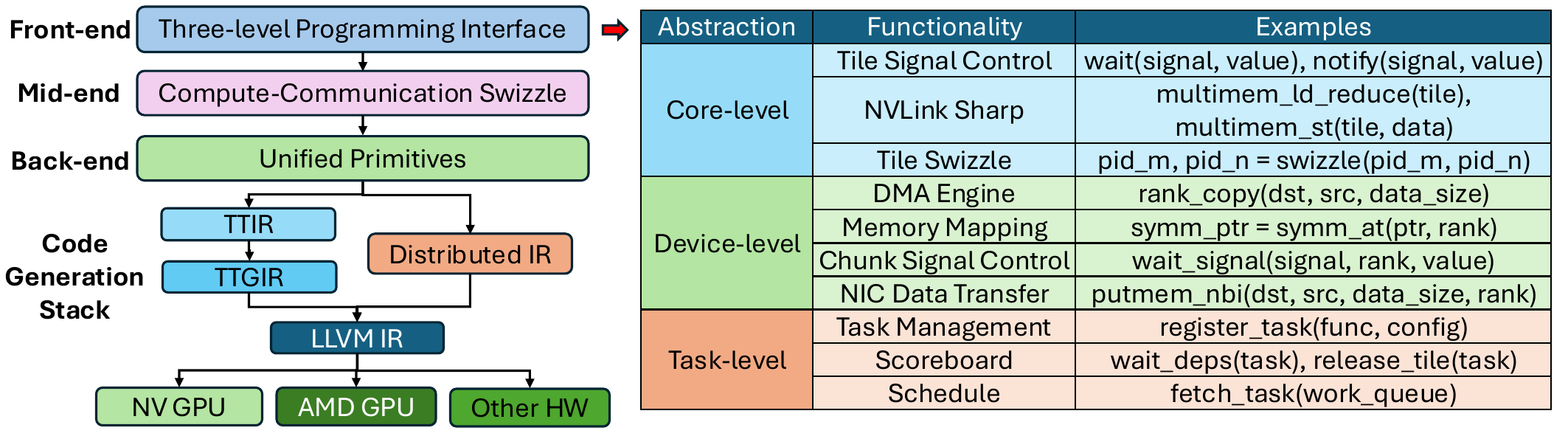}
\end{center}
\caption{\ours{} is composed of front-end interface, mid-end swizzle, and back-end primitives and code generation.}
\label{fig:overview}
\end{figure*}

% \begin{figure*}[!t]
% \begin{center}
% %\framebox[4.0in]{$\;$}
% % \fbox{\rule[-.5cm]{0cm}{4cm} \rule[-.5cm]{4cm}{0cm}}
% \includegraphics[width=\textwidth]{figures/intro.pdf}
% \end{center}
% \caption{\ours{} provides three levels of programming abstractions, which corresponds to the hardware hierarchy of distributed systems.}
% \label{fig:intro}
% \end{figure*}

% We present \ours{}, a flexible and versatile compiler designed for compute-communication overlapping kernel programming. The overview of \ours{} is shown in Figure~\ref{fig:overview}. Figure~\ref{fig:overview} also shows the primitive examples for three levels of programming abstractions. In the following, we first explain the programming abstractions. In Figure~\ref{fig:swizzle}, we show a simplified code example for GEMM+ReduceScatter. Then we explain the swizzle optimizations based on this example. Finally, we illustrate the implementation details of \ours{}.
Followed by the insights discussed in Section~\ref{sec:intro}, we present \ours{}, a unified compiler stack designed to bridge the gap between high-level distributed algorithms and low-level hardware heterogeneity. As illustrated in Figure~\ref{fig:overview}, \ours{} adopts a modular design comprising a hierarchical front-end interface, an optimization-centric mid-end swizzle, and a portable back-end code generation with unified primitives.
% We present \ours{}, a flexible and versatile compiler designed for compute-communication overlapping kernel programming. An overview of \ours{} is provided in Figure~\ref{fig:overview}, which also includes primitive examples for the three levels of programming abstractions. In what follows, we first elaborate on the programming abstractions. Figure~\ref{fig:swizzle} presents a simplified code example for GEMM+ReduceScatter, upon which we then explain the swizzle optimizations. Finally, we detail the implementation specifics of \ours{}.

\subsection{Front-end Three-level Programming Interface}

The core innovation of \ours{} lies in decoupling the logical view of a distributed tensor program from its physical execution. We introduce a three-level tiling abstraction that maps distinct program semantics to the corresponding hardware domains.

\textbf{Core-level Interface:}
At the finest granularity, \ours{} inherits the tile-level semantics from Triton  to manage computation within a single GPU. This level operates on static-shaped tiles (e.g., $128 \times 128$ blocks) to maximize the utilization of specialized compute units like Tensor Cores and TMA engines. By maintaining compatibility with standard Triton syntax, \ours{} allows users to seamlessly reuse existing optimized kernels for computation logic. We show the code to implement a high-performance AllGather+GEMM, GEMM+ReduceScatter, and GEMM+AllReduce overlapping kernels in Appendix~\ref{appendix:code-example}, where the computation kernels only require several lines of code modification (highlighted) to work for distributed systems. Core-level interface mainly contains signal control (\textit{wait, notify}), on-switch computation (\textit{multimem}), and tile swizzling as illustrated in the Table in Figure~\ref{fig:overview}.

\textbf{Device-level Interface:}
Unlike single-device compilers, \ours{} introduces a device-level abstraction to manage data movement across the distributed memory hierarchy. This level operates on chunks, which are coarse-grained data blocks composed of multiple fine-grained tiles. This abstraction is designed with two key features. First, we use DMA-centric semantics. Unlike core-level load/store instructions, device-level primitives (e.g., \textit{putmem}, \textit{getmem}) map directly to asynchronous DMA engines and NICs, bypassing the GPU SMs to save compute resources. Second, we support dynamic shapes. To support dynamic model architectures like MoE, we allow the data chunk size, input size, and output size to be dynamically calculated at runtime. For instance, in an AllToAll operation, the size of chunks transferred between ranks can be determined dynamically based on token routing results.

\textbf{Task-level Interface:}
At the highest level, \ours{} treats the entire distributed workload as a directed acyclic graph (DAG) of tasks. This abstraction allows the compiler to fuse communication and computation kernels into a single MegaKernel. By managing task dependencies and scheduling globally, \ours{} eliminates the overhead of repeated kernel launches and enables persistent residency of kernels on the hardware, ensuring that communication fabrics and compute units are kept constantly busy.
Different from previous MegaKernel work~\cite{mirage, hazymega} that requires CUDA or C++ programming, \ours{} allows users to register native Triton kernels as tasks  and then automatically schedules these kernels at compile time through software-maintained scoreboard. Finally, these Triton kernels are converted to a fused MegaKernel. We show how to register Triton kernels as tasks and the final generated MegaKernel in Appendix~\ref{appendix:code-mega}.

\begin{figure*}[!t]
\begin{center}
%\framebox[4.0in]{$\;$}
% \fbox{\rule[-.5cm]{0cm}{4cm} \rule[-.5cm]{4cm}{0cm}}
\includegraphics[width=\textwidth]{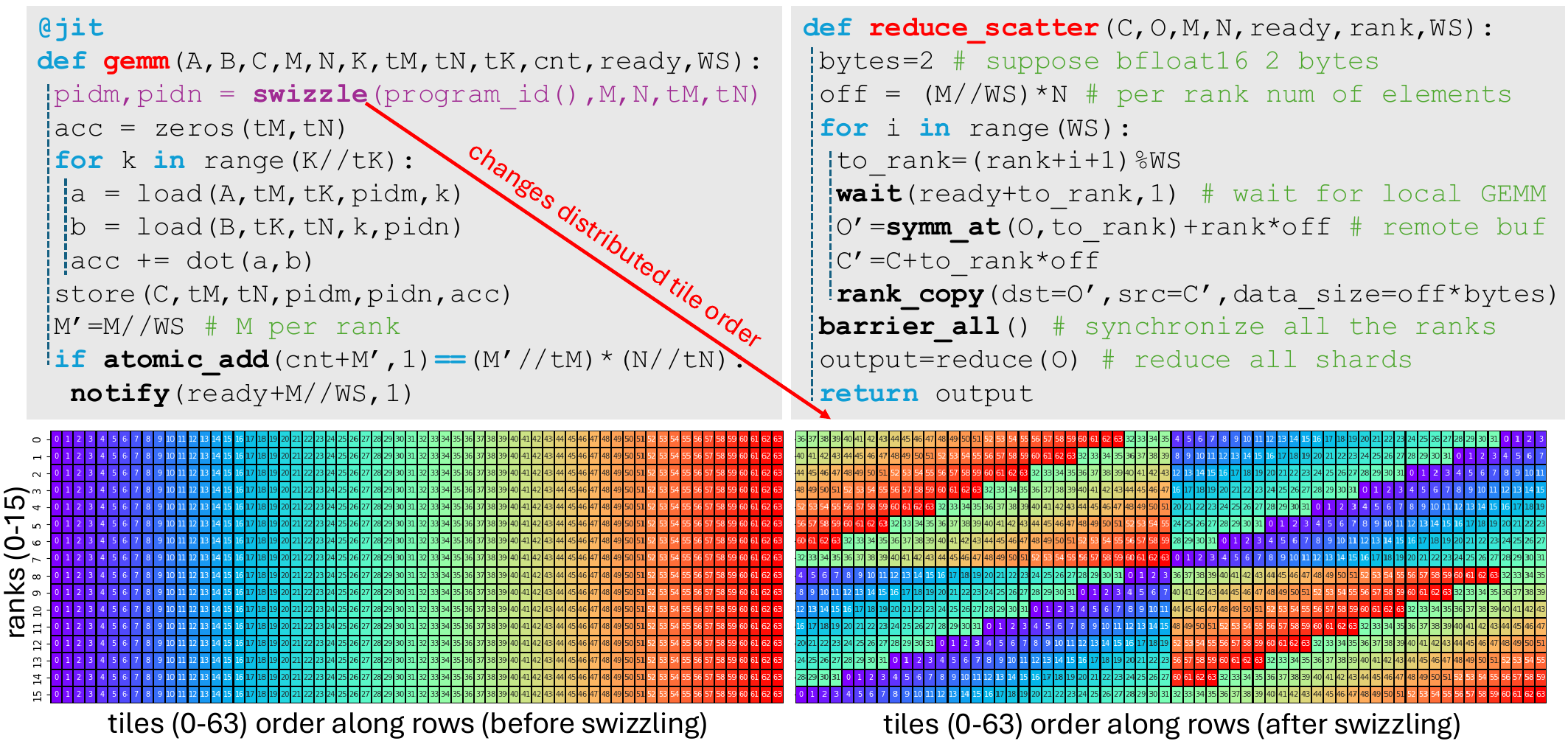}
\end{center}
\caption{Simplified GEMM+ReduceScatter code example and distributed swizzling. The ReduceScatter requires that each GPU rank $i$ takes 4 data chunks ($[4i,4i+4)$) eventually. Tiles are processed from left to right. Without swizzling, all ranks start from data tiles for rank 0 and the other ranks are blocked, while with swizzling, all the ranks can start without blocking. A clearer enlarged swizzle view is presented in the Appendix~\ref{appendix:swizzle} and the full code example is in Appendix~\ref{appendix:code-gemm-rs}.}
\label{fig:swizzle}
\end{figure*}

\subsection{Mid-end Compute-Communication Swizzling}
The Mid-end is responsible for lowering high-level compute-communication algorithm into overlapped versions and applying system-aware optimizations. The most critical optimization in \ours{} is distributed swizzling.

In a distributed settings, the latency of accessing remote memory is orders of magnitude higher than local HBM. Standard sequential execution often leaves compute units idle while waiting for data. \ours{} addresses this by reordering the execution of tiles. We formalize this into two distinct modes handled by the compiler:

\textbf{Gather Mode:}
For operations where computation depends on remote data (e.g., AllGather+GEMM), the compiler schedules remote data fetch requests as early as possible, effectively treating local HBM as a cache for remote memory.

\textbf{Scatter Mode:}
For operations where results must be sent out (e.g., GEMM+ReduceScatter), the compiler prioritizes the computation of tiles destined for the furthest remote nodes, ensuring that long-haul communication is initiated immediately upon data availability.

The swizzling logic is encapsulated within the compiler as a stateless, JIT-compatible utility. Leveraging Triton's JIT infrastructure and \ours{}'s three-level interfaces, we expose swizzling as a pluggable primitive that can be flexibly injected into any kernel. The transformation is defined as:

\begin{align*}
new\_pid & = swizzle\_func(
  old\_pid, \\ & shape\_to\_swizzle, 
  rank, world\_size, block\_size)
\end{align*}

We provide the detailed implementation in Appendix~\ref{appendix:swizzle}. To illustrate its efficacy, Figure~\ref{fig:swizzle} visualizes the execution flow of a GEMM+ReduceScatter workload . In a naive schedule (without swizzling), all ranks sequentially process tiles starting from index 0. Since the data required for tile 0 typically resides on Rank 0, all other ranks suffer from dependency stalls while waiting for data transfer.
Conversely, \ours{} applies a rank-aware offset, which specifically starts execution at $rank\_id \mod{local\_world\_size}$. This ensures that each rank prioritizes tiles for which data is locally available or already arriving, thereby effectively overlapping the GEMM computation with the ReduceScatter communication. Swizzling for imperfect shapes can be harder and we place the discussion in Appendix~\ref{appendix:swizzle}.

\begin{figure*}[!t]
\begin{center}
%\framebox[4.0in]{$\;$}
% \fbox{\rule[-.5cm]{0cm}{4cm} \rule[-.5cm]{4cm}{0cm}}
\includegraphics[width=\textwidth]{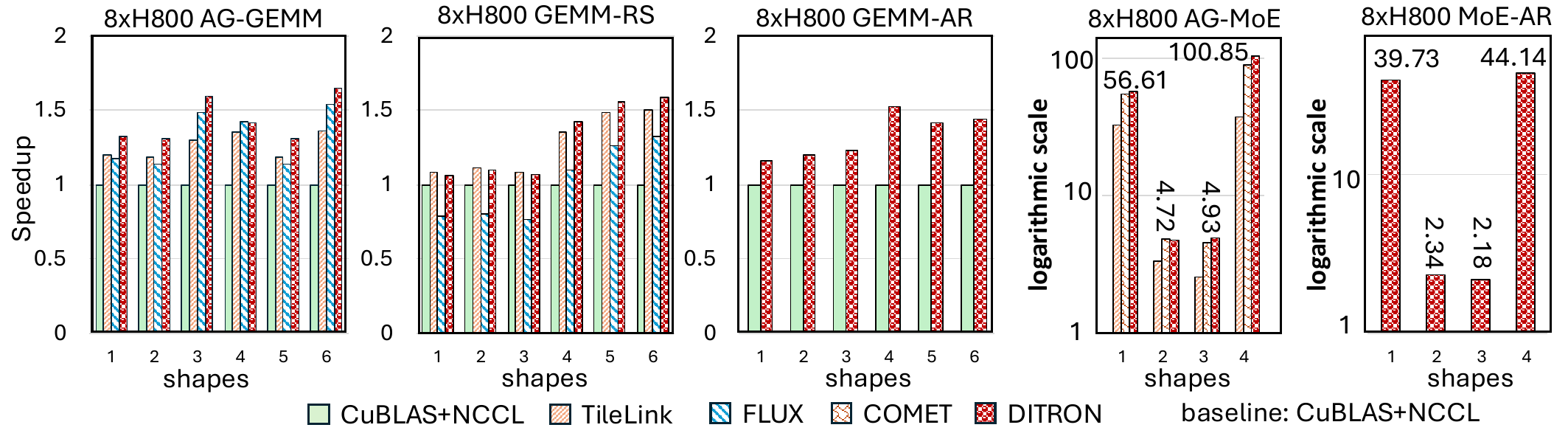}
\end{center}
\caption{Evaluation of single workload on $8\times$ H800 GPUs. Workloads are AllGather+GEMM (AG-GEMM), GEMM+ReduceScatter (GEMM-RS), GEMM+AllReduce (GEMM-AR), AllGather+MoE (AG-MoE), MoE+AllReduce (MoE-AR). A logarithmic scale is used for the speedup of MoE due to its wide span, and the specific values are listed in the Appendix~\ref{appendix:experiments}.}
\label{fig:experiments-h800}
\end{figure*}

\subsection{Back-end Primitives and Code Generation}

\textbf{Primitives:}
To achieve hardware portability (design principle 3), \ours{} abstracts backend-specific communication functionalities into a set of hardware-agnostic primitives compliant with the OpenSHMEM standard. During the code generation phase, \ours{}'s Distributed IR is lowered into LLVM IR. We utilize LLVM's CallExtern capability to link against vendor-specific communication libraries. 

The primitives are divided into three classes: distributed primitives, SIMT primitives, and SHMEM device primitives. Distributed primitives are used to generate low-level code for distributed signal control. SIMT primitives are used for threads cooperation or synchronization. SHMEM device primitives are used for remote data transfer and signal exchange.
In Appendix~\ref{sec:appendix-primitives} we list these primitives in detail.

\textbf{Compiler Stack:}
% \ours{} is implemented with 59k lines of Python code and 7k lines of C++ code, which is mainly composed of a compiler stack and an optimized kernel library.
\ours{} is implemented with around 59,000 lines of Python code and 7,000 lines of C++ code, consisting primarily of a compiler stack and an optimized kernel library.

The compiler stack of \ours{} is illustrated in Figure~\ref{fig:overview}. Users write Triton-like programs using our three levels of tiling interfaces, which cover both computation and communication. The hardware primitives are encapsulated in our Distributed IR for compilation, while standard single-device semantics (e.g., \textit{dot}, \textit{load/store}) in the programs are lowered to standard Triton IR (TTIR) and Triton GPU IR (TTGIR). This Distributed IR with OpenSHMEM semantic is then lowered to assembly with extern symbol linked to hardware-specific low-level library such as NVSHMEM (for NVIDIA) and rocSHMEM (for AMD).
% , which is a communication standard designed for distributed systems. OpenSHMEM has been implemented in efficient low-level communication libraries by NVIDIA (NVSHMEM) and AMD (rocSHMEM), 
To support other hardware, we can also leverage custom SHMEM libraries provided by vendors. 
\ours{} currently support more than five different types of GPUs and different communication fabrics including NVLink, xGMI, PCIe, and IB.
% The distributed IR is then lowered to LLVM IR, where we leverage LLVM’s \textit{CallExtern} capability to invoke external symbols from the SHMEM library.
% The compiler stack of \ours{} is shown in Figure~\ref{fig:overview}. Users program NumPy-like programs using our three levels of abstractions for both computation and communication. 
% We implement distributed IR for compilation.
% The local semantics (such as \textit{dot}, \textit{load/store}, etc.) in the programs are lowered to standard Triton IR (TTIR) and Triton GPU IR (TTGIR), while the distributed semantics are lowered to our distributed IR. The distributed IR follows OpenSHMEM standard, which is a communication standard proposed for distributed systems. OpenSHMEM standard has been implemented into efficient low-level communication libraries by Nvidia (NVSHMEM) and AMD (rocSHMEM). Other hardware can be also supported via customized SHMEM library. Distributed IR is lowered to LLVM IR. We use LLVM \textit{CallExtern} ability to invoke external symbols from the SHMEM library.

\textbf{Optimized Kernel Library:}
Beside compiler stack, we also provide a comprehensive kernel library in \ours{}, which are implemented using the three-level tiling interface provided by \ours{}.
These kernels have been validated in industry LLM training and inference for both numeric precision and performance. We list them in Table~\ref{tab:kernel-list}.

\begin{table}[htbp]
  \centering
  \caption{The Optimized Kernel List in \ours{}}
  \label{tab:kernel-list}
  \small
  \begin{tabular}{|l|p{4cm}|}
    \hline
    \textbf{Kernel Name} & \textbf{Brief Description} \\ \hline
    % \multicolumn{2}{|c|}{\textbf{AllGather \& GEMM}} \\ \hline
    \texttt{ag\_gemm} & Fused AllGather + GEMM \\ \hline
    \texttt{ag\_group\_gemm} & Fused AllGather + Group GEMM \\ \hline
    \texttt{fast\_allgather} & Low-latency AllGather \\ \hline
    \texttt{gemm\_rs} & GEMM + ReduceScatter \\ \hline
    \texttt{gemm\_allreduce} & Fused GEMM + AllReduce \\ \hline
    % \multicolumn{2}{|c|}{\textbf{Attention \& Decoding}} \\ \hline
    \texttt{flash\_decode\_gqa} & GQA Batch Flash Decode \\ \hline
    \texttt{sp\_ag\_attn} & Seq Parallel AllGather Attn \\ \hline
    % \multicolumn{2}{|c|}{\textbf{All-to-All}} \\ \hline
    \texttt{fast\_all\_to\_all} & Optimized All-to-All \\ \hline
    \texttt{a2a\_single\_2d} & 2D Single Kernel All-to-All \\ \hline
    \texttt{a2a\_single\_gemm} & Fused All-to-All + GEMM \\ \hline
    % \multicolumn{2}{|c|}{\textbf{Other}} \\ \hline
    % \texttt{gdn\_fwd} & Gated Delta Rule Forward \\ \hline
    \texttt{ulysses\_comm} & Ulysses SP Pre-Attn Comm \\ \hline
  \end{tabular}
\end{table}

To achieve the best performance, \ours{} incorporates several low-level optimizations to address specific hardware constraints and latency bottlenecks in distributed systems. And we pinpoint three of them:
\begin{itemize}
\item \textbf{Low-Latency (LL) Protocol Integration.} Standard communication protocols often prioritize bandwidth at the cost of synchronization latency. Following the design of NCCL~\cite{nccl}, \ours{} implements a specialized Low-Latency (LL) protocol that bypasses expensive synchronization handshakes. 
% Although this protocol incurs approximately $50\%$ bandwidth overhead due to data padding, it is critical for minimizing latency in small-batch workloads (e.g., batch size 1 inference) where synchronization overhead dominates execution time .
\item \textbf{Device-to-Device (D2D) Copy Fusion.} We observe that discrete data movement calls via driver APIs (e.g., \texttt{cudaMemcpy}) or framework ops (e.g., \texttt{torch.copy}) often introduce launch jitters and non-deterministic SM utilization, creating stragglers that negate the benefits of overlapping. To mitigate this, \ours{} fuses these D2D copy operations directly into the generated communication or computation kernels. This fusion eliminates driver overhead and ensures deterministic resource scheduling.
\item \textbf{PCIe-Aware Synchronization and Topology.} Commodity PCIe interconnects lack hardware guarantees for atomic memory ordering. To deploy \ours{} on PCIe-based GPUs, we implement software barriers using \texttt{volatile} load/store instructions to enforce memory consistency without hardware atomics. 
% Furthermore, recognizing the bandwidth constraints of PCIe, \ours{} adopts Ring-based algorithms for AllGather and ReduceScatter primitives and explicitly offloads cross-UPI traffic to NICs to avoid interconnect contention.
\end{itemize}

%% file: contents/4-experiments.tex
\section{Evaluation}

% \ours{} supports kernel implementation and optimization for various parallel strategies (TP, SP, EP). 
% In experiments, we evaluate \ours{} under different parallel configurations. For inference, where the primary parallel strategy is TP within a single node, we first test 5 distinct workloads for GEMM and MoE. Second, we integrate these into attention modules and FFN modules to assess module-level performance. Third, we evaluate end-to-end inference using Qwen3-32B and LLaMA3-70B. For training, we test TP, SP, and EP kernels implemented by \ours{} across 8 to 128 GPUs, examining both weak and strong scaling. Finally, we use \ours{}'s kernels to scale the training of a Qwen3-like 640B MoE model to 11,520 GPUs. \ours{} also supports AMD GPUs and PCIe GPUs, which is discussed at the end of the evaluation section.
We evaluate \ours{} across diverse parallel configurations. For inference (intra-node TP), we benchmark 5 GEMM/MoE workloads, Attention/FFN modules, and end-to-end Qwen3-32B/LLaMA3-70B models . For training, we analyze weak and strong scaling of TP, SP, and EP kernels on 8–128 GPUs. Finally, we demonstrate support for AMD and PCIe GPUs.
% \ours{} support kernel implementation and optimization for various parallel strategies (TP, SP, EP).
% In experiments, we evaluate \ours{} using different parallel settings. For inference, where the major parallel strategy is TP in single node, we first test 5 different workloads for GEMM and MoE. Second, we integrate them into attention modules and FFN modules to evaluate module-level performance. Third, we test end-to-end inference using Qwen3-32B and LLaMA3-70B.
% For training, we test TP, SP and EP kernels implemented by \ours{} from 8 GPUs to 128 GPUs for both weak scaling and strong scaling.
% Finally, we use \ours{}'s kernels to scale a training task of a Qwen3-like 600B MoE model to over $10,000$ GPUs. \ours{} also supports AMD GPUs and PCIe GPUs, we put the results at the end of evaluation.

% \begin{table}[t]
% \caption{Single Layer Workloads in Experiments}
% \label{table:workloads}
% \begin{center}
% \small
% \begin{tabular}{l|c|l|l|c|l}
% \multicolumn{1}{c}{\bf Name}  &
% \multicolumn{1}{c}{\bf Usage} &
% \multicolumn{1}{c}{\bf Scalability} & \multicolumn{1}{c}{\bf Name}  &
% \multicolumn{1}{c}{\bf Usage} 
% &\multicolumn{1}{c}{\bf Scalability}
% \\ \hline
% AllGather+GEMM  & TP, FSDP   &8-128 GPUs & MoE+ReduceScatter & TP      &8-16 GPUs \\
% GEMM+ReduceScatter  & TP, FSDP    &8-128 GPUs & Dist-Flash-Decode & SP       &8-32 GPUs \\
% GEMM+AllToAll & SP           &8-128 GPUs & EP Dispatch  & EP           &8-128 GPUs \\
% AllGather+MoE & TP         &8-16 GPUs & EP Combine     & EP        &8-128 GPUs \\
% \end{tabular}
% \end{center}
% \end{table}

\begin{figure*}
\begin{center}
%\framebox[4.0in]{$\;$}
% \fbox{\rule[-.5cm]{0cm}{4cm} \rule[-.5cm]{4cm}{0cm}}
\includegraphics[width=\textwidth]{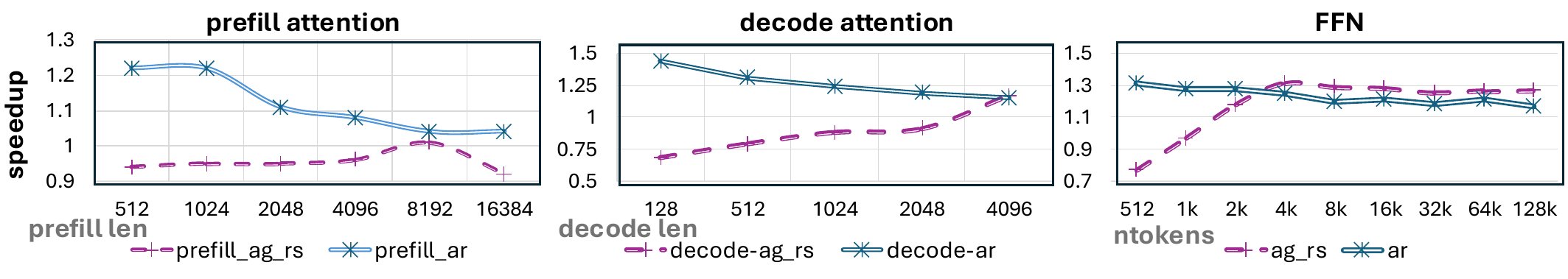}
\end{center}
\caption{Module-level TP evaluation for Qwen3-32B: Results for prefill and decode attention show that \ours{} with AllReduce achieves better speedups over CuBLAS+NCCL compared to using AllGather and ReduceScatter. For MLP, results indicate that the combination of AllGather and ReduceScatter performs better for large shapes, while AllReduce is superior for small shapes.}
\label{fig:module}
\end{figure*}

\begin{figure*}
\begin{center}
%\framebox[4.0in]{$\;$}
% \fbox{\rule[-.5cm]{0cm}{4cm} \rule[-.5cm]{4cm}{0cm}}
\includegraphics[width=\textwidth]{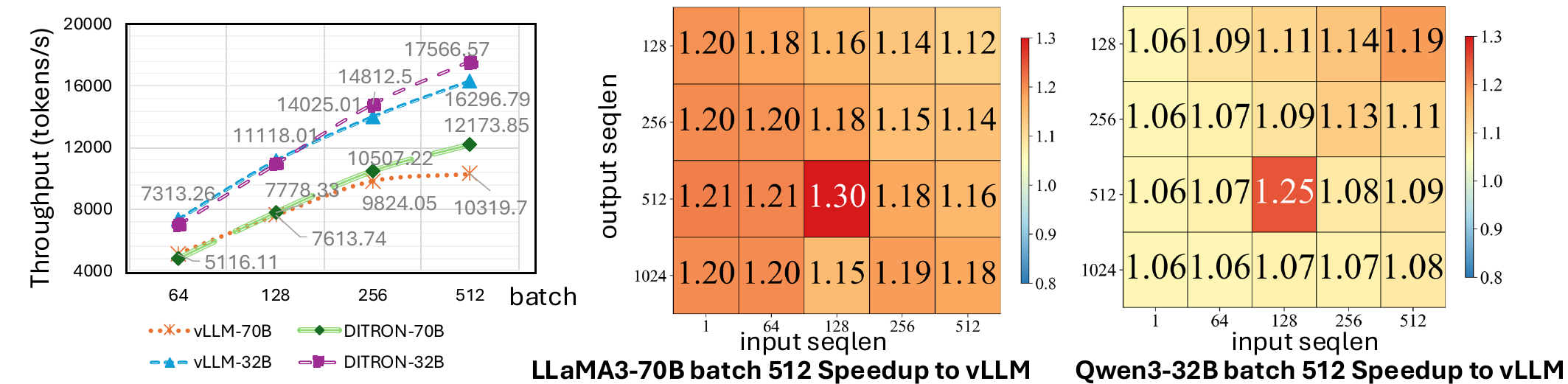}
\end{center}
\caption{End-to-end TP evaluation results for Qwen3-32B and LLaMA3-70B on 8 $\times$ H800s: \ours{} demonstrates advantages over vLLM for large batch sizes and large models.}
\label{fig:e2e}
\end{figure*}

\subsection{Evaluation for Inference}

\textbf{Single Workload Evaluation:} 
We evaluated 5 distinct workloads on 8 $\times$ H800 GPUs across 26 configurations, with each workload featuring multiple shape configurations derived from real-world models such as LLaMA~\citep{llama3}, Mixtral~\citep{mixstral}, GPT~\citep{gpt4}, Qwen~\citep{qwen3}, and DeepSeek~\citep{deepseek-v2}. Detailed shape configurations are provided in Appendix~\ref{appendix:shape}. Our baselines include CuBLAS~\citep{cublas}+NCCL~\citep{nccl} (non-overlapping), TileLink~\citep{tilelink}, FLUX~\citep{flux}, and COMET~\citep{comet}, with results presented in Figure~\ref{fig:experiments-h800}.

% We test 5 different workloads on $8\times$ H800 GPUs with 26 configurations, each workload with several different shape configurations taken from real-world models including LLaMA~\citep{llama3}, Mixtral~\citep{mixstral}, GPT~\citep{gpt4}, Qwen~\citep{qwen3}, and DeepSeek~\citep{deepseek-v2}. The detailed shape configurations are explained in Appendix~\ref{appendix:a}.
% Our baselines are CuBLAS~\citep{cublas}+NCCL~\citep{nccl} (non-overlapping), TileLink~\citep{tilelink}, FLUX~\citep{flux}, and COMET~\citep{comet}.
% The results are shown in Figure~\ref{fig:experiments-h800}.

Overall, for AG-GEMM, the geometric speedup of \ours{} is $1.43\times$ to CuBLAS+NCCL, $1.13\times$ to TileLink, and $1.09\times$ to FLUX. For GEMM-RS, the geometric speedup of \ours{} is $1.27\times$ to CuBLAS+NCCL, $1.02\times$ to TileLink, and $1.30\times$ to FLUX. TileLink and FLUX have no support for AllReduce. For GEMM-AR, the geometric speedup to CuBLAS+NCCL is $1.32\times$. The speedup of AG-GEMM and GEMM-RS mainly comes from the overlapping of communication instead of faster GEMM (actually, we use Triton's GEMM and the GEMM is slightly slower than that of CuBLAS and FLUX), nearly $87.5\%$ communication latency is hidden by computation for large input shapes. The speedup of GEMM-AR comes from faster AllReduce implemented using \ours{}, which supports both one-shot algorithm and two-shot algorithm via multi-memory reduction/broadcast provided by NVLink Sharp.
For AG-MoE, the geometric speedup of \ours{} is $19.18\times$ to CuBLAS+NCCL, $1.89\times$ to TileLink, and $1.06\times$ to COMET. For MoE-AR, the average speedup to CuBLAS+NCCL is $13.89\times$

\begin{figure*}[!t]
\begin{center}
%\framebox[4.0in]{$\;$}
% \fbox{\rule[-.5cm]{0cm}{4cm} \rule[-.5cm]{4cm}{0cm}}
\includegraphics[width=\textwidth]{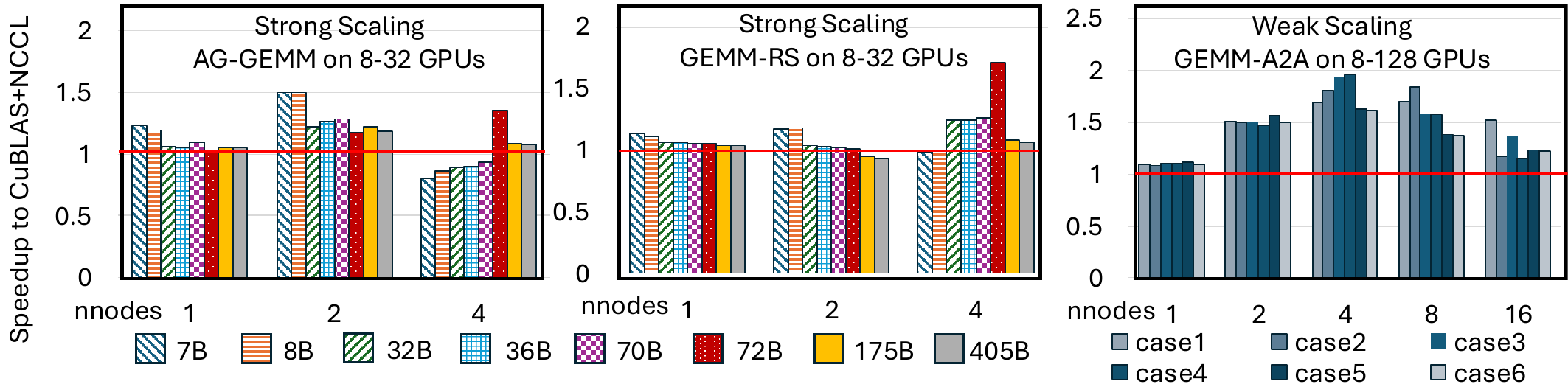}
\end{center}
\caption{Scaling TP and SP workloads on Hopper (96GB HBM) GPU clusters: Results are presented as speedups over CuBLAS+NCCL, with shapes derived from various real-world LLMs (details in Appendix~\ref{appendix:shape}). \ours{} maintains speedups for AG-GEMM across up to 16 GPUs. For 32 GPUs, speedups are achievable only with shapes from large LLMs. The speedup of GEMM-RS remains consistent across 8–32 GPUs. GEMM-A2A exhibits the best weak scaling due to its scalable AllToAll.}
\label{fig:scaling-tp-sp}
\end{figure*}

\begin{figure}
    \centering
    \includegraphics[width=0.45\textwidth]{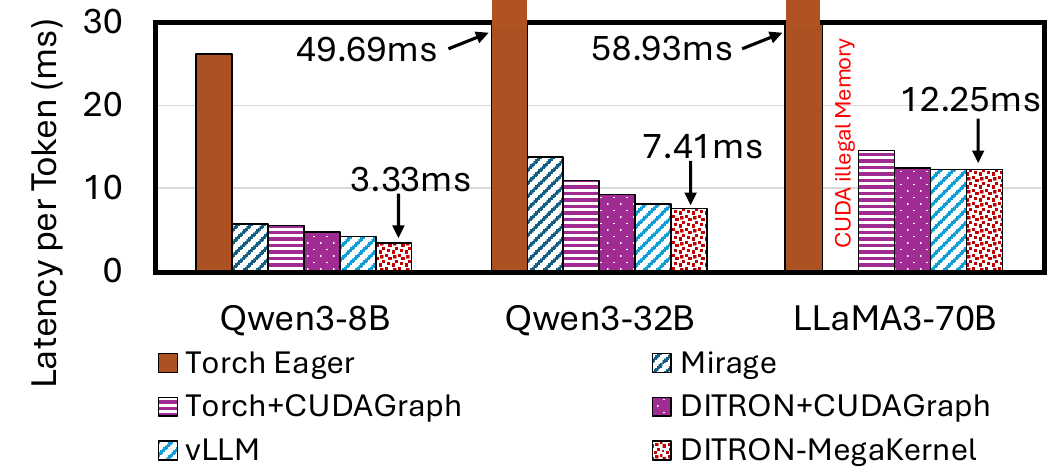}
    \caption{Inference latency for Qwen3-8B, Qwen3-32B, LLaMA-70B using \ours{}'s distributed Megakernel on $8\times$ H800.}
    \label{fig:mega-small}
\end{figure}

\textbf{Module-level Evaluation:} 
We incorporate the aforementioned workloads into attention modules and FFN modules. The QKV projection and output projection in attention modules are replaced by \ours{} kernels. Prefill and decode performance is presented in Figure~\ref{fig:module}. Prefill input lengths range from 512 to 16k, and decode output lengths range from 128 to 4k. The two lines represent the difference in speedups over CuBLAS+NCCL between using AllGather+ReduceScatter and using AllReduce.
% We integrate the above workloads into attention modules and FFN modules. An attention module is composed of QKV projection, RMSNorm, RoPE, self-attention, and output projection workloads, where QKV projection and output projection use \ours{} kernels. We show the prefill and decode performance in Figure~\ref{fig:module}. Prefill input length ranges from 512 to 16k. Decode output length ranges from 128 to 4k. The two lines represent the speedup (to CuBLAS+NCCL) difference of using AllGather+ReduceScatter and non-AllReduce. 
The results indicate that for attention modules, both prefill and decode prefer AllReduce communication because the reduction dimension for GEMM ( head dimension size in attention) is small (128), making GEMM latency not enough to hide communication latency of AllGather or ReduceScatter. The geometric mean speedup of \ours{}'s attention module (using AllReduce) to CuBLAS+NCCL is $1.12\times$ for prefill and $1.26\times$ for decode.
On the other hand, GEMMs in FFN modules use large intermediate size and the latency of GEMMs can hide communication latency when given enough input tokens (large batch or sequence length), so overlapping AllGather and ReduceScatter is better than using AllReduce for sequence length larger than 2k. The speedup for 128k tokens of \ours{} is $1.17\times$ using AllReduce and $1.27\times$ using AllGather+ReduceScatter.

\textbf{End-to-end Evaluation:} We evaluate \ours{} using end-to-end models LLaMA3-70B~\citep{llama3} and Qwen3-32B~\citep{qwen3}. We integrate \ours{} into vLLM~\cite{vllm} and compare the performance with/without \ours{}. The results are shown in Figure~\ref{fig:e2e}. vLLM~\citep{vllm} without \ours{} is still a strong baseline as vLLM natively employs efficient AllReduce kernels designed by experts. For batch size less than 128, vLLM is slightly better than \ours{} in throughput. But for batch sizes larger than 128, \ours{} achieves $5\%-30\%$ speedup to vLLM.
Specially, for batch size 512, we show the detailed speedup under different input lengths and output lengths for LLaMA3-70B and Qwen3-32B in Figure~\ref{fig:e2e}. The speedup translates to 12k tokens/s throughput for LLaMA3-70B and 17k tokens/s throughput for Qwen3-32B.

\textbf{Distributed MegaKernel Evaluation:}
For single batch inference, the hardware resources are usually underutilized, task-level scheduling that produces MegaKernel can eliminate kernel launch overhead, increase SM activity, and improve end-to-end performance. We compare \ours{}'s distributed MegaKernel with PyTorch (Eager mode and CUDAGraph mode), \ours{}+CUDAGraph, and Mirage~\citep{mirage} as shown in Figure~\ref{fig:mega-small}. The geometric speedup  is $6.28\times$ to Torch Eager, $1.73\times$ to Mirage, $1.33\times$ to Torch+CUDAGraph, $1.11\times$ to \ours{}+CUDAGraph, and $1.10\times$ to vLLM.
We also add MegaKernel code examples in Appendix~\ref{appendix:code-mega}.

\subsection{Evaluation for Training}

% \textbf{Scale Single Workload:}
For training, we evaluate both strong scaling and weak scaling performance for TP (AllGather+GEMM, GEMM+ReduceScatter),  SP (GEMM+AllToAll), and EP (MoE Dispatch and MoE Combine) using \ours{}. The results for TP and SP are shown in Figure~\ref{fig:scaling-tp-sp} and the results for EP are shown in Figure~\ref{fig:scaling-ep}. The detailed shape configurations are put in Appendix~\ref{appendix:shape}. More results (on different GPUs) are put in Appendix~\ref{appendix:experiments}.

\begin{figure}
    \centering
    \includegraphics[width=0.45\textwidth]{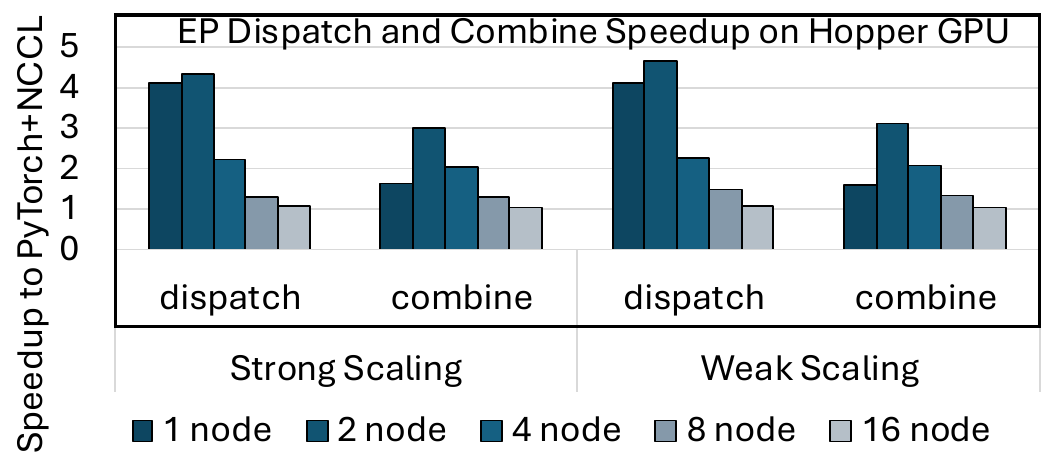}
    \caption{Scaling EP dispatch and combine using \ours{}.}
    \label{fig:scaling-ep}
\end{figure}

TP workloads renders heavy communication among nodes and the low bandwidth of network communication makes scaling non-beneficial. As a result, we only observe speedups of AllGather+GEMM and GEMM+ReduceScatter for 8-32 GPUs (strong scaling, number of total tokens is 32768), which ranges from $0.80\times$ to $1.71\times$ to CuBLAS+NCCL. For cases where the speedup is lower than 1, the main cause is GEMM becomes too small for each GPU after sharding, which cannot hide communication latency. As a comparison, GEMM+AllToAll (GEMM-A2A) gives consistent speedups for weak scaling (sequence length per rank remains unchanged) from 8 GPUs to 128 GPUs, where the shape of GEMM keeps constant.
As for EP (we set 8192 tokens per rank, topk 8, hidden size 7168), dispatch and combine performance remains similar for strong scaling (globally 512 experts in total) and weak scaling (8 experts per rank). The speedup to PyTorch+NCCL implementations ranges from $1.04\times$ to $4.70\times$.

% \textbf{End-to-end Training of MoE Moodels:}
% \ours{} has been validated in industrial training settings on $11,520$ H20 GPUs using a training task that scales a Qwen3-like~\footnote{{Details omitted to protect commercial secrets}} MoE model to around 640B (33B activated). 
% \ours{}'s kernels are bitiwse aligned with native PyTorch and NCCL implementations and t
% The achieved MFU is around 0.29. Compared to native PyTorch implementations, \ours{} helps saving millions of GPU hours per month.

\subsection{Support for Other Platforms}
\ours{} can be transferred to more hardware platforms due to the flexibility of our distributed IR and the compatibility of dependent Triton compiler. For now, we manage to support AMD GPUs and PCIe GPUs. We put our preliminary results in Appendix~\ref{appendix:experiments}. Overall, on AMD GPUs the speedup to RocmBLAS+RCCL ranges from $2\%-38\%$; on PCIe GPUs the average speedup to CuBLAS+NCCL is $8.33\times$.

% \section{Deployment in Front-end Seed Models}

% \textbf{Training Integration.}
In training tasks ranging from models with several billion parameters to hundreds of billions of parameters, different layers are accelerated by \ours{}.
For the Attention module, we employ SP Attention.
We overlap the GEMM and AllToAll communication.
Compared with the original Megatron~\cite{megatron-lm} implementation, the speedup of attention projection part exceeds 20\%.
For the MoE module, we overlap Dispatch, GroupedGEMM, and Combine operations.
Compared with the native Megatron implementation, the end-to-end gain reaches 10\%.
Even compared with the highly tuned handwritten CUDA overlap implementation (FLUX~\cite{flux}), we achieve equivalent performance, while reducing the code length by more than an order of magnitude and cutting the development cycle from months to days.
For the Optimizer, we support optimizations for the Muon Optimizer~\cite{muon}, delivering a speedup of over 20\% for optimization step.
For Pipeline Parallelism, we implement efficient PP communication kernels.
It can saturate the bandwidth using only 8 SMs within a node, and only 1 SM across nodes, while supporting flexible overlap with other layers.
All our adapted kernels are bitwise identical to the native implementations, fully ensuring the accuracy and stability of training.

\textbf{Inference Integration.}
For inference services, we deploy high-performance TP inference services in the cloud.
We accelerate TP inference by overlapping and fusing AllReduce and GEMM operations, supporting various inference scenarios including PCIe and NVLink GPUs, with an end-to-end performance speedup of approximately 20\%.
For edge inference, such as inference on robotic devices, we provide high-performance TP Attention and TP GEMM implementations, achieving an end-to-end speedup of over 30\%.

%% file: contents/5-conclusion.tex
\section{Conclusion}

Distributed inference and training become necessary and require more researchers to be able to program distributed kernels.
This work presents \ours{}, a flexible and versatile distributed compiler with core-level, device-level, and task-level interfaces for overlapping kernels. Researchers can use \ours{} to program efficient distributed kernels for different parallelism with performance comparable to or better than expert-tuned kernels.

%% file: contents/appendix-swizzle.tex
\begin{figure}[!t]
\begin{center}
%\framebox[4.0in]{$\;$}
% \fbox{\rule[-.5cm]{0cm}{4cm} \rule[-.5cm]{4cm}{0cm}}
\includegraphics[width=\textwidth]{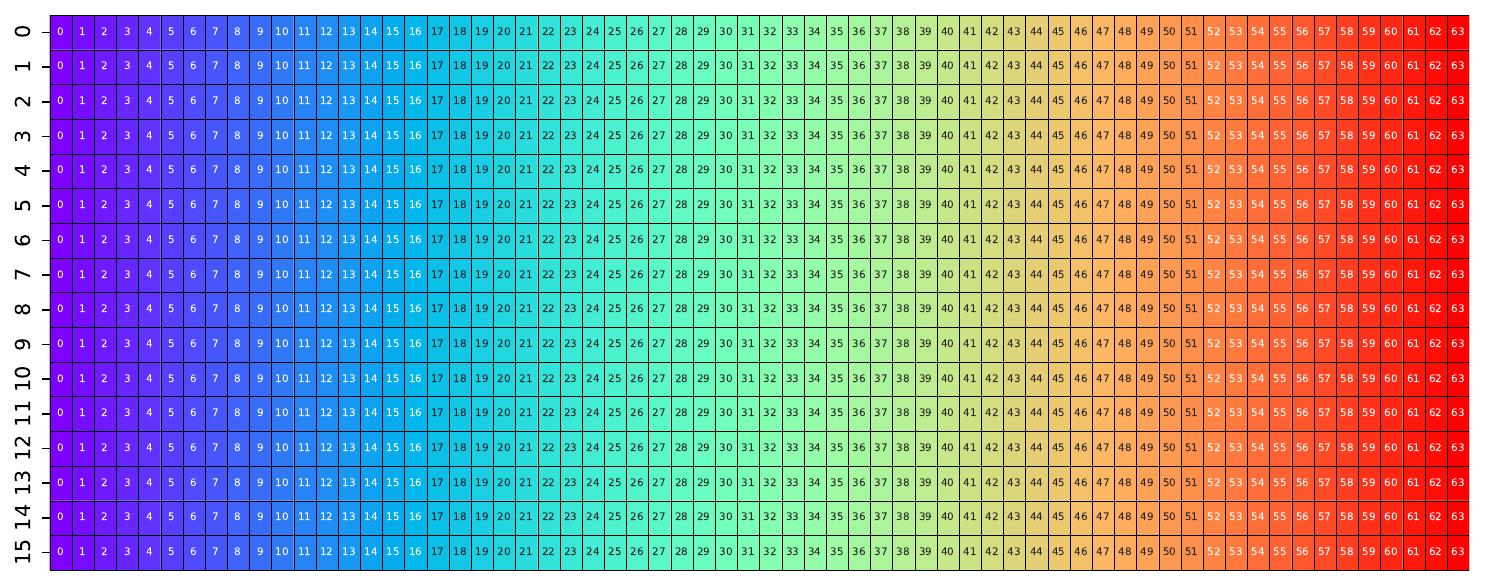}
\end{center}
\caption{Global view of ranks and data tiles without swizzling. Corresponding to Figure~\ref{fig:swizzle}. Each row corresponds to one rank; each column corresponds to one tile. Each rank computes GEMM from the left tile to the right tile. This example is for GEMM+ReduceScatter.}
\label{fig:no-swizzle-perfect-64}
\end{figure}

\begin{figure}[!t]
\begin{center}
%\framebox[4.0in]{$\;$}
% \fbox{\rule[-.5cm]{0cm}{4cm} \rule[-.5cm]{4cm}{0cm}}
\includegraphics[width=\textwidth]{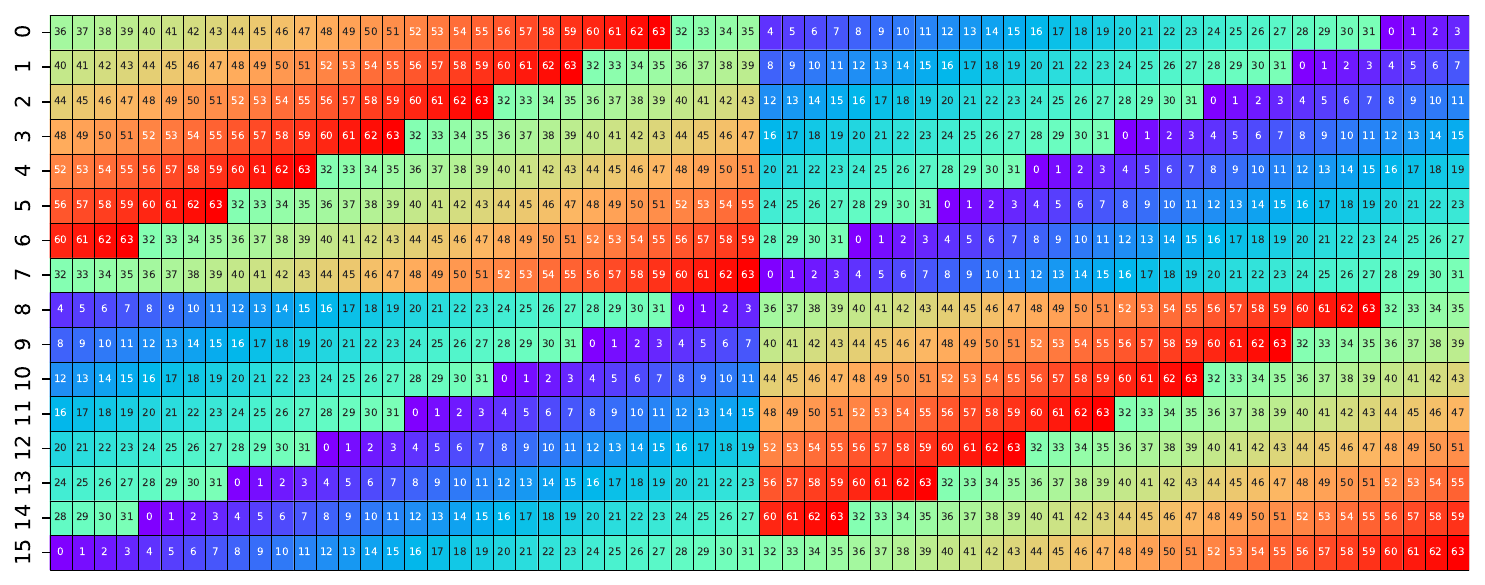}
\end{center}
\caption{Global view of ranks and data tiles with swizzling. Corresponding to Figure~\ref{fig:swizzle}. This example is for GEMM+ReduceScatter.}
\label{fig:swizzle-perfect-64}
\end{figure}

\begin{figure}[!t]
\begin{center}
%\framebox[4.0in]{$\;$}
% \fbox{\rule[-.5cm]{0cm}{4cm} \rule[-.5cm]{4cm}{0cm}}
\includegraphics[width=0.7\textwidth]{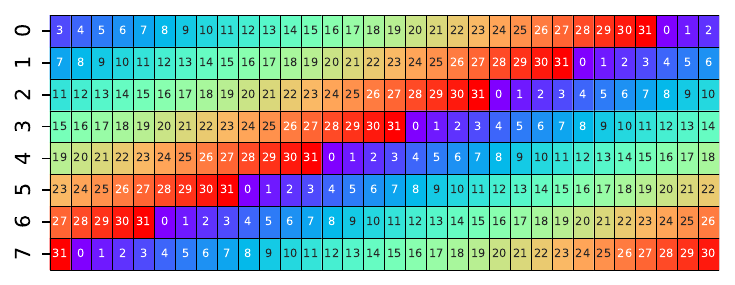}
\end{center}
\caption{Single node global view of swizzling for non-perfect tiling. The tile orders are different from that of perfect tiling. Cross-rank tiles are permuted to left. This example is for GEMM+ReduceScatter.}
\label{fig:swizzle-non-perfect-8gpu-64}
\end{figure}

\begin{figure}[!t]
\begin{center}
%\framebox[4.0in]{$\;$}
% \fbox{\rule[-.5cm]{0cm}{4cm} \rule[-.5cm]{4cm}{0cm}}
\includegraphics[width=\textwidth]{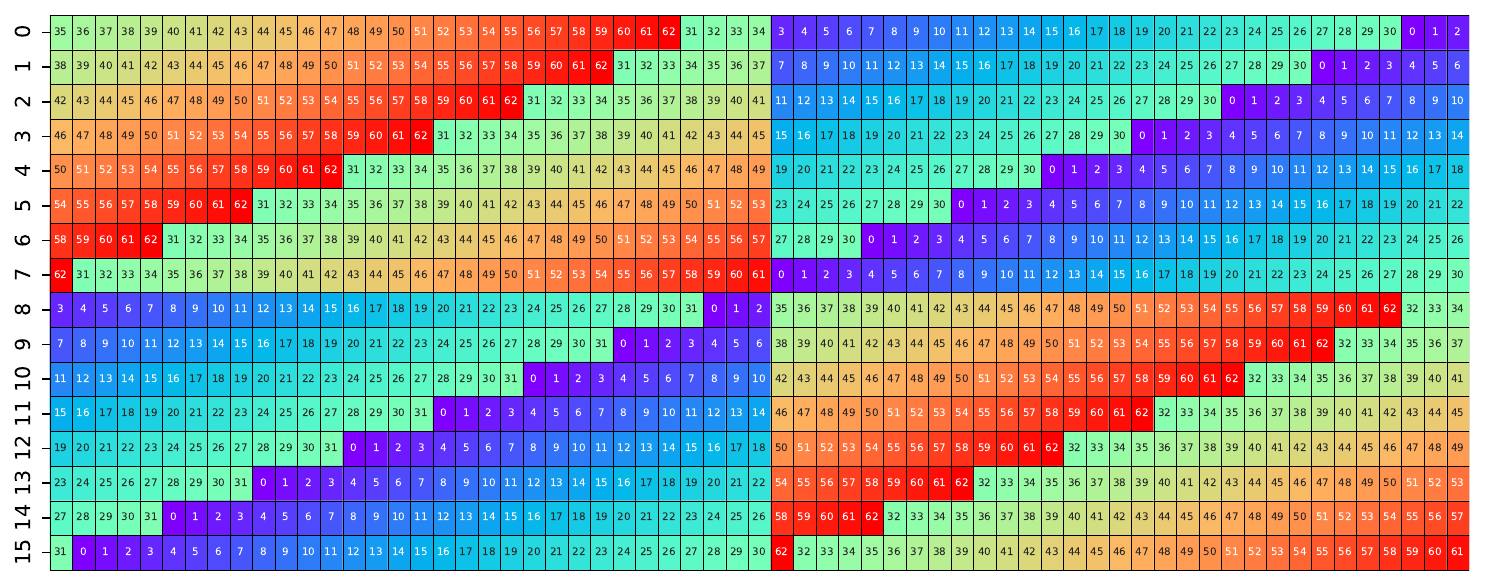}
\end{center}
\caption{Two nodes global view of swizzling for non-perfect tiling. This example is for GEMM+ReduceScatter.}
\label{fig:swizzle-non-perfect-16gpu-64}
\end{figure}

\begin{figure}[!t]
\begin{center}
%\framebox[4.0in]{$\;$}
% \fbox{\rule[-.5cm]{0cm}{4cm} \rule[-.5cm]{4cm}{0cm}}
\includegraphics[width=\textwidth]{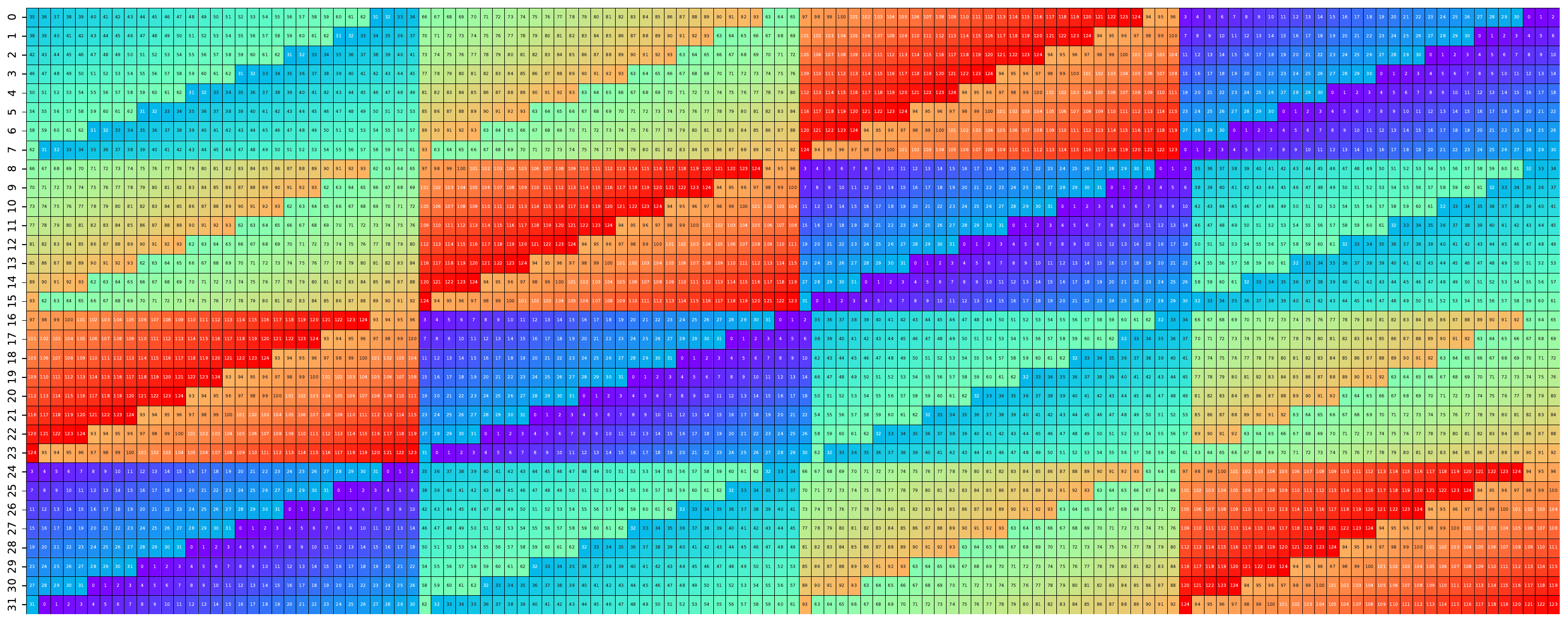}
\end{center}
\caption{Four nodes global view of swizzling for non-perfect tiling. This example is for GEMM+ReduceScatter.}
\label{fig:swizzle-non-perfect-32gpu-64}
\end{figure}

Swizzling optimization has been mentioned in previous work~\citep{coconet, flux, tilelink}. But their swizzling methods are limited to single node and there lack a clear explanation of the swizzling for non-perfect tiling scenarios (which is the most common for LLM training and inference).

Figure~\ref{fig:no-swizzle-perfect-64} and Figure~\ref{fig:swizzle-perfect-64} show an enlarged view of Figure~\ref{fig:swizzle}, which correspond to swizzle optimization for 16 GPUs GEMM+ReduceScatter with perfect shapes (no cross-rank tiles). For this case, we use $M\_per\_rank=1024, N\_per\_rank=256, tile\_M=tile\_N=256$.

For non-perfect shapes, some tiles are needed by multiple ranks for subsequent ReduceScatter. These tiles should be permuted to the left so that they can be computed earlier and transferred to the ranks that need them. Figure~\ref{fig:swizzle-non-perfect-8gpu-64} shows swizzle results for 8 GPUs; Figure~\ref{fig:swizzle-non-perfect-16gpu-64} shows swizzling for 16 GPUs; and Figure~\ref{fig:swizzle-non-perfect-32gpu-64} shows swizzling for 32 GPUs. For this case, we use $M\_per\_rank=997, N\_per\_rank=256, tile\_M=tile\_N=256$.

Besides GEMM+ReduceScatter, we also implement similar swizzling for AllGather+GEMM, and AllGather+MoE.
The swizzle logic is implemented in \ours{} through efficient kernels. We show the core code for swizzle as follows.

First, swizzle code for AllGather+GEMM. \ours{} is implemented based on Triton, so we still use \textit{triton.jit} as compile entry. We extend programming interface with multiple primitives for tile-level or even thread-level control, such as \textit{laneid, warp\_prefix\_sum\_kernel, \_\_ballot\_sync, \_\_shfl\_sync\_i32}.
% \begin{tcblisting}{listing engine=minted,boxrule=0.1mm,
% listing only,left=5mm,
% breakable,
% skin=enhanced,
% minted language=python,
% minted options={fontsize=\tiny,breaklines, breakanywhere, 
% autogobble,linenos,numbersep=2mm}}
\begin{tcblisting}{
    listing engine=listings,    % 引擎改为 listings
    boxrule=0.1mm,
    listing only,
    left=5mm,
    breakable,
    skin=enhanced,
    listing options={           % 原有的参数映射到这里
        language=Python,
        basicstyle=\tiny\ttfamily,
        breaklines=true,
        columns=fullflexible,   % 允许更紧凑的间距
        numbers=left,           % 对应 linenos
        numberstyle=\tiny\color{gray},
        numbersep=2mm,          % 对应 numbersep=2mm
        gobble=0,               % listings 的自动缩进处理（类似 autogobble）
        % 【重点】在这里定义 listings 的转义符
        escapeinside={|}{|},     
        showstringspaces=false
    }
}
@triton.jit(do_not_specialize=["rank"])
def threadblock_swizzle_allgather_gemm_kernel(
    tiled_m,
    M,
    rank,
    WORLD_SIZE: tl.constexpr,
    NNODES: tl.constexpr,
    BLOCK_SIZE_M: tl.constexpr,
    DEBUG: tl.constexpr = False,
):
    LOCAL_WORLD_SIZE = WORLD_SIZE // NNODES
    node_id = rank // LOCAL_WORLD_SIZE
    M_per_rank = M // WORLD_SIZE
    M_per_node = M // NNODES
    node_start = node_id

    lane_id = laneid()

    if lane_id < NNODES:
        n = (lane_id + node_start) % NNODES
        M_node_start = M_per_node * n
        M_node_end = M_per_node * (n + 1)
        tiled_m_node_start = M_node_start // BLOCK_SIZE_M
        prev_tiled_m_node_end = (M_node_start - 1) // BLOCK_SIZE_M
        tiled_m_node_end = (M_node_end - 1) // BLOCK_SIZE_M
        next_tiled_m_node_start = M_node_end // BLOCK_SIZE_M

        if lane_id == 0 and M_node_start != 0:
            if prev_tiled_m_node_end == tiled_m_node_start:
                tiled_m_node_start += 1

        if lane_id == 0 and M_node_end != M:
            if next_tiled_m_node_start == tiled_m_node_end:
                tiled_m_node_end -= 1

        if lane_id != NNODES - 1 and M_node_end != M:
            if next_tiled_m_node_start == tiled_m_node_end:
                tiled_m_node_end -= 1
        if DEBUG and lane_id == NNODES - 1:
            print("tiled_m_node_end", tiled_m_node_end)

        swizzled_tiled_m_size = tiled_m_node_end - tiled_m_node_start + 1
    else:
        swizzled_tiled_m_size = 0

    if DEBUG and lane_id < NNODES:
        print("swizzled_tiled_m_size", swizzled_tiled_m_size, lane_id)
    swizzled_tiled_m_size_accum = (warp_prefix_sum_kernel(swizzled_tiled_m_size, lane_id, NNODES) -
                                   swizzled_tiled_m_size)
    if DEBUG and lane_id < NNODES:
        print("swizzled_tiled_m_size_accum", swizzled_tiled_m_size_accum)

    tiled_m_size_l = __shfl_down_sync_i32(0xFFFFFFFF, swizzled_tiled_m_size, NNODES - node_start)
    tiled_m_size_r = __shfl_up_sync_i32(0xFFFFFFFF, swizzled_tiled_m_size, node_start)
    tiled_m_size = 0
    if lane_id < node_start:
        tiled_m_size = tiled_m_size_l
    elif lane_id < NNODES:
        tiled_m_size = tiled_m_size_r

    if DEBUG and lane_id < NNODES:
        print("tiled_m_size", tiled_m_size)

    tiled_m_size_accum = warp_prefix_sum_kernel(tiled_m_size, lane_id, NNODES) - tiled_m_size
    mask = __ballot_sync(0xFFFFFFFF, tiled_m < swizzled_tiled_m_size_accum)
    n = ffs(mask) - 1 - 1
    if DEBUG and lane_id < NNODES:
        print("tiled_m_size_accum", tiled_m_size_accum)
        print("n", n, tiled_m, swizzled_tiled_m_size_accum, mask)

    # map node
    nid = (n + node_start) % NNODES
    node_offset = __shfl_sync_i32(0xFFFFFFFF, swizzled_tiled_m_size_accum, n)

    tile_size = __shfl_sync_i32(0xFFFFFFFF, swizzled_tiled_m_size, n)

    tiled_m_intra_node = tiled_m - node_offset
    local_rank = rank % LOCAL_WORLD_SIZE
    m_start = M_per_node * nid + M_per_rank * local_rank
    tiled_m_start = tl.cdiv(m_start, BLOCK_SIZE_M)
    swizzled_node_offset = __shfl_sync_i32(0xFFFFFFFF, tiled_m_size_accum, nid)
    rank_offset = max(0, tiled_m_start - swizzled_node_offset)  # this may < 0, bad

    # map rank
    tiled_m_intra_node_new = (tiled_m_intra_node + rank_offset) % tile_size
    return swizzled_node_offset + tiled_m_intra_node_new
\end{tcblisting}

Then, swizzle code for GEMM+ReduceScatter.
% \begin{tcblisting}{listing engine=minted,boxrule=0.1mm,
% listing only,left=5mm,
% breakable,
% skin=enhanced,
% minted language=python,
% minted options={fontsize=\tiny,breaklines, breakanywhere, 
% autogobble,linenos,numbersep=2mm}}
\begin{tcblisting}{
    listing engine=listings,    % 引擎改为 listings
    boxrule=0.1mm,
    listing only,
    left=5mm,
    breakable,
    skin=enhanced,
    listing options={           % 原有的参数映射到这里
        language=Python,
        basicstyle=\tiny\ttfamily,
        breaklines=true,
        columns=fullflexible,   % 允许更紧凑的间距
        numbers=left,           % 对应 linenos
        numberstyle=\tiny\color{gray},
        numbersep=2mm,          % 对应 numbersep=2mm
        gobble=0,               % listings 的自动缩进处理（类似 autogobble）
        % 【重点】在这里定义 listings 的转义符
        escapeinside={|}{|},     
        showstringspaces=false
    }
}
@triton.jit(do_not_specialize=["rank"])
def threadblock_swizzle_gemm_reduce_scatter_kernel(
    tiled_m,
    M,
    rank,
    WORLD_SIZE: tl.constexpr,
    NNODES: tl.constexpr,
    BLOCK_SIZE_M: tl.constexpr,
    DEBUG: tl.constexpr = False,
):
    LOCAL_WORLD_SIZE = WORLD_SIZE // NNODES
    node_id = rank // LOCAL_WORLD_SIZE
    M_per_rank = M // WORLD_SIZE
    M_per_node = M // NNODES
    node_start = node_id + 1

    lane_id = laneid()

    if lane_id < NNODES:
        n = (lane_id + node_start) % NNODES
        M_node_start = M_per_node * n
        M_node_end = M_per_node * (n + 1)
        tiled_m_node_start = M_node_start // BLOCK_SIZE_M
        # if tiled_m_start_node overlaps with previous node, then we need to add 1 to tiled_m_start_node
        prev_tiled_m_node_end = (M_node_start - 1) // BLOCK_SIZE_M
        if lane_id != 0 and M_node_start != 0:
            if prev_tiled_m_node_end == tiled_m_node_start:
                tiled_m_node_start += 1

        tiled_m_node_end = (M_node_end - 1) // BLOCK_SIZE_M
        next_tiled_m_node_start = M_node_end // BLOCK_SIZE_M
        if lane_id == NNODES - 1 and M_node_end != M:
            if next_tiled_m_node_start == tiled_m_node_end:
                tiled_m_node_end -= 1

        swizzled_tiled_m_size = tiled_m_node_end - tiled_m_node_start + 1
    else:
        swizzled_tiled_m_size = 0

    if DEBUG and lane_id < NNODES:
        print("swizzled_tiled_m_size", swizzled_tiled_m_size, lane_id)
    swizzled_tiled_m_size_accum = (warp_prefix_sum_kernel(swizzled_tiled_m_size, lane_id, NNODES) -
                                   swizzled_tiled_m_size)
    if DEBUG and lane_id < NNODES:
        print("swizzled_tiled_m_size_accum", swizzled_tiled_m_size_accum)
    # thread 0 hold node `node_start` size
    # thread 1 hold node `node_start + 1` size
    # ...
    # thread NNODES - node_start hold node `0` size
    # thread NNODES - 1 hold NNODES `node_start - 1` size

    #  => thread 0 want to hold node 0 size

    tiled_m_size_l = __shfl_down_sync_i32(0xFFFFFFFF, swizzled_tiled_m_size, NNODES - node_start)
    tiled_m_size_r = __shfl_up_sync_i32(0xFFFFFFFF, swizzled_tiled_m_size, node_start)
    tiled_m_size = 0
    if lane_id < node_start:
        tiled_m_size = tiled_m_size_l
    elif lane_id < NNODES:
        tiled_m_size = tiled_m_size_r

    if DEBUG and lane_id < NNODES:
        print("tiled_m_size", tiled_m_size)

    tiled_m_size_accum = warp_prefix_sum_kernel(tiled_m_size, lane_id, NNODES) - tiled_m_size
    mask = __ballot_sync(0xFFFFFFFF, tiled_m < swizzled_tiled_m_size_accum)
    n = ffs(mask) - 1 - 1
    if DEBUG and lane_id < NNODES + 1:
        print("tiled_m_size_accum", tiled_m_size_accum)
        print("n", n, tiled_m, swizzled_tiled_m_size_accum, mask)

    # map node
    nid = (n + node_start) % NNODES
    node_offset = __shfl_sync_i32(0xFFFFFFFF, swizzled_tiled_m_size_accum, n)

    tile_size = __shfl_sync_i32(0xFFFFFFFF, swizzled_tiled_m_size, n)

    tiled_m_intra_node = tiled_m - node_offset
    local_rank = rank % LOCAL_WORLD_SIZE
    m_start = M_per_node * nid + M_per_rank * (local_rank + 1)
    tiled_m_start = m_start // BLOCK_SIZE_M
    swizzled_node_offset = __shfl_sync_i32(0xFFFFFFFF, tiled_m_size_accum, nid)
    rank_offset = max(0, tiled_m_start - swizzled_node_offset)  # this may < 0, bad

    # map rank
    tiled_m_intra_node_new = (tiled_m_intra_node + rank_offset) % tile_size
    return swizzled_node_offset + tiled_m_intra_node_new
\end{tcblisting}

At last, the swizzle code for AllGather+MoE.
% \begin{tcblisting}{listing engine=minted,boxrule=0.1mm,
% listing only,left=5mm,
% breakable,
% skin=enhanced,
% minted language=python,
% minted options={fontsize=\tiny,breaklines, breakanywhere, 
% autogobble,linenos,numbersep=2mm}}
\begin{tcblisting}{
    listing engine=listings,    % 引擎改为 listings
    boxrule=0.1mm,
    listing only,
    left=5mm,
    breakable,
    skin=enhanced,
    listing options={           % 原有的参数映射到这里
        language=Python,
        basicstyle=\tiny\ttfamily,
        breaklines=true,
        columns=fullflexible,   % 允许更紧凑的间距
        numbers=left,           % 对应 linenos
        numberstyle=\tiny\color{gray},
        numbersep=2mm,          % 对应 numbersep=2mm
        gobble=0,               % listings 的自动缩进处理（类似 autogobble）
        % 【重点】在这里定义 listings 的转义符
        escapeinside={|}{|},     
        showstringspaces=false
    }
}
@triton.jit(do_not_specialize=["rank"])
def threadblock_swizzle_ag_moe_kernel(
    # input
    ntokens_by_rank_by_expert_ptr,
    # output
    expert_id_ptr,
    tiled_m_ptr,
    segment_start_ptr,
    segment_end_ptr,
    ntiles_ptr,
    # workspace buffer
    ntokens_by_expert_by_rank_acc_ptr,
    ntiles_by_expert_acc_ptr,
    ntiles_by_expert_by_stage_ptr,
    ntiles_by_expert_by_stage_acc_ptr,
    rank,
    N_EXPERTS: tl.constexpr,
    TP_SIZE: tl.constexpr,
    LOCAL_TP_SIZE: tl.constexpr,
    NTILES_NEXT_POW_OF_2: tl.constexpr,
    BLOCK_SIZE_M: tl.constexpr,
    DEBUG: tl.constexpr = False,
):
    """
    tile_index = g(expert_id, stage, index): if tile_index is grouped by expert then stage

    but how to map (stage, index) => tile_index_in_expert?

    first map tile_index_in_expert => (stage, off_in_expert_by_stage, index_in_expert_in_stage)

    tile_index => (expert_id, tile_index_in_expert)                 : ntokens grouped by expert_id by rank
               => (expert_id, stage)                                : get_stage_id
               => (expert_id, off_in_expert_by_stage)               : cumsum by stage
               => (expert_id, off_in_expert_by_stage, index_in_expert_in_stage)  : atomic_add for index_in_expert_in_stage
               => tile_index_new : by function g

    tile_index -> (expert_id, rank_offset) grouped by (expert_id, rank)
    rank_offset -> segment_start, segment_end, stage

    so we can remap tile_index as:
    tile_index_new = g(f(tile_index))
    """
    thread_idx = tid(0)
    N_EXPERTS_NEXT_POW_OF_2: tl.constexpr = next_power_of_2(N_EXPERTS)
    TP_SIZE_NEXT_POW_OF_2: tl.constexpr = next_power_of_2(TP_SIZE)
    offs_by_expert = tl.arange(0, N_EXPERTS_NEXT_POW_OF_2)
    mask_by_expert = offs_by_expert < N_EXPERTS
    offs_by_rank = tl.arange(0, TP_SIZE_NEXT_POW_OF_2)
    mask_by_rank = (offs_by_rank < TP_SIZE)
    offs_by_expert_by_rank = offs_by_expert[:, None] * TP_SIZE + offs_by_rank[None, :]
    mask_by_expert_by_rank = mask_by_expert[:, None] & mask_by_rank[None, :]
    offs_by_rank_by_expert = offs_by_rank[:, None] * N_EXPERTS + offs_by_expert[None, :]
    mask_by_rank_by_expert = mask_by_rank[:, None] & mask_by_expert[None, :]
    ntokens_by_rank_by_expert = tl.load(ntokens_by_rank_by_expert_ptr + offs_by_rank_by_expert,
                                        mask=mask_by_rank_by_expert)

    ntokens_by_expert_by_rank = ntokens_by_rank_by_expert.T
    ntokens_by_expert_by_rank_acc = tl.cumsum(ntokens_by_expert_by_rank, axis=1)
    ntokens_by_expert = tl.sum(ntokens_by_rank_by_expert, axis=0)
    ntiles_by_expert = tl.cdiv(ntokens_by_expert, BLOCK_SIZE_M)
    ntiles_by_expert_acc = tl.cumsum(ntiles_by_expert, axis=0)
    ntiles = tl.sum(ntiles_by_expert)

    tl.store(ntokens_by_expert_by_rank_acc_ptr + offs_by_expert_by_rank, ntokens_by_expert_by_rank_acc,
             mask=mask_by_expert_by_rank)
    tl.store(ntiles_by_expert_acc_ptr + offs_by_expert, ntiles_by_expert_acc)

    # # for each tiled_m in expert eid => stage id / segment_start / segmeng_end / tiled_m
    tile_index = tl.arange(0, NTILES_NEXT_POW_OF_2)
    mask_tile_idx = tile_index < ntiles

    __syncthreads()
    # tile_index -> (expert_id, offset_by_expert, tile_index_in_expert) -> stage, segment_start, segment_end
    expert_id, off_by_expert, off_in_expert = bisect_right_with_offset_kernel(ntiles_by_expert_acc_ptr, tile_index,
                                                                              N_EXPERTS)

    stage, segment_start, segment_end = get_tile_stage(
        off_in_expert,
        rank,
        TP_SIZE,
        LOCAL_TP_SIZE,
        BLOCK_SIZE_M,
        ntokens_by_expert_by_rank_acc_ptr + expert_id * TP_SIZE,
    )

    # histogram by expert by stage
    tl.store(ntiles_by_expert_by_stage_ptr + offs_by_expert_by_rank, 0, mask=mask_by_expert_by_rank)
    __syncthreads()
    off_in_expert_in_stage = tl.atomic_add(ntiles_by_expert_by_stage_ptr + expert_id * TP_SIZE + stage, 1,
                                           sem="relaxed", scope="gpu", mask=mask_tile_idx)
    __syncthreads()

    # do some cumsum
    ntiles_by_expert_by_stage = tl.load(ntiles_by_expert_by_stage_ptr + offs_by_expert_by_rank,
                                        mask=mask_by_expert_by_rank, other=0)
    ntiles_by_expert_by_stage_acc = tl.cumsum(ntiles_by_expert_by_stage, axis=1)
    __syncthreads()
    tl.store(
        ntiles_by_expert_by_stage_acc_ptr + offs_by_expert_by_rank,
        ntiles_by_expert_by_stage_acc,
        mask=mask_by_expert_by_rank,
    )
    __syncthreads()

    off_in_expert_by_stage = tl.where(
        stage == 0,
        0,
        tl.load(ntiles_by_expert_by_stage_acc_ptr + expert_id * TP_SIZE + stage - 1, mask=mask_tile_idx, other=0),
    )

    tile_index_by_expert_by_stage = off_by_expert + off_in_expert_by_stage + off_in_expert_in_stage
    __syncthreads()

    tl.store(expert_id_ptr + tile_index_by_expert_by_stage, expert_id, mask=mask_tile_idx)
    tl.store(tiled_m_ptr + tile_index_by_expert_by_stage, tile_index, mask=mask_tile_idx)
    tl.store(segment_start_ptr + tile_index_by_expert_by_stage, segment_start, mask=mask_tile_idx)
    tl.store(segment_end_ptr + tile_index_by_expert_by_stage, segment_end, mask=mask_tile_idx)
    thread_idx = tid(0)
    if thread_idx == 0:
        tl.store(ntiles_ptr, ntiles)
    if DEBUG and thread_idx < ntiles:
        print("expert_id", expert_id)
\end{tcblisting}

%% file: contents/appendix-shape.tex
\subsection{Evaluation Setup}
The H800 clusters used in our evaluation are equipped with NVLink of uni-directional bandwidth of 200 GB/s. There are 8 NICs in one node and each GPU can communicate with another GPU in other nodes at a uni-directional bandwidth of 50 GB/s.
The Hopper 96GB HBM clusters used in our evaluation are equipped with NVLink of uni-directional bandwidth of 450 GB/s. There are 4 NICs in one node and each GPU can communicate with another GPU in other nodes at a uni-directional bandwidth of 25 GB/s.

\begin{table}[htbp]
  \centering
  \caption{Shapes for AllGather+GEMM on H800.}
  \label{table:ag-gemm-h800}  % 用于文中引用的标签
  \begin{tabular}{l|l|c|c|c}  % l=左对齐，c=居中对齐；4列对应4个字段
    \hline
    No. & From Model & M    & N     & K      \\
    \hline
    1 & LLaMA3-7B     & 8192 & 11008 & 4096   \\
    2 & LLaMA3.1-8B & 8192 & 14336 & 4096   \\
    3 & LLaMA3.1-70B& 8192 & 28672 & 8192   \\
    4 & LLaMA3.1-405B& 8192 & 53248 & 16384  \\
    5 & Mistral-7B   & 8192 & 14336 & 4096   \\
    6 & Qwen2-72B    & 8192 & 29568 & 8192   \\
    \hline
  \end{tabular}
\end{table}

\begin{table}[htbp]
  \centering
  \caption{Shapes for GEMM+ReduceScatter and GEMM+AllReduce on H800.}
  \label{table:gemm-ar-h800}
  \begin{tabular}{l|l|c|c|c}
    \hline
    No. & From Model & M    & N      & K     \\ \hline  % 交换N列与K列的顺序
    1 & LLaMA3-7B     & 8192 & 4096   & 11008 \\ 
    2 & LLaMA3.1-8B & 8192 & 4096   & 14336 \\ 
    3 & LLaMA3.1-70B& 8192 & 8192   & 28672 \\ 
    4 & LLaMA3.1-405B& 8192 & 16384  & 53248 \\ 
    5 & Mistral-7B   & 8192 & 4096   & 14336 \\ 
    6 & Qwen2-72B    & 8192 & 8192   & 29568 \\ 
    \hline
  \end{tabular}
\end{table}

\begin{table}[htbp]
  \centering
  \caption{Shapes for AllGather+MoE and MoE+AllReduce on H800.}
  \label{table:moe-shape}
  \begin{tabular}{l|l|c|c|c|c|c}
    \hline
    No. & From Model & num\_tokens & hidden\_dim & intermediate\_size & num\_experts & topk \\ \hline
    1 & Qwen1.5-MoE-A2.7B & 8192        & 2048        &  1408               & 60           & 4    \\
    2 & Mixtral-8x7B & 8192        & 14336       & 4096               & 8            & 2    \\
    3 & Mixtral-8x22B & 8192        & 16384       & 6144               & 8            & 2    \\
    4 & DeepSeek-MoE & 8192        & 1408        & 2048               & 64           & 6    \\
    \hline
  \end{tabular}
\end{table}

\begin{table}[htbp]
  \centering
  \caption{Shapes for strong scaling AllGather+GEMM.}
  \label{table:shape-strong-scale-ag-gemm}
  \begin{tabular}{c|c|c|c|c}
    \hline
    Label & From Model & M & N & K \\ 
    \hline
    7B & LLaMA3-7B & 32768 & 11008 & 4096 \\ 
    \hline
    8B & LLaMA3.1-8B & 32768 & 14336 & 4096 \\ 
    \hline
    32B & Qwen3-32B & 32768 & 25600 & 5120 \\ 
    \hline
    36B & Seed-OSS-36B & 32768 & 27648 & 5120 \\ 
    \hline
    70B & LLaMA3.1-70B & 32768 & 28672 & 8192 \\ 
    \hline
    72B & Qwen2-72B & 32768 & 29568 & 8192 \\ 
    \hline
    175B & GPT-3-175B & 32768 & 49152 & 12288 \\ 
    \hline
    405B & LLaMA3.1-405B & 32768 & 53248 & 16384 \\ 
    \hline
  \end{tabular}
\end{table}

\begin{table}[htbp]
  \centering
  \caption{Shapes for strong scaling GEMM+ReduceScatter.}
  \label{table:shape-strong-scale-gemm-rs}
  \begin{tabular}{c|c|c|c|c}
    \hline
    1 & From Model & M & N & K \\ 
    \hline
    7B & LLaMA3-7B & 32768 & 4096 & 11008 \\ 
    \hline
    8B & LLaMA3.1-8B & 32768 & 4096 & 14336 \\ 
    \hline
    32B & Qwen3-32B & 32768 & 5120 & 25600 \\ 
    \hline
    36B & Seed-OSS-36B & 32768 & 5120 & 27648 \\ 
    \hline
    70B & LLaMA3.1-70B & 32768 & 8192 & 28672 \\ 
    \hline
    72B & Qwen2-72B & 32768 & 8192 & 29568 \\ 
    \hline
    175B & GPT-3-175B & 32768 & 12288 & 49152 \\ 
    \hline
    405B & LLaMA3.1-405B & 32768 & 16384 & 53248 \\ 
    \hline
  \end{tabular}
\end{table}

\begin{table}[htbp]
  \centering
  \caption{Shapes for strong scaling GEMM+AllToAll.}
  \label{table:shape-strong-scale-gemm-a2a}
  \begin{tabular}{c|c|c|c}
    \hline
    Label & seqlen & heads & head\_size \\ 
    \hline
    case 1 & 262144 & 64 & 128 \\ 
    \hline
    case 2 & 491520 & 64 & 128 \\ 
    \hline
    case 3 & 262144 & 32 & 128 \\ 
    \hline
    case 4 & 491520 & 32 & 128 \\ 
    \hline
    case 5 & 262144 & 16 & 128 \\ 
    \hline
    case 6 & 491520 & 16 & 128 \\ 
    \hline
  \end{tabular}
\end{table}

\subsection{Workload Configuration}
\label{appendix:shape}

The input shapes for AllGather+GEMM used in Figure~\ref{fig:experiments-h800} are listed in Table~\ref{table:ag-gemm-h800}.
The input shapes for GEMM+ReduceScatter and GEMM+AllReduce used in Figure~\ref{fig:experiments-h800} are listed in Table~\ref{table:gemm-ar-h800}.
The input shapes for AllGather+MoE and MoE+AllReduce used in Figure~\ref{fig:experiments-h800} are listed in Table~\ref{table:moe-shape}.

For scaling experiments, the shapes for strong scaling AllGather+GEMM are shown in Table~\ref{table:shape-strong-scale-ag-gemm}. The shapes remain the same from 8 GPUs to 32 GPUs.
The shapes for weak scaling are the same as those in strong scaling except that the value of $M$ varies with the number of GPUs. We use $M=8192$ for 8 GPUs, $M=16384$ for 16 GPUs, and $M=32768$ for 32 GPUs.
The shapes for strong scaling GEMM+ReduceScatter are shown in Table~\ref{table:shape-strong-scale-gemm-rs}. The shapes remain the same for 8 GPUs and 32 GPUs.
For weak scaling, the configurations for $M$ are the same as weak scaling AllGather+GEMM.

The shapes for strong scaling GEMM+AllToAll are shown in Table~\ref{table:shape-strong-scale-gemm-a2a}. The shapes remain the same from 8 GPUs to 128 GPUs.
The shapes for weak scaling are the same as those in strong scaling except that the seqlen varies with the number of GPUs.
For 8 GPUs, seqlens are 65536 and 122880; for 16 GPUs, they are 131072 and 245760; for 32 GPUs, they are 262144 and 491520; for 64 GPUs, they are 524288 and 983040; for 128 GPUs, they are 1048576 and 1966080.

%% file: contents/appendix-experiments.tex
\subsection{Detailed Evaluation Results for MoE}
\label{appendix:result-moe}

The detailed results for AllGather+MoE in Figure~\ref{fig:experiments-h800} are shown in Table~\ref{table:results-ag-moe}. The detailed results for MoE+AllReduce in Figure~\ref{fig:experiments-h800} are shown in Table~\ref{table:results-moe-ar}. The shapes for each case are shown in Table~\ref{table:moe-shape}.

\begin{table}[htbp]
  \centering
  \caption{Latency of AllGather+MoE (ms).}
  \label{table:results-ag-moe}
  \begin{tabular}{c|c|c|c|c}
    \hline
    & CuBLAS+NCCL & TileLink & COMET & \ours{} \\ \hline
    1 & 18.12       & 0.56     & 0.33  & 0.32   \\ \hline
    2 & 7.12        & 2.14     & 1.48  & 1.51   \\ \hline
    3 & 8.49        & 3.30     & 1.85  & 1.72   \\ \hline
    4 & 24.27       & 0.65     & 0.28  & 0.24   \\ \hline
  \end{tabular}
\end{table}

\begin{table}[htbp]
  \centering
  \caption{Latency of MoE+AllReduce (ms).}
  \label{table:results-moe-ar}
  \begin{tabular}{c|c|c|c}
    \hline
    & CuBLAS+NCCL & TileLink & \ours{} \\ \hline
    1 & 22.04       & 1.44     & 0.445  \\ \hline
    2 & 7.99        & 3.70     & 2.603  \\ \hline
    3 & 8.97        & 4.32     & 3.289  \\ \hline
    4 & 28.64       & 0.84     & 0.441  \\ \hline
  \end{tabular}
\end{table}

\begin{table}[htbp]
  \centering
  \caption{Strong scaling results (speedup to CuBLAS+NCCL) for AllGather+GEMM on H800.}
  \label{table:strong-scale-ag-gemm-h800}
  \begin{tabular}{c|c|c|c}
    \hline
    From Model & 1 Node & 2 Node & 4 Node \\ 
    \hline
    LLaMA3-7B & 1.31 & 1.11 & 0.63 \\ 
    \hline
    LLaMA3.1-8B & 1.35 & 1.12 & 0.61 \\ 
    \hline
    Qwen3-32B & 1.59 & 1.32 & 0.65 \\ 
    \hline
    Seed-OSS-36B & 1.64 & 1.32 & 0.64 \\ 
    \hline
    LLaMA3.1-70B & 1.67 & 1.36 & 0.77 \\ 
    \hline
    Qwen2-72B & 1.67 & 1.39 & 1.03 \\ 
    \hline
    GPT-3-175B & 1.46 & 1.68 & 1.00 \\ 
    \hline
    LLaMA3.1-405B & 1.44 & 1.76 & 1.02 \\ 
    \hline
  \end{tabular}
\end{table}

\begin{table}[htbp]
  \centering
  \caption{Weak scaling results (speedup to CuBLAS+NCCL) for AllGather+GEMM on Hopper 96GB HBM GPUs}
  \label{table:weak-scale-ag-gemm-h20}
  \begin{tabular}{c|c|c|c}
    \hline
    From Model & 1 Node & 2 Node & 4 Node \\ 
    \hline
    LLaMA3-7B & 1.23 & 1.50 & 0.80 \\ 
    \hline
    LLaMA3.1-8B & 1.19 & 1.50 & 0.86 \\ 
    \hline
    Qwen3-32B & 1.06 & 1.22 & 0.89 \\ 
    \hline
    Seed-OSS-36B & 1.05 & 1.26 & 0.89 \\ 
    \hline
    LLaMA3.1-70B & 1.10 & 1.29 & 0.93 \\ 
    \hline
    Qwen2-72B & 1.01 & 1.18 & 1.36 \\ 
    \hline
    GPT-3-175B & 1.05 & 1.23 & 1.09 \\ 
    \hline
    LLaMA3.1-405B & 1.05 & 1.19 & 1.08 \\ 
    \hline
  \end{tabular}
\end{table}

\begin{table}[htbp]
  \centering
  \caption{Weak scaling results (speedup to CuBLAS+NCCL) for AllGather+GEMM on H800}
  \label{table:weak-scale-ag-gemm-h800}
  \begin{tabular}{c|c|c|c}
    \hline
    From Model & 1 Node & 2 Node & 4 Node \\ 
    \hline
    LLaMA3-7B & 1.30 & 0.98 & 0.63 \\ 
    \hline
    LLaMA3.1-8B & 1.30 & 1.02 & 0.61 \\ 
    \hline
    Qwen3-32B & 1.53 & 1.14 & 0.65 \\ 
    \hline
    Seed-OSS-36B & 1.57 & 1.15 & 0.64 \\ 
    \hline
    LLaMA3.1-70B & 1.61 & 1.25 & 0.77 \\ 
    \hline
    Qwen2-72B & 1.66 & 1.29 & 1.03 \\ 
    \hline
    GPT-3-175B & 1.43 & 1.58 & 1.00 \\ 
    \hline
    LLaMA3.1-405B & 1.34 & 1.63 & 1.02 \\ 
    \hline
  \end{tabular}
\end{table}

\subsection{Detailed Evaluation Results for Scaling}

In Figure~\ref{fig:scaling-tp-sp} we shave shown strong scaling results for AllGather+GEMM on Hopper 96GB GPUs. Here we show strong scaling results on H800 and weak scaling results on both Hopper 96GB GPUs and H800 in Table~\ref{table:strong-scale-ag-gemm-h800}, Table~\ref{table:weak-scale-ag-gemm-h20}, and Table~\ref{table:weak-scale-ag-gemm-h800}.

\begin{figure*}[!t]
\begin{center}
%\framebox[4.0in]{$\;$}
% \fbox{\rule[-.5cm]{0cm}{4cm} \rule[-.5cm]{4cm}{0cm}}
\includegraphics[width=\textwidth]{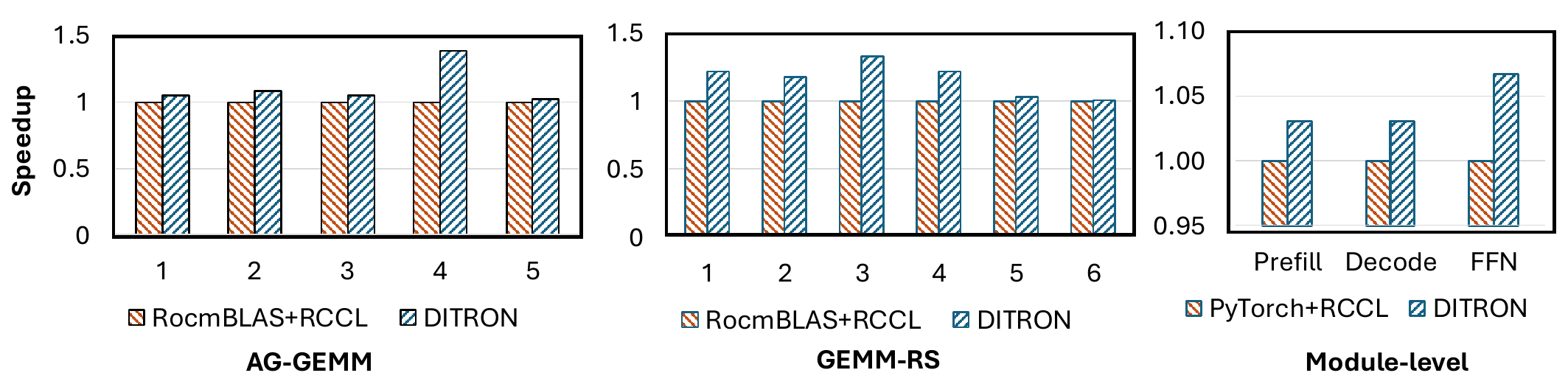}
\end{center}
\caption{Evaluation results on $8\times$AMD GPUs. AllGather+GEMM and GEMM+ReduceScatter shapes are in Table~\ref{table:ag-gemm-h800} and Table~\ref{table:gemm-ar-h800}. Module-level evaluation uses configurations from Qwen3-32B.}
\label{fig:amd-perf}
\end{figure*}

\subsection{Evaluation Results on AMD GPU}

\ours{} can be deployed to AMD GPUs through RocSHMEM~\citep{rocshmem} and Triton for AMD. For evaluation, we use shapes from Table~\ref{table:ag-gemm-h800} and Table~\ref{table:gemm-ar-h800}.
Our baseline is RocmBLAS+RCCL.
Overall, the speedup ranges from $1.02\times$ to $1.38\times$.
The geometric speedup to baseline is $1.11\times$ for AllGather+GEMM and $1.16\times$ to GEMM+ReduceScatter.
For module-level evaluation, we use Qwen3-32B. The speedup for prefill attention is $1.03\times$, the speedup for decode attention is $1.03\times$, and the speedup for FFN is $1.07\times$.

\begin{table}[htbp]
  \centering
  \caption{GEMM+RS performance on $8\times$PCIe GPUs.}
  \label{tab:l20_gemm_rs}
  \begin{tabular}{cccccc}
    \toprule
    M     & N     & K      & CuBLAS+NCCL & DITRON & Speedup \\
    \midrule
    8192  & 8192  & 29568  & 18.63 ms       & 7.34 ms & $2.54\times$   \\
    \bottomrule
  \end{tabular}
\end{table}

\begin{table}[htbp]
  \centering
  \caption{All2All performance on $8\times$PCIe GPU.}
  \label{tab:l20_all2all}
  \begin{tabular}{ccccccc}
    \toprule
    num\_tokens & hidden\_size & num\_experts & topk & NCCL  & DITRON & speedup \\
    \midrule
    8       & 3584         & 128          & 8    & 0.42 ms  & 0.09 ms & $4.51\times$   \\
    \bottomrule
  \end{tabular}
\end{table}

\begin{table}[htbp]
  \centering
  \caption{AG+MoE performance on $8\times$PCIe GPU (shape configurations see Table~\ref{table:moe-shape}).}
  \label{tab:l20_ag_moe}
  \begin{tabular}{ccccc}
    \toprule
    No. & num\_tokens     & CuBLAS+NCCL & DITRON & speedup \\
    \midrule
    1       & 2048  & 21.10 ms       & 0.58 ms & $36.26\times$   \\
    2       & 2048  & 43.91 ms       & 2.67 ms & $16.47\times$   \\
    3       & 2048  & 51.11 ms      & 3.44 ms & $14.86\times$   \\
    4       & 2048  & 21.43 ms       & 0.43 ms & $49.84\times$   \\
    \bottomrule
  \end{tabular}
\end{table}

\begin{table}[htbp]
  \centering
  \small
  \caption{MoE+RS performance on $8\times$PCIe GPU.}
  \label{tab:l20_moe_rs}
  \begin{tabular}{cccccccc}
    \toprule
    num\_tokens     & hidden\_size & intermediate\_size & num\_experts & topk & CuBLAS+NCCL & DITRON & speedup \\
    \midrule
    8192  & 2048         & 1536               & 32           & 2    & 15.80 ms      & 1.71 ms & $9.26\times$  \\
    \bottomrule
  \end{tabular}
\end{table}

\begin{table}[htbp]
  \centering
  \small
  \caption{MoE+AR performance on $8\times$PCIe GPU.}
  \label{tab:l20_moe_ar}
  \begin{tabular}{ccccccccc}
    \toprule
    num\_tokens     & hidden\_size & intermediate\_size & num\_experts & topk & CuBLAS+NCCL & DITRON & speedup \\
    \midrule
    8192  & 2048         & 1536               & 32           & 2    & 18.19 ms      & 6.54 ms & $2.78\times$  \\
    \bottomrule
  \end{tabular}
\end{table}

\begin{table}[htbp]
  \centering
  \caption{A2A+GEMM performance on $8\times$PCIe GPU.}
  \label{tab:l20_a2a_gemm}
  \begin{tabular}{ccccccc}
    \toprule
    DType & M     & N     & K      & CuBLAS+NCCL & DITRON & speedup \\
    \midrule
    Int8   & 7168  & 9216  & 3072   & 7.11 ms       & 2.15 ms & $3.31\times$ \\
    FP8    & 7168  & 9216  & 3072   & 8.06  ms      & 2.17 ms & $3.71\times$ \\
    \bottomrule
  \end{tabular}
\end{table}

\subsection{Evaluation Results on PCIe GPU}
We evaluate 6 different workloads on PCIe GPUs, including GEMM+ReduceScatter (Table~\ref{tab:l20_gemm_rs}), AllToAll (Table~\ref{tab:l20_all2all}), AllGather+MoE (Table~\ref{tab:l20_ag_moe}), MoE+ReduceScatter (Table~\ref{tab:l20_moe_rs}), MoE+AllReduce (Table~\ref{tab:l20_moe_ar}), and AllToAll+GEMM (Table~\ref{tab:l20_a2a_gemm}). The geometric mean speedup of all the workloads is $8.33\times$.

%% file: contents/appendix-code.tex
In this section, we provide some simplified code examples, including All-Gather GEMM, GEMM Reduce-Scatter and GEMM All-Reduce. For each example, the code is generally divided into three main parts: the host-side launcher, the producer kernel, and the consumer kernel.

\subsection{All-Gather GEMM}
\label{appendix:code-ag-gemm}

Here we provide the code example for intra-node All-Gather GEMM. The following is the host-side Python function for launching the All-Gather GEMM operation. It orchestrates the producer (communication) and consumer (computation) kernels.

\begin{tcblisting}{
listing engine=listings,    % 切换引擎
    boxrule=0.1mm,
    listing only,
    left=5mm,
    breakable,
    skin=enhanced,
    % 原有的 minted options 全部合并到这里或由上面的 lstset 控制
    listing options={
        language=Python,
        basicstyle=\tiny\ttfamily,
        breaklines=true
    }
}
def ag_gemm(a, b, ctx: AllGatherGEMMTensorParallelContext, use_cooperative=True):
    """allgather gemm
    Allgather global matrix A and do matmul with local matrix B, produces local matrix C

    Args:
        a (torch.Tensor<float>): local matmul A matrix. shape: [M_per_rank, K]
        b (torch.Tensor<float>): local matmul B matrix. shape: [N_per_rank, K]
        ctx: (AllGatherGEMMTensorParallelContext, Optional): if not provided, created immediately

    Returns:
        C (torch.Tensor<float>): local matmul C matrix. shape: [M, N_per_rank]
    """

    assert a.shape[1] == b.shape[
        1], f"tensor_B should has shape (col_major) [{b.shape[0]}, {a.shape[1]}], but get [{b.shape}]"
    assert a.dtype == b.dtype, f"Dtype of input and weight must be same: tensor_A dtype {a.dtype}, tensor_B dtype {b.dtype}"

    M_per_rank, K = a.shape
    N_per_rank, _ = b.shape

    assert a.shape[0] * ctx.num_ranks <= ctx.max_M and a.shape[1] == ctx.K, f"Shape of tensor_A must not exceed the maxmize M of ctx: tensor_A shape [{a.shape}], ctx shape [{ctx.max_M},{ctx.K}]"
    assert b.shape[0] == ctx.N_per_rank, f"N_per_rank of tensor_B must match that of ctx: tensor_B shape [{b.shape[0]}], ctx shape [{ctx.N_per_rank}]"
    assert ctx.tensor_dtype == a.dtype, f"dtype of ctx must match that of ctx: tensor_A dtype {a.dtype}, ctx dtype {ctx.tensor_dtype}"

    C = torch.empty([ctx.num_ranks * M_per_rank, N_per_rank], dtype=a.dtype, device=a.device)

    local_copy_and_barrier_all(ctx.local_rank, ctx.rank, ctx.num_ranks, a, ctx.symm_workspace, ctx.symm_comm_buf, ctx.symm_barrier, M_per_rank, K, ctx.phase, is_internode=False, use_cooperative=use_cooperative)
    ctx.phase += 2

    current_stream = torch.cuda.current_stream()
    ctx.ag_intranode_stream.wait_stream(current_stream)

    cp_engine_producer_all_gather_full_mesh_pull(ctx.rank, ctx.num_ranks, a, ctx.symm_workspaces, ctx.symm_barriers, ctx.ag_intranode_stream, for_correctness=ctx.for_correctness, all_gather_method=ctx.all_gather_method)

    M_per_rank, K = a.shape
    M = M_per_rank * ctx.num_ranks
    grid = lambda META: (triton.cdiv(M, META["BLOCK_SIZE_M"]) * triton.cdiv(ctx.N_per_rank, META["BLOCK_SIZE_N"]), )
    kernel_consumer_gemm[grid](ctx.symm_workspace[:M], b, C, M, ctx.N_per_rank, ctx.K, ctx.symm_workspace.stride(0), ctx.symm_workspace.stride(1), b.stride(1), b.stride(0), c.stride(0), C.stride(1), ctx.rank, ctx.num_ranks, ctx.symm_barrier, ctx.BLOCK_M, ctx.BLOCK_N, ctx.BLOCK_K, ctx.GROUP_SIZE_M, num_stages=ctx.stages, num_warps=ctx.warps)
    current_stream.wait_stream(ctx.ag_intranode_stream)

    return C
\end{tcblisting}

The producer kernel below is responsible for the All-Gather communication. It copies the local data tensor to the symmetric memory (\texttt{remote\_tensor\_buffers}) of other GPUs and then sets a signal (\texttt{\_set\_signal\_cuda}) to notify the consumer kernels that the data is ready.

% \begin{tcblisting}{listing engine=minted,boxrule=0.1mm,
% listing only,left=5mm,
% breakable,
% skin=enhanced,
% minted language=python,
% minted options={fontsize=\tiny,breaklines, breakanywhere, 
% autogobble,linenos,numbersep=2mm}}
% \begin{tcblisting}{
%     listing engine=minted,
%     boxrule=0.1mm,
%     listing only,
%     left=5mm,
%     breakable,
%     skin=enhanced,
%     minted language=python,
%     minted options={
%         fontsize=\tiny,
%         breaklines, 
%         breakanywhere, 
%         autogobble,
%         linenos,
%         numbersep=2mm,
%         % 【重点】在这里定义转义符，意思是 |...| 中间的是 LaTeX 命令 
%         escapeinside=|| 
%     }
% }
\begin{tcblisting}{
    listing engine=listings,    % 引擎改为 listings
    boxrule=0.1mm,
    listing only,
    left=5mm,
    breakable,
    skin=enhanced,
    listing options={           % 原有的参数映射到这里
        language=Python,
        basicstyle=\tiny\ttfamily,
        breaklines=true,
        columns=fullflexible,   % 允许更紧凑的间距
        numbers=left,           % 对应 linenos
        numberstyle=\tiny\color{gray},
        numbersep=2mm,          % 对应 numbersep=2mm
        gobble=0,               % listings 的自动缩进处理（类似 autogobble）
        % 【重点】在这里定义 listings 的转义符
        escapeinside={|}{|},     
        showstringspaces=false
    }
}
def cp_engine_producer_all_gather_full_mesh_pull(
    rank,
    num_ranks,
    local_tensor: torch.Tensor,
    remote_tensor_buffers: List[torch.Tensor],
    barrier_buffers: List[torch.Tensor],
    stream: torch.cuda.Stream,
):
    M_per_rank, N = local_tensor.shape

    rank_orders = [(rank + i) % num_ranks for i in range(num_ranks)]

    with torch.cuda.stream(stream):
        for src_rank in rank_orders:
            if src_rank == rank:
                continue
            dst = remote_tensor_buffers[rank][src_rank * M_per_rank:(src_rank + 1) * M_per_rank, :]
            src = remote_tensor_buffers[src_rank][src_rank * M_per_rank:(src_rank + 1) * M_per_rank, :]
            dst.copy_(src)
            |\hlcode{\_set\_signal\_cuda(barrier\_buffers[rank][src\_rank], 1, stream)}|
\end{tcblisting}

The consumer kernel first waits on a signal (\texttt{dl.wait}) indicating that the required data are available. Once the data is ready, it proceeds with the GEMM computation on the data.

% \begin{tcblisting}{
%     listing engine=minted,
%     boxrule=0.1mm,
%     listing only,
%     left=5mm,
%     breakable,
%     skin=enhanced,
%     minted language=python,
%     minted options={
%         fontsize=\tiny,
%         breaklines, 
%         breakanywhere, 
%         autogobble,
%         linenos,
%         numbersep=2mm,
%         % 【重点】在这里定义转义符，意思是 |...| 中间的是 LaTeX 命令 
%         escapeinside=|| 
%     }
% }
\begin{tcblisting}{
    listing engine=listings,    % 引擎改为 listings
    boxrule=0.1mm,
    listing only,
    left=5mm,
    breakable,
    skin=enhanced,
    listing options={           % 原有的参数映射到这里
        language=Python,
        basicstyle=\tiny\ttfamily,
        breaklines=true,
        columns=fullflexible,   % 允许更紧凑的间距
        numbers=left,           % 对应 linenos
        numberstyle=\tiny\color{gray},
        numbersep=2mm,          % 对应 numbersep=2mm
        gobble=0,               % listings 的自动缩进处理（类似 autogobble）
        % 【重点】在这里定义 listings 的转义符
        escapeinside={|}{|},     
        showstringspaces=false
    }
}
@triton.jit(do_not_specialize=["rank"])
def kernel_consumer_gemm(
        a_ptr, b_ptr, c_ptr,
        M, N, K,
        stride_am, stride_ak,
        stride_bk, stride_bn,
        stride_cm, stride_cn, rank, WORLD_SIZE: tl.constexpr, barrier_ptr,
        BLOCK_SIZE_M: tl.constexpr, BLOCK_SIZE_N: tl.constexpr, BLOCK_SIZE_K: tl.constexpr,  #
        GROUP_SIZE_M: tl.constexpr,  #
):
    a_dtype = a_ptr.dtype.element_ty
    b_dtype = b_ptr.dtype.element_ty
    c_dtype = c_ptr.dtype.element_ty
    tl.static_assert(a_dtype == b_dtype, "A and B must have the same dtype")

    pid = tl.program_id(axis=0)
    num_pid_m = tl.cdiv(M, BLOCK_SIZE_M)
    num_pid_n = tl.cdiv(N, BLOCK_SIZE_N)
    num_pid_in_group = GROUP_SIZE_M * num_pid_n
    group_id = pid // num_pid_in_group
    first_pid_m = group_id * GROUP_SIZE_M
    group_size_m = min(num_pid_m - first_pid_m, GROUP_SIZE_M)
    pid_m = first_pid_m + ((pid % num_pid_in_group) % group_size_m)
    pid_n = (pid % num_pid_in_group) // group_size_m

    m_per_rank = M // WORLD_SIZE
    m_offset = m_per_rank * rank
    pid_m_offset = tl.cdiv(m_offset, BLOCK_SIZE_M)
    pid_m = (pid_m + pid_m_offset) % num_pid_m

    offs_am = pid_m * BLOCK_SIZE_M
    rank_beg = offs_am // m_per_rank
    rank_end = (min(offs_am + BLOCK_SIZE_M, M) - 1) // m_per_rank
    |\hlcode{token = dl.wait(barrier\_ptr + rank\_beg, rank\_end - rank\_beg + 1, \textquotesingle gpu \textquotesingle, \textquotesingle acquire \textquotesingle, waitValue=1)}|

    offs_am = (pid_m * BLOCK_SIZE_M + tl.arange(0, BLOCK_SIZE_M)) % M
    offs_bn = (pid_n * BLOCK_SIZE_N + tl.arange(0, BLOCK_SIZE_N)) % N
    offs_k = tl.arange(0, BLOCK_SIZE_K)
    a_ptrs = a_ptr + (offs_am[:, None] * stride_am + offs_k[None, :] * stride_ak)
    b_ptrs = b_ptr + (offs_k[:, None] * stride_bk + offs_bn[None, :] * stride_bn)

    |\hlcode{a\_ptrs = dl.consume\_token(a\_ptrs, token)}|

    if a_dtype == tl.int8:
        accumulator = tl.zeros((BLOCK_SIZE_M, BLOCK_SIZE_N), dtype=tl.int32)
    else:
        accumulator = tl.zeros((BLOCK_SIZE_M, BLOCK_SIZE_N), dtype=tl.float32)
    for k in range(0, tl.cdiv(K, BLOCK_SIZE_K)):
        a = tl.load(a_ptrs, mask=offs_k[None, :] < K - k * BLOCK_SIZE_K, other=0.0)
        b = tl.load(b_ptrs, mask=offs_k[:, None] < K - k * BLOCK_SIZE_K, other=0.0)
        accumulator += tl.dot(a, b)
        a_ptrs += BLOCK_SIZE_K * stride_ak
        b_ptrs += BLOCK_SIZE_K * stride_bk

    offs_cm = pid_m * BLOCK_SIZE_M + tl.arange(0, BLOCK_SIZE_M)
    offs_cn = pid_n * BLOCK_SIZE_N + tl.arange(0, BLOCK_SIZE_N)
    c_ptrs = c_ptr + stride_cm * offs_cm[:, None] + stride_cn * offs_cn[None, :]
    c_mask = (offs_cm[:, None] < M) & (offs_cn[None, :] < N)

    tl.store(c_ptrs, accumulator.to(c_dtype), mask=c_mask)
\end{tcblisting}

\subsection{GEMM Reduce-Scatter}
\label{appendix:code-gemm-rs}

Here we provide the code example for GEMM Reduce-Scatter. The following is the host-side Python function for launching the GEMM Reduce-Scatter operation.

% \begin{tcblisting}{listing engine=minted,boxrule=0.1mm,
% listing only,left=5mm,
% breakable,
% skin=enhanced,
% minted language=python,
% minted options={fontsize=\tiny,breaklines, breakanywhere, 
% autogobble,linenos,numbersep=2mm}}
\begin{tcblisting}{
    listing engine=listings,    % 引擎改为 listings
    boxrule=0.1mm,
    listing only,
    left=5mm,
    breakable,
    skin=enhanced,
    listing options={           % 原有的参数映射到这里
        language=Python,
        basicstyle=\tiny\ttfamily,
        breaklines=true,
        columns=fullflexible,   % 允许更紧凑的间距
        numbers=left,           % 对应 linenos
        numberstyle=\tiny\color{gray},
        numbersep=2mm,          % 对应 numbersep=2mm
        gobble=0,               % listings 的自动缩进处理（类似 autogobble）
        % 【重点】在这里定义 listings 的转义符
        escapeinside={|}{|},     
        showstringspaces=false
    }
}
def gemm_rs_op(input, weight, ctx: GEMMReduceScatterTensorParallelContext, persistent: bool = True,
               fuse_scatter: bool = False):
    world_size = ctx.rs_ctx.world_size
    local_world_size = ctx.rs_ctx.local_world_size
    rs_stream = ctx.rs_stream
    output_dtype = ctx.output_dtype
    num_gemm_sms = ctx.num_gemm_sms
    
    M, local_K = input.shape
    N = weight.shape[0]
    assert N == ctx.rs_ctx.N
    
    assert M % world_size == 0
    assert weight.shape[1] == local_K
    M_per_rank = M // world_size
    current_stream = torch.cuda.current_stream()
    rs_stream.wait_stream(current_stream)
    
    output = torch.empty((M_per_rank, N), dtype=output_dtype, device=input.device)
    workspace = torch.zeros((world_size, ), dtype=torch.int32, device=input.device)
    gemm_out = ctx.get_gemm_out_buf(input)
    scatter_signal = ctx.rs_ctx.scatter_signal_buf
    
    triton_config = triton.Config(
        {
            "BLOCK_SIZE_M": ctx.BLOCK_M, "BLOCK_SIZE_N": ctx.BLOCK_N, "BLOCK_SIZE_K": ctx.BLOCK_K, "GROUP_SIZE_M": ctx.GROUP_M
        }, num_stages=ctx.stages, num_warps=8)
    triton_config = update_triton_config(M, N, local_K, input.dtype, world_size, local_world_size, triton_config)
    gemm_rs_producer(input, weight, gemm_out, scatter_signal, workspace, world_size, local_world_size, fuse_scatter, triton_config)
    
    if not fuse_scatter:
        with torch.cuda.stream(rs_stream):
            reduce_scatter_2d_op(gemm_out, ctx.rs_ctx, output)
        current_stream.wait_stream(rs_stream)
    else:
        nvshmem_barrier_all_on_stream(current_stream)
        ring_reduce(gemm_out, output, ctx.rs_ctx.local_rank, local_world_size)
        nvshmem_barrier_all_on_stream(current_stream)
    return output
\end{tcblisting}

The producer kernel below is responsible for the GEMM computation. When \texttt{FUSE\_SCATTER} is enabled, it directly writes the output to the symmetric memory of the destination rank.

% \begin{tcblisting}{
%     listing engine=minted,
%     boxrule=0.1mm,
%     listing only,
%     left=5mm,
%     breakable,
%     skin=enhanced,
%     minted language=python,
%     minted options={
%         fontsize=\tiny,
%         breaklines, 
%         breakanywhere, 
%         autogobble,
%         linenos,
%         numbersep=2mm,
%         % 【重点】在这里定义转义符，意思是 |...| 中间的是 LaTeX 命令 
%         escapeinside=|| 
%     }
% }
\begin{tcblisting}{
    listing engine=listings,    % 引擎改为 listings
    boxrule=0.1mm,
    listing only,
    left=5mm,
    breakable,
    skin=enhanced,
    listing options={           % 原有的参数映射到这里
        language=Python,
        basicstyle=\tiny\ttfamily,
        breaklines=true,
        columns=fullflexible,   % 允许更紧凑的间距
        numbers=left,           % 对应 linenos
        numberstyle=\tiny\color{gray},
        numbersep=2mm,          % 对应 numbersep=2mm
        gobble=0,               % listings 的自动缩进处理（类似 autogobble）
        % 【重点】在这里定义 listings 的转义符
        escapeinside={|}{|},     
        showstringspaces=false
    }
}
@triton.jit(launch_metadata=_matmul_launch_metadata, repr=_gemm_rs_non_persistent_repr)
def kernel_gemm_rs_producer(
    # Pointers to matrices
    a_ptr,  # [M, K]_Ti
    b_ptr,  # [K, N]_Ti
    c_ptr,  # [M, N]_To
    # Matrix dimensions
    M,
    N,
    K,
    # The stride variables represent how much to increase the ptr by when moving by 1
    # element in a particular dimension. E.g. `stride_am` is how much to increase `a_ptr`
    # by to get the element one row down (A has M rows).
    stride_am,
    stride_ak,  #
    stride_bk,
    stride_bn,  #
    stride_cm,
    stride_cn,
    barrier_ptr,
    counter_ptr,
    FUSE_SCATTER: tl.constexpr,
    LOCAL_WORLD_SIZE: tl.constexpr,
    WORLD_SIZE: tl.constexpr,
    # Meta-parameters
    BLOCK_SIZE_M: tl.constexpr,
    BLOCK_SIZE_N: tl.constexpr,
    BLOCK_SIZE_K: tl.constexpr,  #
    GROUP_SIZE_M: tl.constexpr,
):
    """Kernel for computing the matmul C = A x B.
    A has shape (M, K), B has shape (K, N) and C has shape (M, N)
    """
    tl.static_assert(a_ptr.dtype.is_ptr(), "A should be a pointer")
    tl.static_assert(b_ptr.dtype.is_ptr(), "B should be a pointer")
    tl.static_assert(c_ptr.dtype.is_ptr(), "C should be a pointer")
    a_dtype = a_ptr.dtype.element_ty
    b_dtype = b_ptr.dtype.element_ty
    tl.static_assert(a_dtype == b_dtype, "A and B should have the same dtype")
    
    rank = dl.rank()
    NNODES = WORLD_SIZE // LOCAL_WORLD_SIZE
    
    pid = tl.program_id(axis=0)
    num_pid_m = tl.cdiv(M, BLOCK_SIZE_M)
    num_pid_n = tl.cdiv(N, BLOCK_SIZE_N)
    
    M_per_rank = M // WORLD_SIZE
    pid_m, pid_n = swizzle_2d(pid, num_pid_m, num_pid_n, GROUP_SIZE_M)
    
    if NNODES != 1:  # with complex threadblock swizzle logic
        pid_m = threadblock_swizzle_gemm_reduce_scatter_kernel(pid_m, M, rank, WORLD_SIZE, NNODES, BLOCK_SIZE_M)
    else:
        pid_m_offset = (rank + 1) * M_per_rank // BLOCK_SIZE_M
        pid_m = (pid_m + pid_m_offset) % num_pid_m
    
    # Create pointers for the first blocks of A and B.
    offs_am = (pid_m * BLOCK_SIZE_M + tl.arange(0, BLOCK_SIZE_M)) % M
    offs_bn = (pid_n * BLOCK_SIZE_N + tl.arange(0, BLOCK_SIZE_N)) % N
    offs_k = tl.arange(0, BLOCK_SIZE_K)
    a_ptrs = a_ptr + (offs_am[:, None] * stride_am + offs_k[None, :] * stride_ak)
    b_ptrs = b_ptr + (offs_k[:, None] * stride_bk + offs_bn[None, :] * stride_bn)
    
    # Iterate to compute a block of the C matrix.
    if a_ptr.dtype.element_ty == tl.int8:
        accumulator = tl.zeros((BLOCK_SIZE_M, BLOCK_SIZE_N), dtype=tl.int32)
    else:
        accumulator = tl.zeros((BLOCK_SIZE_M, BLOCK_SIZE_N), dtype=tl.float32)
    
    for k in range(0, tl.cdiv(K, BLOCK_SIZE_K)):
        # Load the next block of A and B, generate a mask by checking the K dimension.
        a = tl.load(a_ptrs, mask=offs_k[None, :] < K - k * BLOCK_SIZE_K, other=0.0)
        b = tl.load(b_ptrs, mask=offs_k[:, None] < K - k * BLOCK_SIZE_K, other=0.0)
        # We accumulate along the K dimension.
        accumulator += tl.dot(a, b)
        # Advance the ptrs to the next K block.
        a_ptrs += BLOCK_SIZE_K * stride_ak
        b_ptrs += BLOCK_SIZE_K * stride_bk
    
    # Write back the block of the output matrix C with masks.
    offs_cm = pid_m * BLOCK_SIZE_M + tl.arange(0, BLOCK_SIZE_M)
    offs_cn = pid_n * BLOCK_SIZE_N + tl.arange(0, BLOCK_SIZE_N)
    c_ptrs = c_ptr + stride_cm * offs_cm[:, None] + stride_cn * offs_cn[None, :]
    out_mask = (offs_cm[:, None] < M) & (offs_cn[None, :] < N)
    
    if not FUSE_SCATTER:
        tl.store(c_ptrs, accumulator, mask=out_mask)
        
        # inc barrier
        segment_start = pid_m * BLOCK_SIZE_M // M_per_rank
        segment_end = (min((pid_m + 1) * BLOCK_SIZE_M, M) - 1) // M_per_rank
        __syncthreads()
        segment = segment_start + tid(axis=0)
        if segment <= segment_end:
            m_start = M_per_rank * segment
            m_end = M_per_rank * (segment + 1) - 1
            tiled_m_start = m_start // BLOCK_SIZE_M
            tiled_m_end = m_end // BLOCK_SIZE_M
            tiled_m_size = tiled_m_end - tiled_m_start + 1
            val = atomic_add(counter_ptr + segment, 1, semantic="release", scope="gpu")
            if val == num_pid_n * tiled_m_size - 1:
                # or use other signal op semantic
                |\hlcode{st(barrier\_ptr + segment, 1, scope=\textquotesingle gpu \textquotesingle, semantic=\textquotesingle release \textquotesingle)}|
    else:
        rank_start = pid_m * BLOCK_SIZE_M // M_per_rank
        rank_end = (min((pid_m + 1) * BLOCK_SIZE_M, M) - 1) // M_per_rank
        for cur_rank in range(rank_start, rank_end + 1):
            m_start = max(M_per_rank * cur_rank, pid_m * BLOCK_SIZE_M)
            m_end = min(M_per_rank * (cur_rank + 1) - 1, (min((pid_m + 1) * BLOCK_SIZE_M, M) - 1))
            remote_c_ptr = dl.symm_at(c_ptr, cur_rank)
            mask_offset = m_start - pid_m * BLOCK_SIZE_M
            remote_offs_cm = m_start % M_per_rank + rank * M_per_rank + tl.arange(0, BLOCK_SIZE_M) - mask_offset
            remote_c_ptrs = remote_c_ptr + stride_cm * remote_offs_cm[:, None] + stride_cn * offs_cn[None, :]
            remote_mask = (offs_cm[:, None] <= m_end) & (offs_cm[:, None] >= m_start) & (offs_cn[None, :] < N)
            tl.store(remote_c_ptrs, accumulator, mask=remote_mask)
\end{tcblisting}

The consumer kernel performs the reduce-scatter communication operation. After the GEMM computation is complete, this kernel reduces the partial results across ranks and scatters them to produce the final output.

% \begin{tcblisting}{
%     listing engine=minted,
%     boxrule=0.1mm,
%     listing only,
%     left=5mm,
%     breakable,
%     skin=enhanced,
%     minted language=python,
%     minted options={
%         fontsize=\tiny,
%         breaklines, 
%         breakanywhere, 
%         autogobble,
%         linenos,
%         numbersep=2mm,
%         % 【重点】在这里定义转义符，意思是 |...| 中间的是 LaTeX 命令 
%         escapeinside=|| 
%     }
% }
\begin{tcblisting}{
    listing engine=listings,    % 引擎改为 listings
    boxrule=0.1mm,
    listing only,
    left=5mm,
    breakable,
    skin=enhanced,
    listing options={           % 原有的参数映射到这里
        language=Python,
        basicstyle=\tiny\ttfamily,
        breaklines=true,
        columns=fullflexible,   % 允许更紧凑的间距
        numbers=left,           % 对应 linenos
        numberstyle=\tiny\color{gray},
        numbersep=2mm,          % 对应 numbersep=2mm
        gobble=0,               % listings 的自动缩进处理（类似 autogobble）
        % 【重点】在这里定义 listings 的转义符
        escapeinside={|}{|},     
        showstringspaces=false
    }
}
def reduce_scatter_multi_node(input: torch.Tensor, ctx: ReduceScatter2DContext, output: Optional[torch.Tensor] = None):
    """
    A hierarchical reduce-scatter implementation that overlaps the intra-node scatter
    with the local reduce and the inter-node p2p(after reduce). It also provides a rank-wise
    signal and supports overlap with gemm.
    """
    M, N = input.shape
    M_per_rank = M // ctx.world_size

    current_stream = torch.cuda.current_stream()
    ctx.reduction_stream.wait_stream(current_stream)

    # directly reduce_scatter to output if nnodes == 1
    out_each_node = output if ctx.nnodes == 1 else None
    if not has_fullmesh_nvlink():
        rs_result_per_node = reduce_scatter_for_each_node_ring(input, ctx, out_each_node)
    else:
        rs_result_per_node = reduce_scatter_for_each_node(input, ctx, out_each_node)

    if ctx.nnodes == 1:
        return rs_result_per_node

    nvshmem_barrier_all_on_stream(current_stream)
    if output is None:
        output = torch.empty((M_per_rank, N), dtype=input.dtype, device=input.device)
    ring_reduce(rs_result_per_node, output, ctx.node_id, ctx.nnodes)
    return output

def reduce_scatter_2d_op(input: torch.Tensor, ctx: ReduceScatter2DContext, output: Optional[torch.Tensor] = None):
    M, N = input.shape
    assert input.dtype == ctx.dtype
    assert ctx.max_M >= M and ctx.N == N
    assert M % ctx.world_size == 0

    nvshmem_barrier_all_on_stream(torch.cuda.current_stream())
    output = reduce_scatter_multi_node(input, ctx, output)
    ctx.reset_barriers()
    return output
\end{tcblisting}

\subsection{GEMM All-Reduce}

Here we provide the example for GEMM All-Reduce. The following is the host-side Python function for launching the All-Reduce GEMM operation.

% \begin{tcblisting}{listing engine=minted,boxrule=0.1mm,
% listing only,left=5mm,
% breakable,
% skin=enhanced,
% minted language=python,
% minted options={fontsize=\tiny,breaklines, breakanywhere, 
% autogobble,linenos,numbersep=2mm}}
\begin{tcblisting}{
    listing engine=listings,    % 引擎改为 listings
    boxrule=0.1mm,
    listing only,
    left=5mm,
    breakable,
    skin=enhanced,
    listing options={           % 原有的参数映射到这里
        language=Python,
        basicstyle=\tiny\ttfamily,
        breaklines=true,
        columns=fullflexible,   % 允许更紧凑的间距
        numbers=left,           % 对应 linenos
        numberstyle=\tiny\color{gray},
        numbersep=2mm,          % 对应 numbersep=2mm
        gobble=0,               % listings 的自动缩进处理（类似 autogobble）
        % 【重点】在这里定义 listings 的转义符
        escapeinside={|}{|},     
        showstringspaces=false
    }
}
def gemm_allreduce_op(ctx: GemmARContext, a, b, gemm_config: triton.Config, copy_to_local=True, USE_MULTIMEM_ST=True):
    
    current_stream = torch.cuda.current_stream()
    ar_stream = ctx.ar_stream
    ar_stream.wait_stream(current_stream)
    
    M, N = a.shape[0], b.shape[0]
    # Check constraints.
    assert a.shape[1] == b.shape[1], "Incompatible dimensions"
    assert a.dtype == b.dtype, "Incompatible dtypes"
    symm_c = ctx.get_gemm_out_buf(a, b)
    symm_ar_out = ctx.symm_ar_out_buf
    gemm_barrier = ctx.gemm_barrier_buf
    multi_st_barrier = ctx.multi_st_barrier_buf
    NUM_COMM_SMS = ctx.NUM_COMM_SMS
    ar_out = torch.empty((M, N), dtype=a.dtype, device=a.device)
    BLOCK_SIZE_M = gemm_config.all_kwargs()["BLOCK_SIZE_M"]
    BLOCK_SIZE_N = gemm_config.all_kwargs()["BLOCK_SIZE_N"]
    # add mask in `consumer_all_reduce` can remove this constraint
    assert N % BLOCK_SIZE_N == 0
    persistent_gemm_notify(a, b, symm_c, gemm_barrier, gemm_config)
    with torch.cuda.stream(ar_stream):
        consumer_all_reduce(symm_c, symm_ar_out, ar_out, gemm_barrier, multi_st_barrier, BLOCK_SIZE_M=BLOCK_SIZE_M,
                           BLOCK_SIZE_N=BLOCK_SIZE_N, NUM_COMM_SMS=NUM_COMM_SMS, USE_MULTIMEM_ST=USE_MULTIMEM_ST)
    current_stream.wait_stream(ar_stream)
    if USE_MULTIMEM_ST and copy_to_local:
        ar_out.copy_(symm_ar_out.reshape(-1)[:M * N].reshape(M, N))
    nvshmem_barrier_all_on_stream(current_stream)
    if USE_MULTIMEM_ST and not copy_to_local:
        return symm_ar_out.reshape(-1)[:M * N].reshape(M, N)
    return ar_out
\end{tcblisting}

The producer kernel below is responsible for the GEMM computation. It performs computation and notifies the consumer kernels when tiles are ready for all-reduce.

% \begin{tcblisting}{
%     listing engine=minted,
%     boxrule=0.1mm,
%     listing only,
%     left=5mm,
%     breakable,
%     skin=enhanced,
%     minted language=python,
%     minted options={
%         fontsize=\tiny,
%         breaklines, 
%         breakanywhere, 
%         autogobble,
%         linenos,
%         numbersep=2mm,
%         % 【重点】在这里定义转义符，意思是 |...| 中间的是 LaTeX 命令 
%         escapeinside=|| 
%     }
% }
\begin{tcblisting}{
    listing engine=listings,    % 引擎改为 listings
    boxrule=0.1mm,
    listing only,
    left=5mm,
    breakable,
    skin=enhanced,
    listing options={           % 原有的参数映射到这里
        language=Python,
        basicstyle=\tiny\ttfamily,
        breaklines=true,
        columns=fullflexible,   % 允许更紧凑的间距
        numbers=left,           % 对应 linenos
        numberstyle=\tiny\color{gray},
        numbersep=2mm,          % 对应 numbersep=2mm
        gobble=0,               % listings 的自动缩进处理（类似 autogobble）
        % 【重点】在这里定义 listings 的转义符
        escapeinside={|}{|},     
        showstringspaces=false
    }
}
@triton.jit(do_not_specialize=[])
def kernel_persistent_gemm_notify(a_ptr, b_ptr, c_ptr, gemm_barrier_ptr,  #
                                  M, N, K,  #
                                  stride_am, stride_ak,  #
                                  stride_bn, stride_bk,  #
                                  stride_cm, stride_cn,  #
                                  BLOCK_SIZE_M: tl.constexpr,  #
                                  BLOCK_SIZE_N: tl.constexpr,  #
                                  BLOCK_SIZE_K: tl.constexpr,  #
                                  GROUP_SIZE_M: tl.constexpr,  #
                                  NUM_GEMM_SMS: tl.constexpr,  #
                                  ):
    start_pid = tl.program_id(axis=0)
    num_pid_m = tl.cdiv(M, BLOCK_SIZE_M)
    num_pid_n = tl.cdiv(N, BLOCK_SIZE_N)
    k_tiles = tl.cdiv(K, BLOCK_SIZE_K)
    num_tiles = num_pid_m * num_pid_n
    
    tile_id_c = start_pid - NUM_GEMM_SMS
    
    offs_k_for_mask = tl.arange(0, BLOCK_SIZE_K)
    num_pid_in_group = GROUP_SIZE_M * num_pid_n
    
    for tile_id in tl.range(start_pid, num_tiles, NUM_GEMM_SMS):
        pid_m, pid_n = _compute_pid(tile_id, num_pid_in_group, num_pid_m, GROUP_SIZE_M, NUM_GEMM_SMS)
        start_m = pid_m * BLOCK_SIZE_M
        start_n = pid_n * BLOCK_SIZE_N
        offs_am = start_m + tl.arange(0, BLOCK_SIZE_M)
        offs_bn = start_n + tl.arange(0, BLOCK_SIZE_N)
        offs_am = tl.where(offs_am < M, offs_am, 0)
        offs_bn = tl.where(offs_bn < N, offs_bn, 0)
        offs_am = tl.max_contiguous(tl.multiple_of(offs_am, BLOCK_SIZE_M), BLOCK_SIZE_M)
        offs_bn = tl.max_contiguous(tl.multiple_of(offs_bn, BLOCK_SIZE_N), BLOCK_SIZE_N)
        
        accumulator = tl.zeros((BLOCK_SIZE_M, BLOCK_SIZE_N), dtype=tl.float32)
        for ki in range(k_tiles):
            offs_k = ki * BLOCK_SIZE_K + tl.arange(0, BLOCK_SIZE_K)
            a_ptrs = a_ptr + (offs_am[:, None] * stride_am + offs_k[None, :] * stride_ak)
            b_ptrs = b_ptr + (offs_bn[:, None] * stride_bn + offs_k[None, :] * stride_bk)
            
            a = tl.load(a_ptrs, mask=offs_k_for_mask[None, :] < K - ki * BLOCK_SIZE_K, other=0.0)
            b = tl.load(b_ptrs, mask=offs_k_for_mask[None, :] < K - ki * BLOCK_SIZE_K, other=0.0)
            accumulator = tl.dot(a, b.T, accumulator)
        
        tile_id_c += NUM_GEMM_SMS
        pid_m, pid_n = _compute_pid(tile_id_c, num_pid_in_group, num_pid_m, GROUP_SIZE_M, NUM_GEMM_SMS)
        offs_cm = pid_m * BLOCK_SIZE_M + tl.arange(0, BLOCK_SIZE_M)
        offs_cn = pid_n * BLOCK_SIZE_N + tl.arange(0, BLOCK_SIZE_N)
        c_ptrs = c_ptr + stride_cm * offs_cm[:, None] + stride_cn * offs_cn[None, :]
        c_mask = (offs_cm[:, None] < M) & (offs_cn[None, :] < N)
        c = accumulator.to(c_ptr.dtype.element_ty)
        tl.store(c_ptrs, c, mask=c_mask)
        __syncthreads()
        
        thread_idx = tid(0)
        gemm_barrier_idx = pid_m * num_pid_n + pid_n
        if thread_idx == 0:
            |\hlcode{st(gemm\_barrier\_ptr + gemm\_barrier\_idx, 1, scope=\textquotesingle gpu \textquotesingle, semantic=\textquotesingle release \textquotesingle)}|
\end{tcblisting}

The consumer kernel performs the all-reduce operation. It waits for GEMM tiles to be ready, then performs reduction across all ranks.

% \begin{tcblisting}{
%     listing engine=minted,
%     boxrule=0.1mm,
%     listing only,
%     left=5mm,
%     breakable,
%     skin=enhanced,
%     minted language=python,
%     minted options={
%         fontsize=\tiny,
%         breaklines, 
%         breakanywhere, 
%         autogobble,
%         linenos,
%         numbersep=2mm,
%         % 【重点】在这里定义转义符，意思是 |...| 中间的是 LaTeX 命令 
%         escapeinside=|| 
%     }
% }
\begin{tcblisting}{
    listing engine=listings,    % 引擎改为 listings
    boxrule=0.1mm,
    listing only,
    left=5mm,
    breakable,
    skin=enhanced,
    listing options={           % 原有的参数映射到这里
        language=Python,
        basicstyle=\tiny\ttfamily,
        breaklines=true,
        columns=fullflexible,   % 允许更紧凑的间距
        numbers=left,           % 对应 linenos
        numberstyle=\tiny\color{gray},
        numbersep=2mm,          % 对应 numbersep=2mm
        gobble=0,               % listings 的自动缩进处理（类似 autogobble）
        % 【重点】在这里定义 listings 的转义符
        escapeinside={|}{|},     
        showstringspaces=false
    }
}
@triton.jit(do_not_specialize=[])
def consumer_all_reduce_kernel(
    symm_input_ptr,
    symm_ar_out_ptr,
    ar_out_ptr,  #
    gemm_barrier_ptr,
    multi_st_barrier_ptr,  #
    M,
    N,  #
    BLOCK_SIZE_M: tl.constexpr,  #
    BLOCK_SIZE_N: tl.constexpr,  #
    NUM_COMM_SMS: tl.constexpr,
    USE_MULTIMEM_ST: tl.constexpr,
):
    |\hlcode{rank = dl.rank()}|
    |\hlcode{world\_size = dl.num\_ranks()}|
    pid = tl.program_id(0)
    num_pid_m = tl.cdiv(M, BLOCK_SIZE_M)
    num_pid_n = tl.cdiv(N, BLOCK_SIZE_N)
    num_tiles = num_pid_m * num_pid_n
    thread_idx = tid(0)
    block_dim = num_warps() * 32
    VEC_SIZE: tl.constexpr = 128 // tl.constexpr(symm_input_ptr.dtype.element_ty.primitive_bitwidth)

    tl.static_assert(BLOCK_SIZE_N % VEC_SIZE == 0)
    VEC_PER_ROW = BLOCK_SIZE_N // VEC_SIZE
    |\hlcode{src\_data\_mc\_ptr = libshmem\_device.remote\_mc\_ptr(libshmem\_device.NVSHMEMX\_TEAM\_NODE, symm\_input\_ptr)}|
    if not USE_MULTIMEM_ST:
        for tile_id in range(pid, num_tiles, NUM_COMM_SMS):
            pid_m = tile_id // num_pid_n
            pid_n = tile_id % num_pid_n
            if thread_idx < world_size:
                |\hlcode{peer\_gemm\_barrier\_ptr = dl.symm\_at(gemm\_barrier\_ptr, thread\_idx)}|
                |\hlcode{while ld(peer\_gemm\_barrier\_ptr + tile\_id, scope=\textquotesingle sys \textquotesingle, semantic=\textquotesingle acquire \textquotesingle) != 1:}|
                    pass
            __syncthreads()
            tile_m = min(M - pid_m * BLOCK_SIZE_M, BLOCK_SIZE_M)
            cur_tile_nelem = tile_m * BLOCK_SIZE_N
            for idx in range(thread_idx, cur_tile_nelem // VEC_SIZE, block_dim):
                row_id = idx // VEC_PER_ROW
                col_id = idx % VEC_PER_ROW
                offset = (row_id + pid_m * BLOCK_SIZE_M) * N + col_id * VEC_SIZE + pid_n * BLOCK_SIZE_N
                val0, val1, val2, val3 = multimem_ld_reduce_v4(src_data_mc_ptr + offset)
                st_v4_b32(ar_out_ptr + offset, val0, val1, val2, val3)
    else:
        |\hlcode{symm\_out\_mc\_ptr = libshmem\_device.remote\_mc\_ptr(libshmem\_device.NVSHMEMX\_TEAM\_NODE, symm\_ar\_out\_ptr)}|
        for tile_id in range(pid + rank * NUM_COMM_SMS, num_tiles, NUM_COMM_SMS * world_size):
            pid_m = tile_id // num_pid_n
            pid_n = tile_id % num_pid_n
            if thread_idx < world_size:
                |\hlcode{peer\_gemm\_barrier\_ptr = dl.symm\_at(gemm\_barrier\_ptr, thread\_idx)}|
                while ld(peer_gemm_barrier_ptr + tile_id, scope="sys", semantic="acquire") != 1:
                    pass
            __syncthreads()
            
            tile_m = min(M - pid_m * BLOCK_SIZE_M, BLOCK_SIZE_M)
            cur_tile_nelem = tile_m * BLOCK_SIZE_N
            for idx in range(thread_idx, cur_tile_nelem // VEC_SIZE, block_dim):
                row_id = idx // VEC_PER_ROW
                col_id = idx % VEC_PER_ROW
                offset = (row_id + pid_m * BLOCK_SIZE_M) * N + col_id * VEC_SIZE + pid_n * BLOCK_SIZE_N
                val0, val1, val2, val3 = multimem_ld_reduce_v4(src_data_mc_ptr + offset)
                multimem_st_v4(symm_out_mc_ptr + offset, val0, val1, val2, val3)
        __syncthreads()
        
        # barrier on all blocks with same pid
        # 0. set barrier to all blocks with same pid on all peer ranks
        if thread_idx < world_size:
            |\hlcode{peer\_ptr = dl.symm\_at(multi\_st\_barrier\_ptr, thread\_idx)}|
            |\hlcode{st(peer\_ptr + rank * NUM\_COMM\_SMS + pid, 1, scope=\textquotesingle sys \textquotesingle, semantic=\textquotesingle release \textquotesingle)}|
        
        # 1. wait barrier
        if thread_idx < world_size:
            multi_st_barrier_idx = thread_idx * NUM_COMM_SMS + pid
            |\hlcode{while ld(multi\_st\_barrier\_ptr + multi\_st\_barrier\_idx, scope=\textquotesingle sys \textquotesingle, semantic=\textquotesingle acquire \textquotesingle) != 1:}|
                pass
            |\hlcode{st(multi\_st\_barrier\_ptr + multi\_st\_barrier\_idx, 0)}|
    
    # Each block can safely reset the part of the barriers it is waiting for.
    # In low latency kernel, we ensure that the gemm barriers used for two consecutive iteration are different,
    # it can be reset without any sync.
    for tile_id in range(pid + rank * NUM_COMM_SMS, num_tiles, NUM_COMM_SMS * world_size):
        peer_gemm_barrier_ptr = dl.symm_at(gemm_barrier_ptr, thread_idx)
        if thread_idx < world_size:
            |\hlcode{st(peer\_gemm\_barrier\_ptr + tile\_id, 0, scope=\textquotesingle sys \textquotesingle, semantic=\textquotesingle relaxed \textquotesingle)}|
\end{tcblisting}

\subsection{Register MegaKernel Task}
\label{appendix:code-mega}

In this section, we use linear (GEMM) as an example to show how to register kernels as tasks.
The following code contains a tile-level GEMM kernel, a chunk-level (multiple tiles) GEMM kernel, and an interface kernel to keep unified function interface for all various tasks.
These code are written by users and only minor modification is required to change users' original kernels.
\begin{tcblisting}{
listing engine=listings,    % 切换引擎
    boxrule=0.1mm,
    listing only,
    left=5mm,
    breakable,
    skin=enhanced,
    % 原有的 minted options 全部合并到这里或由上面的 lstset 控制
    listing options={
        language=Python,
        basicstyle=\tiny\ttfamily,
        breaklines=true
    }
}
@triton.jit
def tile_wise_matmul_compute(tile_id, a_ptr, b_ptr, c_ptr, M, N, K, BLOCK_SIZE_M, BLOCK_SIZE_N, BLOCK_SIZE_K,
                             NUM_STAGES):
    # linear: a (M, K) x b (N, K) -> c (M, N)
    num_pid_n = tl.cdiv(N, BLOCK_SIZE_N)
    k_tiles = tl.cdiv(K, BLOCK_SIZE_K)

    offs_k_for_mask = tl.arange(0, BLOCK_SIZE_K)

    pid_m = tile_id // num_pid_n
    pid_n = tile_id % num_pid_n
    start_m = pid_m * BLOCK_SIZE_M
    start_n = pid_n * BLOCK_SIZE_N
    offs_am = start_m + tl.arange(0, BLOCK_SIZE_M)
    offs_bn = start_n + tl.arange(0, BLOCK_SIZE_N)
    offs_am = tl.where(offs_am < M, offs_am, 0)
    offs_bn = tl.where(offs_bn < N, offs_bn, 0)
    offs_am = tl.max_contiguous(tl.multiple_of(offs_am, BLOCK_SIZE_M), BLOCK_SIZE_M)
    offs_bn = tl.max_contiguous(tl.multiple_of(offs_bn, BLOCK_SIZE_N), BLOCK_SIZE_N)
    accumulator = tl.zeros((BLOCK_SIZE_M, BLOCK_SIZE_N), dtype=tl.float32)
    for ki in tl.range(0, k_tiles, num_stages=NUM_STAGES):
        offs_k = ki * BLOCK_SIZE_K + tl.arange(0, BLOCK_SIZE_K)
        a_ptrs = a_ptr + (offs_am[:, None] * K + offs_k[None, :])
        b_ptrs = b_ptr + (offs_bn[:, None] * K + offs_k[None, :])

        a = tl.load(a_ptrs, mask=offs_k_for_mask[None, :] < K - ki * BLOCK_SIZE_K, other=0.0)
        b = tl.load(b_ptrs, mask=offs_k_for_mask[None, :] < K - ki * BLOCK_SIZE_K, other=0.0)
        accumulator = tl.dot(a, b.T, accumulator)

    offs_cm = pid_m * BLOCK_SIZE_M + tl.arange(0, BLOCK_SIZE_M)
    offs_cn = pid_n * BLOCK_SIZE_N + tl.arange(0, BLOCK_SIZE_N)
    c_ptrs = c_ptr + N * offs_cm[:, None] + offs_cn[None, :]
    c_mask = (offs_cm[:, None] < M) & (offs_cn[None, :] < N)
    c = accumulator.to(c_ptr.dtype.element_ty)
    tl.store(c_ptrs, c, mask=c_mask)

@triton.jit
def tile_range_matmul_compute_and_notify(tile_start, sb_base_ptr, a_ptr, b_ptr, c_ptr, M, N, K, BLOCK_SIZE_M,
                                         BLOCK_SIZE_N, BLOCK_SIZE_K, NUM_STAGES, TILE_READY_SIGNAL, NUM_SMS):
    num_pid_m = tl.cdiv(M, BLOCK_SIZE_M)
    num_pid_n = tl.cdiv(N, BLOCK_SIZE_N)
    num_tiles = num_pid_m * num_pid_n
    for tile_id in tl.range(tile_start, num_tiles, NUM_SMS, flatten=True, warp_specialize=True):
        tile_wise_matmul_compute(tile_id, a_ptr, b_ptr, c_ptr, M, N, K, BLOCK_SIZE_M, BLOCK_SIZE_N, BLOCK_SIZE_K,
                                 NUM_STAGES)
        st(sb_base_ptr + tile_id, TILE_READY_SIGNAL, "gpu", "release")

@triton.jit
def linear_task_compute(task_base_info: TaskBaseInfo, scoreboard: Scoreboard, BLOCK_SIZE_M: tl.constexpr,
                        BLOCK_SIZE_N: tl.constexpr, BLOCK_SIZE_K: tl.constexpr, NUM_STAGES: tl.constexpr,
                        ALIGNMENT_K: tl.constexpr):

    input: TensorDesc = task_base_info.get_tensor(0)
    weight: TensorDesc = task_base_info.get_tensor(1)
    output: TensorDesc = task_base_info.get_tensor(2)

    M = input.size(0)
    K = input.size(1, ALIGNMENT_K)
    N = weight.size(0)
    a_ptr = input.data_ptr(tl.bfloat16)
    b_ptr = weight.data_ptr(tl.bfloat16)
    c_ptr = output.data_ptr(tl.bfloat16)

    tile_id = task_base_info.tile_id_or_start
    tile_wise_matmul_compute(tile_id, a_ptr, b_ptr, c_ptr, M, N, K, BLOCK_SIZE_M, BLOCK_SIZE_N, BLOCK_SIZE_K,
                             NUM_STAGES)
    scoreboard.release_tile(task_base_info, tile_id)
\end{tcblisting}

Then, we can register the kernels as tasks using the following interface. Users are required to describe the problem size, data dependency, and the tiling configurations.
\begin{tcblisting}{
listing engine=listings,    % 切换引擎
    boxrule=0.1mm,
    listing only,
    left=5mm,
    breakable,
    skin=enhanced,
    % 原有的 minted options 全部合并到这里或由上面的 lstset 控制
    listing options={
        language=Python,
        basicstyle=\tiny\ttfamily,
        breaklines=true
    }
}
@registry.register_task(op_type="linear", task_cls=LinearTask, config_factory=linear_config_factory,
                        codegen_func=codegen_linear)
class LinearTaskBuilder(TaskBuilderBase):

    @classmethod
    def get_problem_size(cls, io_tensors: List[List['torch.Tensor']], extra_params: Dict[str, Any]):
        a, b = io_tensors[0]
        M, K = a.shape
        N, K = b.shape
        return (M, N, K)

    @classmethod
    def _build_tasks_impl(cls, device_prop, layer_id: int, dependency: TaskDependency, io_tensors, extra_params,
                          tile_wise=True, config_args={}) -> List[TaskBase]:
        assert tile_wise == True  # noqa: E712
        kernel_config = cls.create_config(**config_args)
        task_id = cls.get_task_id(layer_id)
        BLOCK_SIZE_M = kernel_config.BLOCK_SIZE_M
        BLOCK_SIZE_N = kernel_config.BLOCK_SIZE_N
        M, N, K = cls.get_problem_size(io_tensors, extra_params)
        num_tiles_m = cdiv(M, BLOCK_SIZE_M)
        num_tiles_n = cdiv(N, BLOCK_SIZE_N)
        num_tiles = num_tiles_m * num_tiles_n
        x, w = io_tensors[0]
        y = io_tensors[1][0]

        num_sm = device_prop.NUM_SMS
        tasks = []
        cls.log(
            f"Linear Task: M = {M}, N = {N}, K = {K}, num_tiles = {num_tiles}, num_sm = {num_sm}, tile_wise = {tile_wise}, dependency = {dependency}, BLOCK_SIZE_M ={BLOCK_SIZE_M}, BLOCK_SIZE_N = {BLOCK_SIZE_N}"
        )
        for tm in range(num_tiles_m):
            for tn in range(num_tiles_n):
                tile_id = tm * num_tiles_n + tn
                bm = min(BLOCK_SIZE_M, M - tm * BLOCK_SIZE_M)
                bn = min(BLOCK_SIZE_N, N - tn * BLOCK_SIZE_N)
                x_desc = InputDependencyDesc(x, require_full=False, start_indices=(tm * BLOCK_SIZE_M, 0),
                                             data_sizes=(bm, K))
                w_desc = InputDependencyDesc(w, require_full=False, start_indices=(tn * BLOCK_SIZE_N, 0),
                                             data_sizes=(bn, K))
                y_desc = OutputTilingDesc(tile_sizes=(BLOCK_SIZE_M, BLOCK_SIZE_N),
                                          start_indices=(tm * BLOCK_SIZE_M, tn * BLOCK_SIZE_N))
                inputs_dep = {x: x_desc, w: w_desc}
                outs_tile_mapping = {y: y_desc}
                tasks.append(
                    cls._create_task(layer_id, task_id, tile_id, num_tiles, kernel_config, dependency, io_tensors,
                                     extra_params, inputs_dep, outs_tile_mapping))
        return tasks

    @classmethod
    def build_tasks(cls, device_prop: 'DeviceProp', layer_id: int, dependency: TaskDependency,
                    io_tensors: List[List['torch.Tensor']], extra_params: Dict[str, Any]) -> List[TaskBase]:
        return cls._build_tasks_impl(device_prop, layer_id, dependency, io_tensors, extra_params)
\end{tcblisting}

Using the registration interface, multiple tasks can be added. And \ours{} will \textbf{automatically} generate MegaKernel. Here we show the generated code. Note that the code is auto-generated, not written by users. The task-level primitives such as \textit{fetch\_task} and \textit{scoreboard.wait\_deps} are used here.

\begin{tcblisting}{
listing engine=listings,    % 切换引擎
    boxrule=0.1mm,
    listing only,
    left=5mm,
    breakable,
    skin=enhanced,
    % 原有的 minted options 全部合并到这里或由上面的 lstset 控制
    listing options={
        language=Python,
        basicstyle=\tiny\ttfamily,
        breaklines=true
    }
}
@triton.jit
def FETCH_TASK(work_queues, idx, INT_PER_TASK, NUM_SMS, MAX_NUM_TENSOR_DIMS, ENABLE_RUNTIME_SCHEDUER=False):
    sm_id = tl.program_id(axis=0)
    TASK_TYPE_OFFSET = 0
    LAYER_ID_OFFSET = 1
    TASK_ID_OFFSET = 2
    TILE_ID_OR_START_OFFSET = 3
    DEPEND_ENTRY_START_OFFSET = 4
    DEPEND_ENTRY_END_OFFSET = 5
    IO_TENSORS_OFFSET = 6

    if not ENABLE_RUNTIME_SCHEDUER:
        offset = INT_PER_TASK * NUM_SMS

        task_type = tl.load(work_queues + idx * offset + sm_id * INT_PER_TASK + TASK_TYPE_OFFSET).to(tl.int32)
        layer_id = tl.load(work_queues + idx * offset + sm_id * INT_PER_TASK + LAYER_ID_OFFSET).to(tl.int32)
        task_id = tl.load(work_queues + idx * offset + sm_id * INT_PER_TASK + TASK_ID_OFFSET).to(tl.int32)
        tile_id_or_start = tl.load(work_queues + idx * offset + sm_id * INT_PER_TASK + TILE_ID_OR_START_OFFSET).to(tl.int32)
        depend_entry_start = tl.load(work_queues + idx * offset + sm_id * INT_PER_TASK + DEPEND_ENTRY_START_OFFSET).to(tl.int32)
        depend_entry_end = tl.load(work_queues + idx * offset + sm_id * INT_PER_TASK + DEPEND_ENTRY_END_OFFSET).to(tl.int32)
        io_tensors_ptr = work_queues + idx * offset + sm_id * INT_PER_TASK + IO_TENSORS_OFFSET
    else:
        task_type = tl.load(work_queues + idx * INT_PER_TASK + TASK_TYPE_OFFSET).to(tl.int32)
        layer_id = tl.load(work_queues + idx * INT_PER_TASK + LAYER_ID_OFFSET).to(tl.int32)
        task_id = tl.load(work_queues + idx * INT_PER_TASK + TASK_ID_OFFSET).to(tl.int32)
        tile_id_or_start = tl.load(work_queues + idx * INT_PER_TASK + TILE_ID_OR_START_OFFSET).to(tl.int32)
        depend_entry_start = tl.load(work_queues + idx * INT_PER_TASK + DEPEND_ENTRY_START_OFFSET).to(tl.int32)
        depend_entry_end = tl.load(work_queues + idx * INT_PER_TASK + DEPEND_ENTRY_END_OFFSET).to(tl.int32)
        io_tensors_ptr = work_queues + idx * INT_PER_TASK + IO_TENSORS_OFFSET
    
    task_base_info = TaskBaseInfo(io_tensors_ptr, task_type, layer_id, task_id, tile_id_or_start, depend_entry_start, depend_entry_end, MAX_NUM_TENSOR_DIMS)
    return task_base_info

@triton.jit
def MEGA_TRITON_KERNEL(
    
    work_queue_start, # [1, ] int32, init with zero
    work_queues, # [MAX_INS, NUM_SMS, INS], int32
    num_tasks_per_wq, #[num_sms,]
    scoreboard_ptr,
    task_deps_ptr,  # [num_deps_entry_of_all_tasks, INT_PER_DEPS]

    INT_PER_DEPS: tl.constexpr,
    INT_PER_TASK: tl.constexpr,
    MAX_TASK_ID: tl.constexpr,
    MAX_NUM_TILES_PER_OP: tl.constexpr,
    MAX_NUM_TENSOR_DIMS: tl.constexpr,
    NUM_SMS: tl.constexpr,
    num_warps: tl.constexpr
):

    WARP_SIZE: tl.constexpr = 32
    NUM_THREADS: tl.constexpr = num_warps * WARP_SIZE
    scoreboard = Scoreboard(task_deps_ptr, INT_PER_DEPS, scoreboard_ptr, MAX_TASK_ID, MAX_NUM_TILES_PER_OP, tl.constexpr(1), NUM_THREADS)
    sm_id = tl.program_id(axis=0)

    num_tasks = tl.load(num_tasks_per_wq + sm_id)
    cur_task_idx = 0

    # early exit
    if cur_task_idx >= num_tasks:
        return

    cur_task_base_info = FETCH_TASK(work_queues, cur_task_idx, INT_PER_TASK, NUM_SMS, MAX_NUM_TENSOR_DIMS, ENABLE_RUNTIME_SCHEDUER=False)
    nxt_task_base_info = cur_task_base_info
    nxt_task_idx = cur_task_idx

    while cur_task_idx < num_tasks:

        task_type = cur_task_base_info.task_type

        # task kernel need to set signal for each tile
        
        scoreboard.wait_deps(cur_task_base_info)

        #### run task ####
        task_base_info = cur_task_base_info
        
        if task_type == 0: # RMSNormTask
            rmsnorm_task_compute(task_base_info, scoreboard, RMS_EPS=1e-06, BLOCK_SIZE_N = 2048)

        elif task_type == 6: # QKVProjTask
            fc1_task_compute(task_base_info, scoreboard, BLOCK_SIZE_M=16, BLOCK_SIZE_N=64,
                            BLOCK_SIZE_K=256, NUM_STAGES=5)

        elif task_type == 9: # QKNormRopeUpdateKVCacheTask
            rmsnorm_rope_update_kv_cache_task_compute(
                task_base_info, scoreboard, NUM_Q_HEADS=4, NUM_KV_HEADS=1, Q_HEAD_DIM=128,
                V_HEAD_DIM=128, PAGE_SIZE=1, MAX_NUM_BLOCKS_PER_SEQ=1024,
                Q_RMS_EPS=1e-06, K_RMS_EPS=1e-06
            )

        elif task_type == 2: # AttnSplitTask
            attn_gqa_fwd_batch_decode_split_kv_task_compute(
                task_base_info, scoreboard, SM_SCALE=0.08838834764831845, SOFT_CAP=0.0,
                NUM_Q_HEADS=4, NUM_KV_HEADS=1, Q_HEAD_DIM=128, V_HEAD_DIM=128, PAGE_SIZE=1,
                MAX_NUM_BLOCKS_PER_SEQ=1024, BLOCK_N=64, BLOCK_HEAD_DIM=128,
                BLOCK_DPE=0, BLOCK_DV=128, BLOCK_H=16, NUM_KV_SPLITS=32
            )

        elif task_type == 12: # AttnCombineTask
            attn_gqa_fwd_batch_decode_combine_task_compute(
                task_base_info, scoreboard, NUM_Q_HEADS=4, V_HEAD_DIM=128, BLOCK_DV=128, NUM_KV_SPLITS=32
            )

        elif task_type == 4: # OProjTask
            fc1_task_compute(task_base_info, scoreboard, BLOCK_SIZE_M=16, BLOCK_SIZE_N=128,
                            BLOCK_SIZE_K=64, NUM_STAGES=7)

        elif task_type == 11: # BarrierAllIntraNodeTask
            barrier_all_intra_node_task_compute(task_base_info, scoreboard)

        elif task_type == 7: # AllReduceTask
            allreduce_task_compute(task_base_info, scoreboard, BLOCK_SIZE=1024)

        elif task_type == 8: # AddTask
            add_task_compute(task_base_info, scoreboard, BLOCK_SIZE=512)

        elif task_type == 1: # MLPFC1Task
            fc1_task_compute(task_base_info, scoreboard, BLOCK_SIZE_M=16, BLOCK_SIZE_N=64,
                            BLOCK_SIZE_K=128, NUM_STAGES=6)

        elif task_type == 3: # SiLUMulUpTask
            silu_mul_up_task_compute(task_base_info, scoreboard, BLOCK_SIZE_M=8, BLOCK_SIZE_N=128)

        elif task_type == 5: # MLPFC2Task
            fc1_task_compute(task_base_info, scoreboard, BLOCK_SIZE_M=16, BLOCK_SIZE_N=64,
                            BLOCK_SIZE_K=256, NUM_STAGES=6)

        elif task_type == 10: # LinearTask
            linear_task_compute(task_base_info, scoreboard, BLOCK_SIZE_M=16, BLOCK_SIZE_N=128,
                            BLOCK_SIZE_K=128, NUM_STAGES=4, ALIGNMENT_K=16)

        # nxt task
        
        cur_task_idx = cur_task_idx + 1
        if cur_task_idx < num_tasks:
            cur_task_base_info = FETCH_TASK(work_queues, cur_task_idx, INT_PER_TASK, NUM_SMS, MAX_NUM_TENSOR_DIMS, ENABLE_RUNTIME_SCHEDUER=False)
        
\end{tcblisting}

%% file: contents/appendix-primitives.tex
\section{Hardware Primitives}
\label{sec:appendix-primitives}

We list the unified primitives in \ours{}. These primitives are common among various hardware backends. To port \ours{} to different hardware, the developers only need to instantiate these primitives with corresponding code generation rules.

\begin{table}[h]
\centering
\begin{tabular}{|l|p{5cm}|p{6cm}|}
\hline
\textbf{Primitive Name} & \textbf{Parameters} & \textbf{Description} \\ \hline
\multicolumn{3}{|c|}{\textbf{Distributed Primitives}} \\ \hline
\texttt{wait} & \texttt{barrierPtrs, numBarriers, scope, semantic} & Waits for a signal on barrier pointer(s). \\ \hline
\texttt{consume\_token} & \texttt{value, token} & Enforces ordering dependencies using a token. \\ \hline
\texttt{rank} & \texttt{axis=-1} & Returns the process rank. \\ \hline
\texttt{num\_ranks} & \texttt{axis=-1} & Returns the total number of ranks. \\ \hline
\texttt{symm\_at} & \texttt{ptr, rank} & Returns a pointer to symmetric memory on a remote rank. \\ \hline
\texttt{notify} & \texttt{ptr, rank, signal, ...} & Sends a signal to a remote rank. \\ \hline

\multicolumn{3}{|c|}{\textbf{SIMT Primitives}} \\ \hline
\texttt{simt\_exec\_region} & \textit{None} & Context manager for SIMT execution. \\ \hline
\texttt{extract} & \texttt{input, indices} & Extracts a scalar from a tensor. \\ \hline
\texttt{insert} & \texttt{input, scalar, indices} & Inserts a scalar into a tensor. \\ \hline
\texttt{vector} & \texttt{args} & Creates a vector from values. \\ \hline
\texttt{zeros\_vector} & \texttt{vec\_size, dtype} & Creates a zero-filled vector. \\ \hline

\texttt{tid} & \texttt{axis} & Thread ID (0=x, 1=y, 2=z). \\ \hline
\texttt{ntid} & \texttt{axis} & Block dimension (number of threads). \\ \hline
\texttt{laneid} & \textit{None} & Lane ID within a warp. \\ \hline
\texttt{\_\_syncthreads} & \textit{None} & Synchronizes threads in a block. \\ \hline
\texttt{atomic\_cas} & \texttt{ptr, cmp, val, ...} & Atomic Compare-And-Swap. \\ \hline
\texttt{atomic\_add} & \texttt{ptr, val, ...} & Atomic Add. \\ \hline
\texttt{ld} / \texttt{st} & \texttt{ptr, val, scope, semantic} & Load/Store with specific memory semantics. \\ \hline
\texttt{smid} & \textit{None} & Streaming Multiprocessor ID. \\ \hline
\multicolumn{3}{|c|}{\textbf{SHMEM Device Primitives (NVSHMEM/ROCSHMEM)}} \\ \hline
\texttt{my\_pe} & \textit{None} & Returns current PE (Processing Element) ID. \\ \hline
\texttt{n\_pes} & \textit{None} & Returns total number of PEs. \\ \hline
\texttt{remote\_ptr} & \texttt{local\_ptr, pe} & Resolves a pointer on a remote PE. \\ \hline
\texttt{barrier\_all} & \textit{None} & Synchronizes all PEs. \\ \hline
\texttt{sync\_all} & \textit{None} & Synchronizes all PEs (lighter than barrier). \\ \hline
\texttt{fence} & \textit{None} & Ensures ordering of memory operations. \\ \hline
\texttt{putmem} & \texttt{dest, source, bytes, pe} & Writes data to remote memory (Blocking). \\ \hline
\texttt{putmem\_nbi} & \texttt{dest, source, bytes, pe} & Writes data to remote memory (Non-Blocking). \\ \hline
\texttt{getmem} & \texttt{dest, source, bytes, pe} & Reads data from remote memory (Blocking). \\ \hline
\texttt{getmem\_nbi} & \texttt{dest, source, bytes, pe} & Reads data from remote memory (Non-Blocking). \\ \hline
\texttt{putmem\_signal} & \texttt{dest, src, bytes, sig, ...} & Writes data and signals remote PE. \\ \hline
\end{tabular}
\caption{Unified Primitives in \ours{}}
\label{tab:triton_dist_full_api}
\end{table}

%% file: paper.bbl
\begin{thebibliography}{33}
\providecommand{\natexlab}[1]{#1}
\providecommand{\url}[1]{\texttt{#1}}
\expandafter\ifx\csname urlstyle\endcsname\relax
  \providecommand{\doi}[1]{doi: #1}\else
  \providecommand{\doi}{doi: \begingroup \urlstyle{rm}\Url}\fi

\bibitem[AMD()]{rocshmem}
AMD.
\newblock URL \url{https://github.com/ROCm/rocSHMEM}.

\bibitem[Anthropic(2024)]{claude}
Anthropic.
\newblock Claude 3.5 sonnet.
\newblock \url{https://www.anthropic.com/news/claude-3-5-sonnet}, 2024.

\bibitem[Awad et~al.(2025)Awad, Osama, and Potter]{iris}
Muhammad Awad, Muhammad Osama, and Brandon Potter.
\newblock Iris: First-class multi-{GPU} programming experience in {Triton}, 2025.

\bibitem[Chang et~al.(2024)Chang, Bao, Hou, Jiang, Zheng, Zhong, Zhang, Song, Jiang, Lin, Jin, and Liu]{flux}
Li{-}Wen Chang, Wenlei Bao, Qi~Hou, Chengquan Jiang, Ningxin Zheng, Yinmin Zhong, Xuanrun Zhang, Zuquan Song, Ziheng Jiang, Haibin Lin, Xin Jin, and Xin Liu.
\newblock {FLUX:} fast software-based communication overlap on gpus through kernel fusion.
\newblock \emph{CoRR}, abs/2406.06858, 2024.
\newblock \doi{10.48550/ARXIV.2406.06858}.
\newblock URL \url{https://doi.org/10.48550/arXiv.2406.06858}.

\bibitem[Comanici et~al.(2025)Comanici, Bieber, Schaekermann, Pasupat, Sachdeva, Dhillon, Blistein, Ram, Zhang, Rosen, et~al.]{gemini}
Gheorghe Comanici, Eric Bieber, Mike Schaekermann, Ice Pasupat, Noveen Sachdeva, Inderjit Dhillon, Marcel Blistein, Ori Ram, Dan Zhang, Evan Rosen, et~al.
\newblock Gemini 2.5: Pushing the frontier with advanced reasoning, multimodality, long context, and next generation agentic capabilities.
\newblock \emph{arXiv preprint arXiv:2507.06261}, 2025.

\bibitem[DeepSeek{-}AI et~al.(2024{\natexlab{a}})DeepSeek{-}AI, Liu, Feng, Wang, Wang, Liu, Zhao, Deng, Ruan, Dai, Guo, Yang, Chen, Ji, Li, Lin, Luo, Hao, Chen, Li, Zhang, Xu, Yang, Zhang, Ding, Xin, Gao, Li, Qu, Cai, Liang, Guo, Ni, Li, Chen, Yuan, Qiu, Song, Dong, Gao, Guan, Wang, Zhang, Xu, Xia, Zhao, Zhang, Li, Wang, Zhang, Zhang, Tang, Li, Tian, Huang, Wang, Zhang, Zhu, Chen, Du, Chen, Jin, Ge, Pan, Xu, Chen, Li, Lu, Zhou, Chen, Wu, Ye, Ma, Wang, Zhou, Yu, Zhou, Zheng, Wang, Pei, Yuan, Sun, Xiao, Zeng, An, Liu, Liang, Gao, Zhang, Li, Jin, Wang, Bi, Liu, Wang, Shen, Chen, Chen, Nie, and Sun]{deepseek-v2}
DeepSeek{-}AI, Aixin Liu, Bei Feng, Bin Wang, Bingxuan Wang, Bo~Liu, Chenggang Zhao, Chengqi Deng, Chong Ruan, Damai Dai, Daya Guo, Dejian Yang, Deli Chen, Dongjie Ji, Erhang Li, Fangyun Lin, Fuli Luo, Guangbo Hao, Guanting Chen, Guowei Li, Hao Zhang, Hanwei Xu, Hao Yang, Haowei Zhang, Honghui Ding, Huajian Xin, Huazuo Gao, Hui Li, Hui Qu, J.~L. Cai, Jian Liang, Jianzhong Guo, Jiaqi Ni, Jiashi Li, Jin Chen, Jingyang Yuan, Junjie Qiu, Junxiao Song, Kai Dong, Kaige Gao, Kang Guan, Lean Wang, Lecong Zhang, Lei Xu, Leyi Xia, Liang Zhao, Liyue Zhang, Meng Li, Miaojun Wang, Mingchuan Zhang, Minghua Zhang, Minghui Tang, Mingming Li, Ning Tian, Panpan Huang, Peiyi Wang, Peng Zhang, Qihao Zhu, Qinyu Chen, Qiushi Du, R.~J. Chen, R.~L. Jin, Ruiqi Ge, Ruizhe Pan, Runxin Xu, Ruyi Chen, S.~S. Li, Shanghao Lu, Shangyan Zhou, Shanhuang Chen, Shaoqing Wu, Shengfeng Ye, Shirong Ma, Shiyu Wang, Shuang Zhou, Shuiping Yu, Shunfeng Zhou, Size Zheng, Tao Wang, Tian Pei, Tian Yuan, Tianyu Sun, W.~L. Xiao, Wangding Zeng, Wei An, Wen
  Liu, Wenfeng Liang, Wenjun Gao, Wentao Zhang, X.~Q. Li, Xiangyue Jin, Xianzu Wang, Xiao Bi, Xiaodong Liu, Xiaohan Wang, Xiaojin Shen, Xiaokang Chen, Xiaosha Chen, Xiaotao Nie, and Xiaowen Sun.
\newblock Deepseek-v2: {A} strong, economical, and efficient mixture-of-experts language model.
\newblock \emph{CoRR}, abs/2405.04434, 2024{\natexlab{a}}.
\newblock \doi{10.48550/ARXIV.2405.04434}.
\newblock URL \url{https://doi.org/10.48550/arXiv.2405.04434}.

\bibitem[DeepSeek{-}AI et~al.(2024{\natexlab{b}})DeepSeek{-}AI, Liu, Feng, Xue, Wang, Wu, Lu, Zhao, Deng, Zhang, Ruan, Dai, Guo, Yang, Chen, Ji, Li, Lin, Dai, Luo, Hao, Chen, Li, Zhang, Bao, Xu, Wang, Zhang, Ding, Xin, Gao, Li, Qu, Cai, Liang, Guo, Ni, Li, Wang, Chen, Chen, Yuan, Qiu, Li, Song, Dong, Hu, Gao, Guan, Huang, Yu, Wang, Zhang, Xu, Xia, Zhao, Wang, Zhang, Li, Wang, Zhang, Zhang, Tang, Li, Tian, Huang, Wang, Zhang, Wang, Zhu, Chen, Du, Chen, Jin, Ge, Zhang, Pan, Wang, Xu, Zhang, Chen, Li, Lu, Zhou, Chen, Wu, Ye, Ye, Ma, Wang, Zhou, Yu, Zhou, Pan, Wang, Yun, Pei, Sun, Xiao, and Zeng]{deepseek-v3}
DeepSeek{-}AI, Aixin Liu, Bei Feng, Bing Xue, Bingxuan Wang, Bochao Wu, Chengda Lu, Chenggang Zhao, Chengqi Deng, Chenyu Zhang, Chong Ruan, Damai Dai, Daya Guo, Dejian Yang, Deli Chen, Dongjie Ji, Erhang Li, Fangyun Lin, Fucong Dai, Fuli Luo, Guangbo Hao, Guanting Chen, Guowei Li, H.~Zhang, Han Bao, Hanwei Xu, Haocheng Wang, Haowei Zhang, Honghui Ding, Huajian Xin, Huazuo Gao, Hui Li, Hui Qu, J.~L. Cai, Jian Liang, Jianzhong Guo, Jiaqi Ni, Jiashi Li, Jiawei Wang, Jin Chen, Jingchang Chen, Jingyang Yuan, Junjie Qiu, Junlong Li, Junxiao Song, Kai Dong, Kai Hu, Kaige Gao, Kang Guan, Kexin Huang, Kuai Yu, Lean Wang, Lecong Zhang, Lei Xu, Leyi Xia, Liang Zhao, Litong Wang, Liyue Zhang, Meng Li, Miaojun Wang, Mingchuan Zhang, Minghua Zhang, Minghui Tang, Mingming Li, Ning Tian, Panpan Huang, Peiyi Wang, Peng Zhang, Qiancheng Wang, Qihao Zhu, Qinyu Chen, Qiushi Du, R.~J. Chen, R.~L. Jin, Ruiqi Ge, Ruisong Zhang, Ruizhe Pan, Runji Wang, Runxin Xu, Ruoyu Zhang, Ruyi Chen, S.~S. Li, Shanghao Lu, Shangyan Zhou,
  Shanhuang Chen, Shaoqing Wu, Shengfeng Ye, Shengfeng Ye, Shirong Ma, Shiyu Wang, Shuang Zhou, Shuiping Yu, Shunfeng Zhou, Shuting Pan, T.~Wang, Tao Yun, Tian Pei, Tianyu Sun, W.~L. Xiao, and Wangding Zeng.
\newblock Deepseek-v3 technical report.
\newblock \emph{CoRR}, abs/2412.19437, 2024{\natexlab{b}}.
\newblock \doi{10.48550/ARXIV.2412.19437}.
\newblock URL \url{https://doi.org/10.48550/arXiv.2412.19437}.

\bibitem[Dubey et~al.(2024)Dubey, Jauhri, Pandey, Kadian, Al{-}Dahle, Letman, Mathur, Schelten, Yang, Fan, Goyal, Hartshorn, Yang, Mitra, Sravankumar, Korenev, Hinsvark, Rao, Zhang, Rodriguez, Gregerson, Spataru, Rozi{\`{e}}re, Biron, Tang, Chern, Caucheteux, Nayak, Bi, Marra, McConnell, Keller, Touret, Wu, Wong, Ferrer, Nikolaidis, Allonsius, Song, Pintz, Livshits, Esiobu, Choudhary, Mahajan, Garcia{-}Olano, Perino, Hupkes, Lakomkin, AlBadawy, Lobanova, Dinan, Smith, Radenovic, Zhang, Synnaeve, Lee, Anderson, Nail, Mialon, Pang, Cucurell, Nguyen, Korevaar, Xu, Touvron, Zarov, Ibarra, Kloumann, Misra, Evtimov, Copet, Lee, Geffert, Vranes, Park, Mahadeokar, Shah, van~der Linde, Billock, Hong, Lee, Fu, Chi, Huang, Liu, Wang, Yu, Bitton, Spisak, Park, Rocca, Johnstun, Saxe, Jia, Alwala, Upasani, Plawiak, Li, Heafield, Stone, and et~al.]{llama3}
Abhimanyu Dubey, Abhinav Jauhri, Abhinav Pandey, Abhishek Kadian, Ahmad Al{-}Dahle, Aiesha Letman, Akhil Mathur, Alan Schelten, Amy Yang, Angela Fan, Anirudh Goyal, Anthony Hartshorn, Aobo Yang, Archi Mitra, Archie Sravankumar, Artem Korenev, Arthur Hinsvark, Arun Rao, Aston Zhang, Aur{\'{e}}lien Rodriguez, Austen Gregerson, Ava Spataru, Baptiste Rozi{\`{e}}re, Bethany Biron, Binh Tang, Bobbie Chern, Charlotte Caucheteux, Chaya Nayak, Chloe Bi, Chris Marra, Chris McConnell, Christian Keller, Christophe Touret, Chunyang Wu, Corinne Wong, Cristian~Canton Ferrer, Cyrus Nikolaidis, Damien Allonsius, Daniel Song, Danielle Pintz, Danny Livshits, David Esiobu, Dhruv Choudhary, Dhruv Mahajan, Diego Garcia{-}Olano, Diego Perino, Dieuwke Hupkes, Egor Lakomkin, Ehab AlBadawy, Elina Lobanova, Emily Dinan, Eric~Michael Smith, Filip Radenovic, Frank Zhang, Gabriel Synnaeve, Gabrielle Lee, Georgia~Lewis Anderson, Graeme Nail, Gr{\'{e}}goire Mialon, Guan Pang, Guillem Cucurell, Hailey Nguyen, Hannah Korevaar, Hu~Xu, Hugo
  Touvron, Iliyan Zarov, Imanol~Arrieta Ibarra, Isabel~M. Kloumann, Ishan Misra, Ivan Evtimov, Jade Copet, Jaewon Lee, Jan Geffert, Jana Vranes, Jason Park, Jay Mahadeokar, Jeet Shah, Jelmer van~der Linde, Jennifer Billock, Jenny Hong, Jenya Lee, Jeremy Fu, Jianfeng Chi, Jianyu Huang, Jiawen Liu, Jie Wang, Jiecao Yu, Joanna Bitton, Joe Spisak, Jongsoo Park, Joseph Rocca, Joshua Johnstun, Joshua Saxe, Junteng Jia, Kalyan~Vasuden Alwala, Kartikeya Upasani, Kate Plawiak, Ke~Li, Kenneth Heafield, Kevin Stone, and et~al.
\newblock The llama 3 herd of models.
\newblock \emph{CoRR}, abs/2407.21783, 2024.
\newblock \doi{10.48550/ARXIV.2407.21783}.
\newblock URL \url{https://doi.org/10.48550/arXiv.2407.21783}.

\bibitem[Gond et~al.(2025)Gond, Kwatra, and Ramjee]{tokenweave}
Raja Gond, Nipun Kwatra, and Ramachandran Ramjee.
\newblock Tokenweave: Efficient compute-communication overlap for distributed llm inference, 2025.
\newblock URL \url{https://arxiv.org/abs/2505.11329}.

\bibitem[Google(2025)]{pallas}
Google.
\newblock {Pallas}, 2025.
\newblock URL \url{https://docs.jax.dev/en/latest/pallas/index.html}.

\bibitem[Jangda et~al.(2022)Jangda, Huang, Liu, Sabet, Maleki, Miao, Musuvathi, Mytkowicz, and Saarikivi]{coconet}
Abhinav Jangda, Jun Huang, Guodong Liu, Amir Hossein~Nodehi Sabet, Saeed Maleki, Youshan Miao, Madanlal Musuvathi, Todd Mytkowicz, and Olli Saarikivi.
\newblock Breaking the computation and communication abstraction barrier in distributed machine learning workloads.
\newblock In Babak Falsafi, Michael Ferdman, Shan Lu, and Thomas~F. Wenisch, editors, \emph{{ASPLOS} '22: 27th {ACM} International Conference on Architectural Support for Programming Languages and Operating Systems, Lausanne, Switzerland, 28 February 2022 - 4 March 2022}, pages 402--416. {ACM}, 2022.
\newblock \doi{10.1145/3503222.3507778}.
\newblock URL \url{https://doi.org/10.1145/3503222.3507778}.

\bibitem[Jiang et~al.(2024)Jiang, Sablayrolles, Roux, Mensch, Savary, Bamford, Chaplot, de~Las~Casas, Hanna, Bressand, Lengyel, Bour, Lample, Lavaud, Saulnier, Lachaux, Stock, Subramanian, Yang, Antoniak, Scao, Gervet, Lavril, Wang, Lacroix, and Sayed]{mixstral}
Albert~Q. Jiang, Alexandre Sablayrolles, Antoine Roux, Arthur Mensch, Blanche Savary, Chris Bamford, Devendra~Singh Chaplot, Diego de~Las~Casas, Emma~Bou Hanna, Florian Bressand, Gianna Lengyel, Guillaume Bour, Guillaume Lample, L{\'{e}}lio~Renard Lavaud, Lucile Saulnier, Marie{-}Anne Lachaux, Pierre Stock, Sandeep Subramanian, Sophia Yang, Szymon Antoniak, Teven~Le Scao, Th{\'{e}}ophile Gervet, Thibaut Lavril, Thomas Wang, Timoth{\'{e}}e Lacroix, and William~El Sayed.
\newblock Mixtral of experts.
\newblock \emph{CoRR}, abs/2401.04088, 2024.
\newblock \doi{10.48550/ARXIV.2401.04088}.
\newblock URL \url{https://doi.org/10.48550/arXiv.2401.04088}.

\bibitem[Kwon et~al.(2023)Kwon, Li, Zhuang, Sheng, Zheng, Yu, Gonzalez, Zhang, and Stoica]{vllm}
Woosuk Kwon, Zhuohan Li, Siyuan Zhuang, Ying Sheng, Lianmin Zheng, Cody~Hao Yu, Joseph Gonzalez, Hao Zhang, and Ion Stoica.
\newblock Efficient memory management for large language model serving with pagedattention.
\newblock In Jason Flinn, Margo~I. Seltzer, Peter Druschel, Antoine Kaufmann, and Jonathan Mace, editors, \emph{Proceedings of the 29th Symposium on Operating Systems Principles, {SOSP} 2023, Koblenz, Germany, October 23-26, 2023}, pages 611--626. {ACM}, 2023.
\newblock \doi{10.1145/3600006.3613165}.
\newblock URL \url{https://doi.org/10.1145/3600006.3613165}.

\bibitem[Liao et~al.(2025)Liao, Qin, Wang, Golden, Kuchnik, Yetim, Ang, Fu, He, Hsia, Jiang, Li, Pashkevich, Puvvada, Shi, Steiner, Xiao, Yan, Yu, Fang, Zainul-Abedin, Singh, Yu, Chi, Huang, Zhang, Weller, Marine, Cook, Wu, and Liu]{kernelevolve}
Gang Liao, Hongsen Qin, Ying Wang, Alicia Golden, Michael Kuchnik, Yavuz Yetim, Jia~Jiunn Ang, Chunli Fu, Yihan He, Samuel Hsia, Zewei Jiang, Dianshi Li, Uladzimir Pashkevich, Varna Puvvada, Feng Shi, Matt Steiner, Ruichao Xiao, Nathan Yan, Xiayu Yu, Zhou Fang, Abdul Zainul-Abedin, Ketan Singh, Hongtao Yu, Wenyuan Chi, Barney Huang, Sean Zhang, Noah Weller, Zach Marine, Wyatt Cook, Carole-Jean Wu, and Gaoxiang Liu.
\newblock Kernelevolve: Scaling agentic kernel coding for heterogeneous ai accelerators at meta, 2025.
\newblock URL \url{https://arxiv.org/abs/2512.23236}.

\bibitem[Liu et~al.(2025)Liu, Su, Yao, Jiang, Lai, Du, Qin, Xu, Lu, Yan, Chen, Zheng, Liu, Liu, Yin, He, Zhu, Wang, Wang, Dong, Zhang, Kang, Zhang, Xu, Zhang, Wu, Zhou, and Yang]{muon}
Jingyuan Liu, Jianlin Su, Xingcheng Yao, Zhejun Jiang, Guokun Lai, Yulun Du, Yidao Qin, Weixin Xu, Enzhe Lu, Junjie Yan, Yanru Chen, Huabin Zheng, Yibo Liu, Shaowei Liu, Bohong Yin, Weiran He, Han Zhu, Yuzhi Wang, Jianzhou Wang, Mengnan Dong, Zheng Zhang, Yongsheng Kang, Hao Zhang, Xinran Xu, Yutao Zhang, Yuxin Wu, Xinyu Zhou, and Zhilin Yang.
\newblock Muon is scalable for {LLM} training.
\newblock \emph{CoRR}, abs/2502.16982, 2025.
\newblock \doi{10.48550/ARXIV.2502.16982}.
\newblock URL \url{https://doi.org/10.48550/arXiv.2502.16982}.

\bibitem[Narayanan et~al.(2021)Narayanan, Shoeybi, Casper, LeGresley, Patwary, Korthikanti, Vainbrand, Kashinkunti, Bernauer, Catanzaro, Phanishayee, and Zaharia]{megatron-lm}
Deepak Narayanan, Mohammad Shoeybi, Jared Casper, Patrick LeGresley, Mostofa Patwary, Vijay Korthikanti, Dmitri Vainbrand, Prethvi Kashinkunti, Julie Bernauer, Bryan Catanzaro, Amar Phanishayee, and Matei Zaharia.
\newblock Efficient large-scale language model training on {GPU} clusters using megatron-lm.
\newblock In Bronis~R. de~Supinski, Mary~W. Hall, and Todd Gamblin, editors, \emph{International Conference for High Performance Computing, Networking, Storage and Analysis, {SC} 2021, St. Louis, Missouri, USA, November 14-19, 2021}, page~58. {ACM}, 2021.
\newblock \doi{10.1145/3458817.3476209}.
\newblock URL \url{https://doi.org/10.1145/3458817.3476209}.

\bibitem[{NVIDIA}(2018)]{nvswitch}
{NVIDIA}.
\newblock Nvidia nvswitch: Technical overview.
\newblock Technical report, NVIDIA, 2018.
\newblock URL \url{https://images.nvidia.com/content/pdf/nvswitch-technical-overview.pdf}.

\bibitem[NVIDIA(2022)]{cublas}
NVIDIA.
\newblock {cuBLAS}, 2022.
\newblock URL \url{https://developer.nvidia.com/cublas}.

\bibitem[{NVIDIA}(2023)]{hopper}
{NVIDIA}.
\newblock Hopper architecture whitepaper.
\newblock Technical report, NVIDIA, 2023.
\newblock URL \url{https://resources.nvidia.com/en-us-tensor-core/gtc22-whitepaper-hopper}.

\bibitem[NVIDIA(2024)]{nccl}
NVIDIA.
\newblock Nvidia collective communications library.
\newblock \url{https://developer.nvidia.com/nccl}, 2024.

\bibitem[OpenAI(2023)]{gpt4}
OpenAI.
\newblock {GPT-4} technical report.
\newblock \emph{CoRR}, abs/2303.08774, 2023.
\newblock \doi{10.48550/ARXIV.2303.08774}.
\newblock URL \url{https://doi.org/10.48550/arXiv.2303.08774}.

\bibitem[Qwen-Team(2024)]{qwen-max}
Qwen-Team.
\newblock Qwen2.5 technical report.
\newblock \emph{arXiv preprint arXiv:2412.15115}, 2024.

\bibitem[Rivi{\`{e}}re et~al.(2024)Rivi{\`{e}}re, Pathak, Sessa, Hardin, Bhupatiraju, Hussenot, Mesnard, Shahriari, Ram{\'{e}}, Ferret, Liu, Tafti, Friesen, Casbon, Ramos, Kumar, Lan, Jerome, Tsitsulin, Vieillard, Stanczyk, Girgin, Momchev, Hoffman, Thakoor, Grill, Neyshabur, Bachem, Walton, Severyn, Parrish, Ahmad, Hutchison, Abdagic, Carl, Shen, Brock, Coenen, Laforge, Paterson, Bastian, Piot, Wu, Royal, Chen, Kumar, Perry, Welty, Choquette{-}Choo, Sinopalnikov, Weinberger, Vijaykumar, Rogozinska, Herbison, Bandy, Wang, Noland, Moreira, Senter, Eltyshev, Visin, Rasskin, Wei, Cameron, Martins, Hashemi, Klimczak{-}Plucinska, Batra, Dhand, Nardini, Mein, Zhou, Svensson, Stanway, Chan, Zhou, Carrasqueira, Iljazi, Becker, Fernandez, van Amersfoort, Gordon, Lipschultz, Newlan, Ji, Mohamed, Badola, Black, Millican, McDonell, Nguyen, Sodhia, Greene, Sj{\"{o}}sund, Usui, Sifre, Heuermann, Lago, and McNealus]{gemma2}
Morgane Rivi{\`{e}}re, Shreya Pathak, Pier~Giuseppe Sessa, Cassidy Hardin, Surya Bhupatiraju, L{\'{e}}onard Hussenot, Thomas Mesnard, Bobak Shahriari, Alexandre Ram{\'{e}}, Johan Ferret, Peter Liu, Pouya Tafti, Abe Friesen, Michelle Casbon, Sabela Ramos, Ravin Kumar, Charline~Le Lan, Sammy Jerome, Anton Tsitsulin, Nino Vieillard, Piotr Stanczyk, Sertan Girgin, Nikola Momchev, Matt Hoffman, Shantanu Thakoor, Jean{-}Bastien Grill, Behnam Neyshabur, Olivier Bachem, Alanna Walton, Aliaksei Severyn, Alicia Parrish, Aliya Ahmad, Allen Hutchison, Alvin Abdagic, Amanda Carl, Amy Shen, Andy Brock, Andy Coenen, Anthony Laforge, Antonia Paterson, Ben Bastian, Bilal Piot, Bo~Wu, Brandon Royal, Charlie Chen, Chintu Kumar, Chris Perry, Chris Welty, Christopher~A. Choquette{-}Choo, Danila Sinopalnikov, David Weinberger, Dimple Vijaykumar, Dominika Rogozinska, Dustin Herbison, Elisa Bandy, Emma Wang, Eric Noland, Erica Moreira, Evan Senter, Evgenii Eltyshev, Francesco Visin, Gabriel Rasskin, Gary Wei, Glenn Cameron, Gus
  Martins, Hadi Hashemi, Hanna Klimczak{-}Plucinska, Harleen Batra, Harsh Dhand, Ivan Nardini, Jacinda Mein, Jack Zhou, James Svensson, Jeff Stanway, Jetha Chan, Jin~Peng Zhou, Joana Carrasqueira, Joana Iljazi, Jocelyn Becker, Joe Fernandez, Joost van Amersfoort, Josh Gordon, Josh Lipschultz, Josh Newlan, Ju{-}yeong Ji, Kareem Mohamed, Kartikeya Badola, Kat Black, Katie Millican, Keelin McDonell, Kelvin Nguyen, Kiranbir Sodhia, Kish Greene, Lars~Lowe Sj{\"{o}}sund, Lauren Usui, Laurent Sifre, Lena Heuermann, Leticia Lago, and Lilly McNealus.
\newblock Gemma 2: Improving open language models at a practical size.
\newblock \emph{CoRR}, abs/2408.00118, 2024.
\newblock \doi{10.48550/ARXIV.2408.00118}.
\newblock URL \url{https://doi.org/10.48550/arXiv.2408.00118}.

\bibitem[Shah et~al.(2025)Shah, Jangda, Li, Rocha, Hwang, Jose, Musuvathi, Saarikivi, Cheng, Zhou, Dathathri, Maleki, and Yang]{mscclpp}
Aashaka Shah, Abhinav Jangda, Binyang Li, Caio Rocha, Changho Hwang, Jithin Jose, Madan Musuvathi, Olli Saarikivi, Peng Cheng, Qinghua Zhou, Roshan Dathathri, Saeed Maleki, and Ziyue Yang.
\newblock Msccl++: Rethinking gpu communication abstractions for cutting-edge ai applications, 2025.
\newblock URL \url{https://arxiv.org/abs/2504.09014}.

\bibitem[Spector et~al.(2025)Spector, Juravsky, Sul, Dugan, Lim, Fu, Arora, and R{\'e}]{hazymega}
Benjamin Spector, Jordan Juravsky, Stuart Sul, Owen Dugan, Dylan Lim, Dan Fu, Simran Arora, and Chris R{\'e}.
\newblock Look ma, no bubbles! designing a low-latency megakernel for {LLAMA}-1{B}.
\newblock \url{https://hazyresearch.stanford.edu/blog/2025-05-27-no-bubbles}, 2025.
\newblock Hazy Research Blog.

\bibitem[TileLang-Team(2025)]{tilelang}
TileLang-Team.
\newblock Tilelang, 2025.
\newblock URL \url{https://github.com/tile-ai/tilelang}.

\bibitem[Tillet et~al.(2019)Tillet, Kung, and Cox]{triton}
Philippe Tillet, Hsiang{-}Tsung Kung, and David~D. Cox.
\newblock Triton: an intermediate language and compiler for tiled neural network computations.
\newblock In Tim Mattson, Abdullah Muzahid, and Armando Solar{-}Lezama, editors, \emph{Proceedings of the 3rd {ACM} {SIGPLAN} International Workshop on Machine Learning and Programming Languages, MAPL@PLDI 2019, Phoenix, AZ, USA, June 22, 2019}, pages 10--19. {ACM}, 2019.
\newblock \doi{10.1145/3315508.3329973}.
\newblock URL \url{https://doi.org/10.1145/3315508.3329973}.

\bibitem[Wang et~al.(2023)Wang, Wei, Sabne, Davis, Ilbeyi, Hechtman, Chen, Murthy, Maggioni, Zhang, Kumar, Guo, Xu, and Zhou]{dist-enisum}
Shibo Wang, Jinliang Wei, Amit Sabne, Andy Davis, Berkin Ilbeyi, Blake Hechtman, Dehao Chen, Karthik~Srinivasa Murthy, Marcello Maggioni, Qiao Zhang, Sameer Kumar, Tongfei Guo, Yuanzhong Xu, and Zongwei Zhou.
\newblock Overlap communication with dependent computation via decomposition in large deep learning models.
\newblock In Tor~M. Aamodt, Natalie D.~Enright Jerger, and Michael~M. Swift, editors, \emph{Proceedings of the 28th {ACM} International Conference on Architectural Support for Programming Languages and Operating Systems, Volume 1, {ASPLOS} 2023, Vancouver, BC, Canada, March 25-29, 2023}, pages 93--106. {ACM}, 2023.
\newblock \doi{10.1145/3567955.3567959}.
\newblock URL \url{https://doi.org/10.1145/3567955.3567959}.

\bibitem[Wu et~al.(2025)Wu, Cheng, Liu, Shi, Ji, Ao, Velliengiri, Miao, Padon, and Jia]{mirage}
Mengdi Wu, Xinhao Cheng, Shengyu Liu, Chunan Shi, Jianan Ji, Kit Ao, Praveen Velliengiri, Xupeng Miao, Oded Padon, and Zhihao Jia.
\newblock Mirage: A multi-level superoptimizer for tensor programs.
\newblock In \emph{19th USENIX Symposium on Operating Systems Design and Implementation (OSDI 25)}, Boston, MA, July 2025. USENIX Association.

\bibitem[Yang et~al.(2025)Yang, Li, Yang, Zhang, Hui, Zheng, Yu, Gao, Huang, Lv, Zheng, Liu, Zhou, Huang, Hu, Ge, Wei, Lin, Tang, Yang, Tu, Zhang, Yang, Yang, Zhou, Zhou, Lin, Dang, Bao, Yang, Yu, Deng, Li, Xue, Li, Zhang, Wang, Zhu, Men, Gao, Liu, Luo, Li, Tang, Yin, Ren, Wang, Zhang, Ren, Fan, Su, Zhang, Zhang, Wan, Liu, Wang, Cui, Zhang, Zhou, and Qiu]{qwen3}
An~Yang, Anfeng Li, Baosong Yang, Beichen Zhang, Binyuan Hui, Bo~Zheng, Bowen Yu, Chang Gao, Chengen Huang, Chenxu Lv, Chujie Zheng, Dayiheng Liu, Fan Zhou, Fei Huang, Feng Hu, Hao Ge, Haoran Wei, Huan Lin, Jialong Tang, Jian Yang, Jianhong Tu, Jianwei Zhang, Jian Yang, Jiaxi Yang, Jingren Zhou, Jingren Zhou, Junyang Lin, Kai Dang, Keqin Bao, Kexin Yang, Le~Yu, Lianghao Deng, Mei Li, Mingfeng Xue, Mingze Li, Pei Zhang, Peng Wang, Qin Zhu, Rui Men, Ruize Gao, Shixuan Liu, Shuang Luo, Tianhao Li, Tianyi Tang, Wenbiao Yin, Xingzhang Ren, Xinyu Wang, Xinyu Zhang, Xuancheng Ren, Yang Fan, Yang Su, Yichang Zhang, Yinger Zhang, Yu~Wan, Yuqiong Liu, Zekun Wang, Zeyu Cui, Zhenru Zhang, Zhipeng Zhou, and Zihan Qiu.
\newblock Qwen3 technical report.
\newblock \emph{CoRR}, abs/2505.09388, 2025.
\newblock \doi{10.48550/ARXIV.2505.09388}.
\newblock URL \url{https://doi.org/10.48550/arXiv.2505.09388}.

\bibitem[Zhang et~al.(2025)Zhang, Zheng, Lin, Jiang, Bao, Jiang, Hou, Cui, Zheng, Chang, Chen, and Liu]{comet}
Shulai Zhang, Ningxin Zheng, Haibin Lin, Ziheng Jiang, Wenlei Bao, Chengquan Jiang, Qi~Hou, Weihao Cui, Size Zheng, Li{-}Wen Chang, Quan Chen, and Xin Liu.
\newblock Comet: Fine-grained computation-communication overlapping for mixture-of-experts.
\newblock \emph{CoRR}, abs/2502.19811, 2025.
\newblock \doi{10.48550/ARXIV.2502.19811}.
\newblock URL \url{https://doi.org/10.48550/arXiv.2502.19811}.

\bibitem[Zhao et~al.(2025)Zhao, Zhou, Zhang, Deng, Xu, Liu, Yu, Li, and Zhao]{deepep}
Chenggang Zhao, Shangyan Zhou, Liyue Zhang, Chengqi Deng, Zhean Xu, Yuxuan Liu, Kuai Yu, Jiashi Li, and Liang Zhao.
\newblock Deepep: an efficient expert-parallel communication library.
\newblock \url{https://github.com/deepseek-ai/DeepEP}, 2025.

\bibitem[Zheng et~al.(2025)Zheng, Fang, Zheng, Hou, Bao, Zheng, Jiang, Wang, Ye, Lin, Chang, and Liu]{tilelink}
Size Zheng, Jin Fang, Xuegui Zheng, Qi~Hou, Wenlei Bao, Ningxin Zheng, Ziheng Jiang, Dongyang Wang, Jianxi Ye, Haibin Lin, Li-Wen Chang, and Xin Liu.
\newblock Tilelink: Generating efficient compute-communication overlapping kernels using tile-centric primitives, 2025.
\newblock URL \url{https://arxiv.org/abs/2503.20313}.

\end{thebibliography}
